\documentclass{article}

\PassOptionsToPackage{numbers}{natbib}
\usepackage[preprint]{neurips_2021}

\usepackage[utf8]{inputenc} 
\usepackage[T1]{fontenc}    
\usepackage{booktabs}       
\usepackage{amsfonts}       
\usepackage{nicefrac}       
\usepackage{microtype}      
\usepackage{amsmath}
\usepackage{graphicx}
\usepackage{tipa}
\usepackage{multirow}
\usepackage{longtable}

\usepackage{subcaption}
\usepackage{color}
\usepackage[table]{xcolor}
\usepackage{threeparttable}
\usepackage{diagbox}
\usepackage{enumitem}
\usepackage{tablefootnote}
\usepackage{hyperref}
\usepackage{url}            
\usepackage{xurl}
\usepackage{breakurl}

\usepackage[edges]{forest}
\usepackage{tikz}
\usepackage{tikz-qtree}
\usetikzlibrary{fadings}
\usetikzlibrary{mindmap,trees}
\usetikzlibrary{arrows,automata,shapes,positioning,shadows,trees}
\usetikzlibrary{shapes,snakes,shadows}
\usetikzlibrary{shapes.arrows}
\usetikzlibrary{calc,shapes, positioning}
\usetikzlibrary{decorations.text}
\usetikzlibrary{matrix,chains,positioning,decorations.pathreplacing,arrows}
\usetikzlibrary{bayesnet}
\usetikzlibrary{shadows.blur}
\usetikzlibrary{shapes.geometric}

\tikzstyle{edge}=[-latex',draw=black!90,shorten <=1pt,shorten >=1pt]
\tikzstyle{redge}=[latex'-,draw=black!90,shorten <=1pt,shorten >=1pt]
\tikzstyle{dedge}=[latex'-latex',draw=black!90,shorten <=1pt,shorten >=1pt]
\tikzstyle{block}=[draw, text width=5em,align=center,shape=rectangle, rounded corners, , align=center]
\tikzstyle{nobox}=[align=center]

\definecolor{emb}{RGB}{209,228,252}
\definecolor{hidden-blue}{RGB}{194,232,247}
\definecolor{hidden-orange}{RGB}{243,202,120}
\definecolor{hidden-yellow}{RGB}{242,244,193}
\definecolor{output-purple}{RGB}{219,203,231}
\definecolor{output-green}{RGB}{204,231,207}
\definecolor{hiddendraw}{RGB}{205, 44, 36}

\tikzstyle{mybox}=[
    rectangle,
    draw=hiddendraw,
    rounded corners,
    text opacity=1,
    minimum height=1.5em,
    minimum width=5em,
    inner sep=2pt,
    align=center,
    fill opacity=.2,
    ]

\tikzstyle{emb-purple}=[
    rectangle,
    draw=output-purple!50!purple,
    fill=output-purple,
    text opacity=1,
    minimum height=1.5em,
    minimum width=1.5em,
    inner sep=0pt,
    align=center,
    fill opacity=.9,
    ]

\tikzstyle{emb-blue}=[
    rectangle,
    draw=emb!50!blue,
    fill=emb,
    text opacity=1,
    minimum height=1.5em,
    minimum width=1.5em,
    inner sep=0pt,
    align=center,
    fill opacity=.9,
]

\definecolor{colone}{RGB}{178, 34, 34}
\definecolor{coltwo}{RGB}{106, 90, 205}
\definecolor{colthree}{RGB}{255, 250, 205}
\definecolor{colfour}{RGB}{0, 139, 69}
\definecolor{colfive}{RGB}{245,238,197}
\definecolor{colsix}{RGB}{243,235,179}
\definecolor{colseven}{RGB}{241,231,163}
\title{A Survey on Neural Speech Synthesis}

\author{
Xu Tan\thanks{Corresponding author: Xu Tan, \texttt{xuta@microsoft.com}}, ~Tao Qin, ~Frank Soong, ~Tie-Yan Liu \\
\texttt{\{xuta,taoqin,frankkps,tyliu\}@microsoft.com} \\
Microsoft Research Asia \\
}

\begin{document}

\maketitle

\begin{abstract}
Text to speech (TTS), or speech synthesis, which aims to synthesize intelligible and natural speech given text, is a hot research topic in speech, language, and machine learning communities and has broad applications in the industry. As the development of deep learning and artificial intelligence, neural network-based TTS has significantly improved the quality of synthesized speech in recent years. In this paper, we conduct a comprehensive survey on neural TTS, aiming to provide a good understanding of current research and future trends. We focus on the key components in neural TTS, including text analysis, acoustic models, and vocoders, and several advanced topics, including fast TTS, low-resource TTS, robust TTS, expressive TTS, and adaptive TTS, etc. We further summarize resources related to TTS (e.g., datasets, opensource implementations) and discuss future research directions. This survey can serve both academic researchers and industry practitioners working on TTS. 
\end{abstract}

\section{Introduction}
Text to speech (TTS), also known as speech synthesis, which aims to synthesize intelligible and natural speech from text~\cite{taylor2009text}, has broad applications in human communication~\cite{adler1991understanding} and has long been a research topic in artificial intelligence, natural language and speech processing~\cite{russell2002artificial,manning1999foundations,jurafsky2000speech}. Developing a TTS system requires knowledge about languages and human speech production, and involves multiple disciplines including linguistics~\cite{de2011course}, acoustics~\cite{kinsler1999fundamentals}, digital signal processing~\cite{stanley1988digital}, and machine learning~\cite{bishop2006pattern,jordan2015machine}. 

As the development of deep learning~\cite{lecun2015deep,goodfellow2016deep}, neural network-based TTS has thrived, and a large amount of research work comes out focusing on different aspects of neural TTS~\cite{zen2013statistical,oord2016wavenet,wang2017tacotron,shen2018natural,kalchbrenner2018efficient,ping2018deep,li2019neural,ren2019fastspeech}. Consequently, the quality of synthesized speech has been largely improved in recent years. Understanding the current research status and figuring out unsolved research problems are very helpful for people working on TTS. While there are multiple survey papers on statistical parametric speech synthesis~\cite{black2007statistical,zen2009statistical,tokuda2013speech,zen2015acoustic} and neural TTS~\cite{tabet2011speech,mali2014survey,siddhi2017survey,ning2019review,hsu2019towards,panda2020survey,mu2021review}, a comprehensive survey on the basics and the recent developments of neural TTS is still necessary since the topics in this area are diverse and evolve quickly. In this paper, we conduct a deep and comprehensive survey on neural TTS\footnote{This survey paper is originated from our TTS tutorials, including TTS tutorial at ISCSLP 2021 (\url{https://tts-tutorial.github.io/iscslp2021/}) and TTS tutorial at IJCAI 2021 (\url{https://tts-tutorial.github.io/ijcai2021/}).} \footnote{Readers can use this Github page (\url{https://github.com/tts-tutorial/survey}) to check updates and initiate discussions on this survey paper.}. 

In the following subsections, we first briefly review the history of TTS technologies, then introduce some basic knowledge of neural TTS, and finally outline this survey.

\subsection{History of TTS Technology}
People have tried to build machines to synthesize human speech dating back to the 12th century~\cite{wiki_Speech_synthesis}. In the 2nd half of the 18th century, the Hungarian scientist, Wolfgang von Kempelen, had constructed a speaking machine with a series of bellows, springs, bagpipes and resonance boxes to produce some simple words and short sentences~\cite{dudley1950speaking}. The first speech synthesis system that built upon computer came out in the latter half of the 20th century~\cite{wiki_Speech_synthesis}. The early computer-based speech synthesis methods include articulatory synthesis~\cite{coker1976model,shadle2001prospects}, formant synthesis~\cite{seeviour1976automatic,allen1979mitalk,klatt1980software,klatt1987review}, and concatenative synthesis~\cite{olive1977rule,moulines1990pitch,sagisaka1992atr,hunt1996unit,black1998festival}. Later, as the development of statistics machine learning, statistical parametric speech synthesis (SPSS) is proposed~\cite{yoshimura1999simultaneous,tokuda2000speech,zen2009statistical,tokuda2013speech}, which predicts parameters such as spectrum, fundamental frequency and duration for speech synthesis. From 2010s, neural network-based speech synthesis~\cite{zen2013statistical,qian2014training,fan2014tts,zen2015unidirectional,wang2016first,li2018emphasis,oord2016wavenet,wang2017tacotron} has gradually become the dominant methods and achieved much better voice quality.

\paragraph{Articulatory Synthesis}
Articulatory synthesis~\cite{coker1976model,shadle2001prospects} produces speech by simulating the behavior of human articulator such as lips, tongue, glottis and moving vocal tract. Ideally, articulatory synthesis can be the most effective method for speech synthesis since it is the way how human generates speech. However, it is very difficult to model these articulator behaviors in practice. For example, it is hard to collect the data for articulator simulation. Therefore, the speech quality by articulatory synthesis is usually worse than that by later formant synthesis and concatenative synthesis.

\paragraph{Formant Synthesis}
Formant synthesis~\cite{seeviour1976automatic,allen1979mitalk,klatt1980software} produces speech based on a set of rules that control a simplified source-filter model. These rules are usually developed by linguists to mimic the formant structure and other spectral properties of speech as closely as possible. The speech is synthesized by an additive synthesis module and an acoustic model with varying parameters like fundamental frequency, voicing, and noise levels. The formant synthesis can produce highly intelligible speech with moderate computation resources that are well-suited for embedded systems, and does not rely on large-scale human speech corpus as in concatenative synthesis. However, the synthesized speech sounds less natural and has artifacts. Moreover, it is difficult to specify rules for synthesis. 

\paragraph{Concatenative Synthesis}
Concatenative synthesis~\cite{olive1977rule,moulines1990pitch,sagisaka1992atr,hunt1996unit,black1998festival} relies on the concatenation of pieces of speech that are stored in a database. Usually, the database consists of speech units ranging from whole sentence to syllables that are recorded by voice actors. In inference, the concatenative TTS system searches speech units to match the given input text, and produces speech waveform by concatenating these units together. Generally speaking, concatenative TTS can generate audio with high intelligibility and authentic timbre close to the original voice actor. However, concatenative TTS requires huge recording database in order to cover all possible combinations of speech units for spoken words. Another drawback is that the generated voice is less natural and emotional, since concatenation can result in less smoothness in stress, emotion, prosody, etc.

\paragraph{Statistical Parametric Synthesis}
To address the drawbacks of concatenative TTS, statistical parametric speech synthesis (SPSS) is proposed~\cite{yoshimura1999simultaneous,tokuda2000speech,yoshimura2002simultaneous,zen2009statistical,tokuda2013speech}. The basic idea is that instead of direct generating waveform through concatenation, we can first generate the acoustic parameters~\cite{fukada1992adaptive,tokuda1994mel,kawahara1999restructuring} that are necessary to produce speech and then recover speech from the generated acoustic parameters using some algorithms~\cite{imai1983mel,imai1983cepstral,kawahara2006straight,morise2016world}. SPSS usually consists of three components: a text analysis module, a parameter prediction module (acoustic model), and a vocoder analysis/synthesis module (vocoder). The text analysis module first processes the text, including text normalization~\cite{sproat2001normalization}, grapheme-to-phoneme conversion~\cite{bisani2008joint}, word segmentation, etc, and then extracts the linguistic features, such as phonemes, duration and POS tags from different granularities. The acoustic models (e.g., hidden Markov model (HMM) based) are trained with the paired linguistic features and parameters (acoustic features), where the acoustic features include fundamental frequency, spectrum or cepstrum~\cite{fukada1992adaptive,tokuda1994mel}, etc, and are extracted from the speech through vocoder analysis~\cite{imai1983mel,kawahara2006straight,morise2016world}. The vocoders synthesize speech from the predicted acoustic features. SPSS has several advantages over previous TTS systems: 1) naturalness, the audio is more natural; 2) flexibility, it is convenient to modify the parameters to control the generate speech; 3) low data cost, it requires less recordings than concatenative synthesis. However, SPSS also has its drawbacks: 1) the generated speech has lower intelligibility due to artifacts such as muffled, buzzing or noisy audio; 2) the generated voice is still robotic and can be easily differentiated from human recording speech.

In near 2010s, as neural network and deep learning have achieved rapid progress, some works first introduce deep neural network into SPSS, such as deep neural network (DNN) based~\cite{zen2013statistical,qian2014training} and recurrent neural network (RNN) based~\cite{fan2014tts,zen2015acoustic,zen2015unidirectional}. However, these models replace HMM with neural networks and still predict the acoustic features from linguistic features, which follow the paradigm of SPSS. Later, \citet{wang2016first} propose to directly generate acoustic features from phoneme sequence instead of linguistic features, which can be regarded as the first exploration for end-to-end\footnote{The term ``end-to-end'' in TTS has a vague meaning. In early studies, ``end-to-end'' refers to that the text-to-spectrogram model is end-to-end, but still uses a separate waveform synthesizer (vocoder). It can also broadly refer to the neural based TTS models which do not use complicated linguistic or acoustic features. For example, WaveNet~\cite{oord2016wavenet} does not use acoustic features but directly generate waveform from linguistic features, and Tacotron~\cite{wang2017tacotron} does not use linguistic features but directly generate spectrogram from character or phoneme. However, the strict end-to-end model refers to directly generating waveform from text. Therefore, in this paper we use ``end-to-end'', ``more end-to-end'' and ``fully end-to-end'' to differentiate the degree of end-to-end for TTS models.} speech synthesis. In this survey, we focus on neural based speech synthesis, and mostly on end-to-end models. Since later SPSS also uses neural networks as the acoustic models, we briefly describe these models but do not dive deep into the details. 

\begin{figure}[h!]
  \centering
  \includegraphics[scale=0.5,trim=0cm 0cm 0cm 1cm,clip=true]{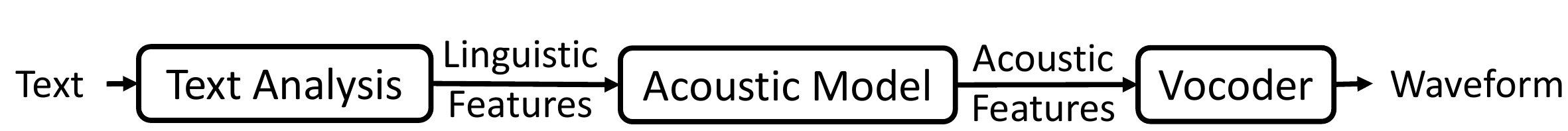}
  \caption{The three key components in neural TTS.}
  \label{fig_tts_module}
\end{figure}

\paragraph{Neural Speech Synthesis}
As the development of deep learning, neural network-based TTS (neural TTS for short) is proposed, which adopts (deep) neural networks as the model backbone for speech synthesis. Some early neural models are adopted in SPSS to replace HMM for acoustic modeling. Later, WaveNet~\cite{oord2016wavenet} is proposed to directly generate waveform from linguistic features, which can be regarded as the first modern neural TTS model. Other models like DeepVoice 1/2~\cite{arik2017deep,gibiansky2017deep} still follow the three components in statistical parametric synthesis, but upgrade them with the corresponding neural network based models. Furthermore, some end-to-end models (e.g., Tacotron 1/2~\cite{wang2017tacotron,shen2018natural}, Deep Voice 3~\cite{ping2018deep}, and FastSpeech 1/2~\cite{ren2019fastspeech,ren2021fastspeech}) are proposed to simplify text analysis modules and directly take character/phoneme sequences as input, and simplify acoustic features with mel-spectrograms. Later, fully end-to-end TTS systems are developed to directly generate waveform from text, such as ClariNet~\cite{ping2018clarinet}, FastSpeech 2s~\cite{ren2021fastspeech} and EATS~\cite{donahue2020end}. Compared to previous TTS systems based on concatenative synthesis and statistical parametric synthesis, the advantages of neural network based speech synthesis include high voice quality in terms of both intelligibility and naturalness, and less requirement on human preprocessing and feature development. 

\subsection{Organization of This Survey}
\begin{figure*}[thp]
 \centering
\begin{forest}
  forked edges,
  for tree={
  grow=east,
  reversed=true,  
  anchor=base west,
  parent anchor=east,
  child anchor=west,
  base=left,
  font=\normalsize,
  rectangle,
  draw=hiddendraw,
  rounded corners,
  align=left,
  minimum width=2.5em,
  inner xsep=4pt,
  inner ysep=0pt,
  },
  where level=1{text width=14em,align=center,font=\normalsize}{},
  where level=2{text width=15.5em,align=left,font=\normalsize}{},
    [TTS Survey
        [Sec.~\ref{sec_fundamental}: Key Components in TTS
            [Sec.~\ref{sec_funda_taxonomy}: Main Taxonomy
            ]
            [Sec.~\ref{sec_funda_ta}: Text Analysis
            ]
            [Sec.~\ref{sec_funda_am}: Acoustic Model
            ]
            [Sec.~\ref{sec_funda_voc}: Vocoder
            ]
            [Sec.~\ref{sec_funda_e2e}: Towards Fully E2E TTS 
            ]
            [Sec.~\ref{sec_funda_other}: Other Taxonomies
            ]
        ]
        [Sec.~\ref{sec_advanced}: Advanced Topics in TTS
            [Sec.~\ref{sec_advance_taxonomy}: Background and Taxonomy
            ]
            [Sec.~\ref{sec_advance_fast}: Fast TTS
            ]
            [Sec.~\ref{sec_advance_low}: Low-Resource TTS 
            ]
            [Sec.~\ref{sec_advance_robust}: Robust TTS 
            ]
            [Sec.~\ref{sec_advance_express}: Expressive TTS 
            ]
            [Sec.~\ref{sec_advance_adapt}: Adaptive TTS 
            ]
        ]
    ]
\end{forest}
\caption{Organization of this survey paper.}
\label{org_survey_paper}
\end{figure*}
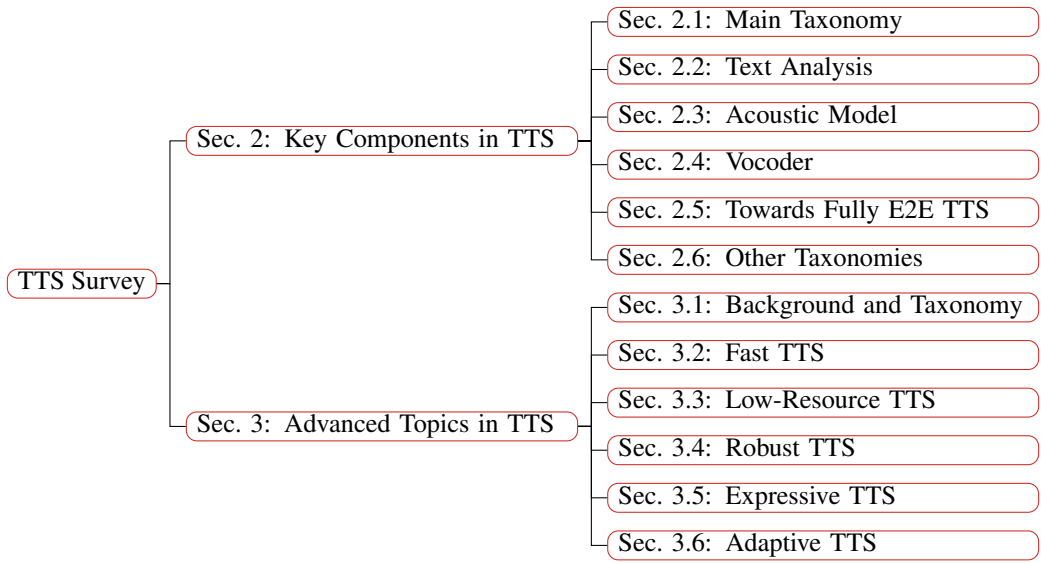

In this paper, we mainly review research works on neural TTS, which consists of two parts, as shown in Figure~\ref{org_survey_paper}.

\tikzstyle{leaf}=[mybox,minimum height=1.2em,
fill=hidden-orange!50, text width=5em,  text=black,align=left,font=\footnotesize,
inner xsep=4pt,
inner ysep=1pt,
]

\paragraph{Key Components in TTS} A modern TTS system consists of three basic components\footnote{Although some end-to-end models do not explicitly use text analysis (e.g., Tacotron 2~\cite{shen2018natural}), acoustic models (e.g., WaveNet~\cite{oord2016wavenet}), or vocoders (e.g., Tacotron~\cite{wang2017tacotron}, and some systems only use a single end-to-end model (e.g., FastSpeech 2s~\cite{ren2021fastspeech}), using these components are still popular in current TTS research and product.}: a text analysis module, an acoustic model, and a vocoder. As shown in Figure~\ref{fig_tts_module}, the text analysis module converts a text sequence into linguistic features, the acoustic models generate acoustic features from linguistic features, and then the vocoders synthesize waveform from acoustic features. We review the research on the three components of neural TTS in Section~\ref{sec_fundamental}. Specifically, we first introduce the main taxonomy for the basic components of neural TTS in Section~\ref{sec_funda_taxonomy}, and then introduce the works on text analysis, acoustic models, and vocoders in Section~\ref{sec_funda_ta}, Section~\ref{sec_funda_am}, and Section~\ref{sec_funda_voc} respectively. We further introduce the research towards fully end-to-end TTS in Section~\ref{sec_funda_e2e}. Although we mainly review the research works according to the taxonomy of key components in neural TTS, we also describe several other taxonomies, including the way of sequence generation (autoregressive or non-autoregressive), different generative models, and different network structures in Section~\ref{sec_funda_other}. Besides, we also illustrate the time evolution of some representative TTS works in Section~\ref{sec_funda_other}.

\paragraph{Advanced Topics in TTS} Besides the key components of neural TTS, we further review several advanced topics in neural TTS, which push the frontier of TTS research and address practical challenges in TTS product. For example, as TTS is a typical sequence to sequence generation task and the output sequence is usually very long, how to speed up the autoregressive generation and reduce the model size for fast speech synthesis are hot research topics (Section~\ref{sec_advance_fast}). A good TTS system should generate both natural and intelligible speech and a lot of TTS research works aim to improve the intelligibility and naturalness of speech synthesis. For example, in low-resource scenarios where the data to train a TTS model is insufficient, the synthesized speech may be of both low intelligibility and naturalness. Therefore, a lot of works aim to build data-efficient TTS models under low-resource settings (Section~\ref{sec_advance_low}). Since TTS models are facing robustness issues where word skipping and repeating problems in generated speech affect the speech quality, a lot of works aim to improve the robustness of speech synthesis (Section~\ref{sec_advance_robust}). To improve the naturalness and expressiveness, a lot of works model, control, and transfer the style/prosody of speech in order to generate expressive speech (Section~\ref{sec_advance_express}). Adapting a TTS model to support the voice of any target speakers is very helpful for the broad usage of TTS. Therefore, efficient voice adaptation with limited adaptation data and parameters is critical for practical TTS applications (Section~\ref{sec_advance_adapt}). 

To further enrich this survey, we summarize TTS related resources including open-source implementations, corpora, and other useful resources in Section~\ref{sec_resource}. We summarize this survey and discuss future research directions in Section~\ref{sec_summary}.

\tikzstyle{leaf}=[mybox,minimum height=1.2em,
fill=hidden-orange!50, text width=5em,  text=black,align=left,font=\footnotesize,
inner xsep=4pt,
inner ysep=1pt,
]

\begin{figure*}[thp]
 \centering
\begin{subfigure}[b]{1.0\textwidth}
 \centering
\begin{forest}
  forked edges,
  for tree={
  grow=east,
  reversed=true,  
  anchor=base west,
  parent anchor=east,
  child anchor=west,
  base=left,
  font=\normalsize,
  rectangle,
  draw=hiddendraw,
  rounded corners,
  align=left,
  minimum width=2.5em,
  inner xsep=4pt,
  inner ysep=0pt,
  },
  where level=1{text width=6.9em,font=\normalsize}{},
  where level=2{text width=9em,font=\normalsize}{},
  where level=3{font=\footnotesize,yshift=0.25pt}{},
    [TTS
        [Text Analysis
            [Char$\rightarrow$Linguistic
                [TN~\cite{sproat2016rnn,mansfield2019neural,zhang2019neural}{,} G2P~\cite{yao2015sequence,sun2019token,sun2019knowledge}\\
                Prosody Prediction~\cite{sridhar2007exploiting,jeon2009automatic,qian2010automatic,pan2019mandarin} \\
                \textcolor{black}{Unified Model}~\cite{pan2020unified,zhang2020unified}\\
                DeepVoice 1/2~\cite{arik2017deep,gibiansky2017deep},
                leaf,text width=15em
                ]
            ]
        ]
        [Acoustic Model
            [Linguistic$\rightarrow$Acoustic
                [
                 HMM-based~\cite{yoshimura1999simultaneous,tokuda2000speech,yoshimura2002simultaneous,tokuda2013speech} \\ 
                 DNN based~\cite{zen2013statistical,qian2014training} \\
                 RNN based~\cite{fan2014tts,zen2015acoustic}{,} 
                 \textcolor{black}{Emphasis}~\cite{li2018emphasis},leaf,text width=15em
                ]
            ]
            [Char/Phone$\rightarrow$Acoustic
                [ARST~\cite{wang2016first}{,}
                \textcolor{black}{DeepVoice 3}~\cite{ping2018deep}\\
                 \textcolor{black}{Tacotron 1/2}~\cite{wang2017tacotron,shen2018natural}{,} DurIAN~\cite{yu2020durian}\\
                 \textcolor{black}{FastSpeech 1/2}~\cite{ren2019fastspeech,ren2021fastspeech}{,} DCTTS~\cite{tachibana2018efficiently}\\
                 TransformerTTS~\cite{li2019neural}{,} VoiceLoop~\cite{taigman2018voiceloop}\\
                 ParaNet~\cite{peng2020non}{,} Glow-TTS~\cite{kim2020glow}\\
                 Grad-TTS~\cite{popov2021grad}{,} PriorGrad~\cite{lee2021priorgrad},leaf,text width=15em
                ]
            ]
        ]
        [Vocoder
            [Vocoder in SPSS            
                [\textcolor{black}{STRAIGHT}~\cite{kawahara2006straight}{,}
                 \textcolor{black}{WORLD}~\cite{morise2016world},leaf,text width=15em
                ]
            ]
            [Linguistic$\rightarrow$Wav
                [\textcolor{black}{WaveNet}~\cite{oord2016wavenet}{,}
                 \textcolor{black}{Par.WaveNet}~\cite{oord2018parallel}\\
                 \textcolor{black}{WaveRNN}~\cite{kalchbrenner2018efficient}{,}
                 \textcolor{black}{GAN-TTS}~\cite{binkowski2019high},leaf,text width=15em
                ]
            ]
            [Acoustic$\rightarrow$Wav
                [\textcolor{black}{LPCNet}~\cite{valin2019lpcnet}{,}
                 \textcolor{black}{WaveGlow}~\cite{prenger2019waveglow}\\
                 FloWaveNet~\cite{kim2019flowavenet}{,} MelGAN~\cite{kumar2019melgan}\\ \textcolor{black}{Par.WaveGAN}~\cite{oord2018parallel}{,} HiFi-GAN~\cite{kong2020hifi}\\
                 DiffWave~\cite{kong2020diffwave}{,} \textcolor{black}{WaveGrad}~\cite{chen2020wavegrad}\\
                 ,leaf,text width=15em
                ]
            ]
        ]
        [Fully E2E Model
            [Char/Phone$\rightarrow$Wav
                [\textcolor{black}{Char2Wav}~\cite{sotelo2017char2wav}{,}
                 \textcolor{black}{FastSpeech 2s}~\cite{ren2021fastspeech}\\
                 \textcolor{black}{ClariNet}~\cite{ping2018clarinet}{,}
                 \textcolor{black}{EATS}~\cite{donahue2020end}{,} VITS~\cite{kim2021conditional}\\
                 \textcolor{black}{Wave-Tacotron}~\cite{weiss2020wave}{,} EfficientTTS~\cite{miao2020efficienttts},leaf,text width=15em
                ]
            ]
        ]
    ]
\end{forest}
\caption{A taxonomy of neural TTS.}
\label{main_taxonomy_of_tts}
\end{subfigure}
\vspace{0.5cm}
\begin{subfigure}[b]{1.0\textwidth}
 \centering
  \includegraphics[scale=0.85]{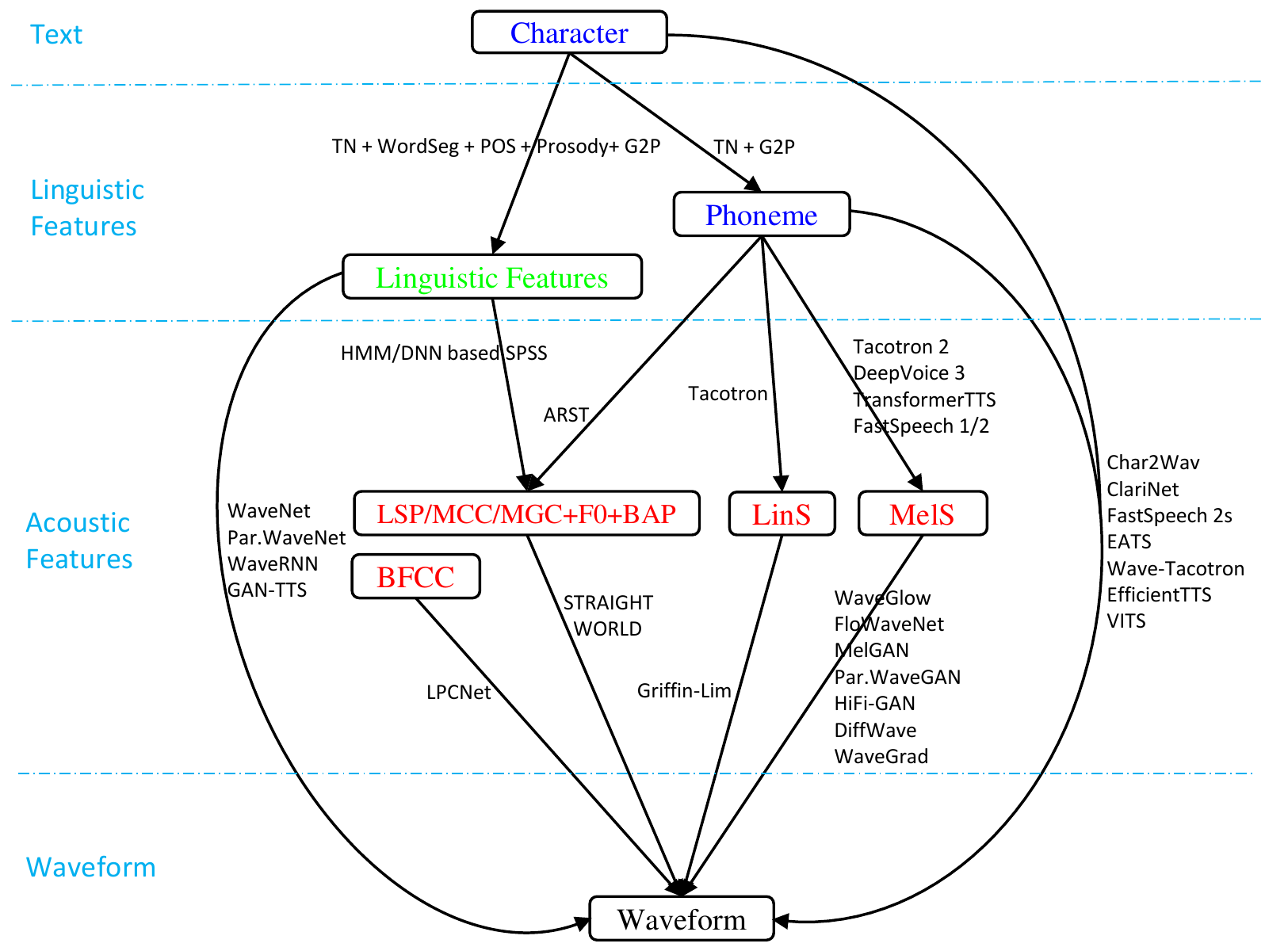}
  \caption{The data flows from text to waveform.}
  \label{fig_tts_data_flow}
\end{subfigure}
\caption{A taxonomy of neural TTS from the perspectives of key components and data flows.}
\label{main_taxonomy}
\end{figure*}

\section{Key Components in TTS}
\label{sec_fundamental}

In this section, we review the research works from the perspective of the key components (text analysis, acoustic models, and vocoders) in neural TTS. We first introduce the main taxonomy under this perspective in Section~\ref{sec_funda_taxonomy}, and then introduce the three TTS components in Section~\ref{sec_funda_ta}, Section~\ref{sec_funda_am}, and Section~\ref{sec_funda_voc}, respectively. Furthermore, we review the works towards fully end-to-end TTS in Section~\ref{sec_funda_e2e}. Besides the main taxonomy, we also introduce more taxonomies such as autoregressive/non-autoregressive sequence generation, generative model, network structure, as well as the timeline of representative research works on TTS in Section~\ref{sec_funda_other}.

\subsection{Main Taxonomy}
\label{sec_funda_taxonomy}
We categorize the works on neural TTS mainly from the perspective of basic TTS components: text analysis, acoustic models, vocoders\footnote{Note that some neural TTS models such as WaveNet~\cite{oord2016wavenet} and WaveRNN~\cite{kalchbrenner2018efficient} are first introduced to directly generate waveform from linguistic features. From this perspective, WaveNet can be regarded as a combination of an acoustic model and a vocoder. Following works usually leverage WaveNet and WaveRNN as a vocoder by taking mel-spectrograms as input to generate waveform. Therefore, we categorize WaveNet/WaveRNN into vocoders and introduce in Section~\ref{sec_funda_voc}.}, and fully end-to-end models, as shown in Figure~\ref{main_taxonomy_of_tts}. We find this taxonomy is consistent with the data conversion flow from text to waveform: 1) Text analysis converts character into phoneme or linguistic features; 2) Acoustic models generate acoustic features, from either linguistic features or characters/phonemes; 3) Vocoders generate waveform from either linguistic features or acoustic features; 4) Fully end-to-end models directly convert characters/phonemes into waveform. 

We re-organize the TTS works according to the data flow from text to waveform, as shown in Figure~\ref{fig_tts_data_flow}. There are several data representations in the process of text to speech conversion: 1) Characters, which are the raw format of text. 2) Linguistic features, which are obtained through text analysis and contain rich context information about pronunciation and prosody. Phonemes are one of the most important elements in linguistic features, and are usually used alone to represent text in neural based TTS models. 3) Acoustic features, which are abstractive representations of speech waveform. In statistical parametric speech synthesis~\cite{yoshimura1999simultaneous,tokuda2000speech,yoshimura2002simultaneous,zen2009statistical,tokuda2013speech}, LSP (line spectral pairs)~\cite{itakura1975line}, MCC (mel-cepstral coefficients)~\cite{fukada1992adaptive}, MGC (mel-generalized coefficients)~\cite{tokuda1994mel}, F0 and BAP (band aperiodicities)~\cite{kawahara1999restructuring,kawahara2001aperiodicity} are used as acoustic features, which can be easily converted into waveform through vocoders such as STRAIGHT~\cite{kawahara2006straight} and WORLD~\cite{morise2016world}. In neural based end-to-end TTS models, mel-spectrograms or linear-spectrograms are usually used as acoustic features, which are converted into waveform using neural based vocoders. 4) Waveform, the final format of speech. As can be seen from Figure~\ref{fig_tts_data_flow}, there can be different data flows from text to waveform, including: 1) character $\rightarrow$ linguistic features $\rightarrow$ acoustic features $\rightarrow$ waveform; 2) character $\rightarrow$ phoneme $\rightarrow$ acoustic features $\rightarrow$ waveform; 3) character $\rightarrow$ linguistic features $\rightarrow$ waveform; 4) character $\rightarrow$ phoneme $\rightarrow$ acoustic features $\rightarrow$ waveform; 5) character $\rightarrow$ phoneme $\rightarrow$ waveform, or character $\rightarrow$ waveform.


\subsection{Text Analysis}
\label{sec_funda_ta}
Text analysis, also called frontend in TTS, transforms input text into linguistic features that contain rich information about pronunciation and prosody to ease the speech synthesis. In statistic parametric synthesis, text analysis is used to extract a sequence of linguistic feature vectors~\cite{tokuda2013speech}, and contains several functionalities such as text normalization~\cite{sproat2016rnn,zhang2020hybrid}, word segmentation~\cite{xue2003chinese}, part-of-speech (POS) tagging~\cite{schlunz2010effects}, prosody prediction~\cite{chu2001locating}, and grapheme-to-phoneme conversion~\cite{yao2015sequence}. In end-to-end neural TTS, due to the large modeling capacity of neural based models, the character or phoneme sequences are directly taken as input for synthesis, and thus the text analysis module is largely simplified. In this scenario, text normalization is still needed to get standard word format from character input, and grapheme-to-phoneme conversion is further needed to get phonemes from standard word format. Although some TTS models claim fully end-to-end synthesis that directly generates waveform from text, text normalization is still needed to handle raw text with any possible non-standard formats for practical usage. Besides, some end-to-end TTS models incorporate conventional text analysis functions. For example, Char2Wav~\cite{sotelo2017char2wav} and DeepVoice 1/2~\cite{arik2017deep,gibiansky2017deep} implement the character-to-linguistic feature conversion into its pipeline, purely based on neural networks, and some works\cite{stanton2018predicting} explicitly predict prosody features with text encoder. In the remaining of this subsection, we first introduce the typical tasks for text analysis in statistic parametric synthesis, and then discuss the development of text analysis in end-to-end TTS models.

\begin{table}[h!]
\small
	\caption{Typical tasks in text analysis (i.e., TTS fronend, character$\rightarrow$linguistic).}
	\centering
	\begin{tabular}{ l  l }
		\toprule
		Task & Research Work   \\
		\midrule
		Text Normalization & Rule-based~\cite{sproat2001normalization}, Neural-based~\cite{sproat2016rnn,mansfield2019neural,yolchuyeva2018text,zhang2019neural}, Hybrid~\cite{zhang2020hybrid} \\
		Word Segmentation & \cite{xue2003chinese,zheng2013deep,pei2014max} \\
		POS Tagging & \cite{schlunz2010effects,sun2011improved,mamateli2011morphological,zheng2013deep,janicki2004application} \\
		Prosody Prediction &  \cite{chu2001locating,ying2001rnn,sridhar2007exploiting,levow2008automatic,jeon2009automatic,sun2009chinese,qian2010automatic,ding2015automatic,zhang2016mandarin,lu2019self,lu2019implementing,ai2020neural} \\
		Grapheme to Phoneme & N-gram~\cite{chen2003conditional,bisani2008joint}, Neural-based~\cite{yao2015sequence,rao2015grapheme,chae2018convolutional,sun2019token} \\
		-~- Polyphone Disambiguation & \cite{zhang2002efficient,xu2004grapheme,mao2007inequality,shan2016bi,sun2019knowledge,cai2019polyphone,park2020g2pm}   \\
		\bottomrule
	\end{tabular}
	\vspace{0.3cm}
	\label{tab_text_analysis_works}
\end{table}

We summarize some typical tasks in text analysis in Table~\ref{tab_text_analysis_works}, and introduce some representative works for each task as follows.
\begin{itemize}[leftmargin=*]
    \item Text normalization. The raw written text (non-standard words) should be converted into spoken-form words through text normalization, which can make the words easy to pronounce for TTS models. For example, the year ``1989'' is normalized into ``nineteen eighty nine'', ``Jan. 24'' is normalized into ``Janunary twenty-fourth''. Early works on text normalization are rule based~\cite{sproat2001normalization}, and then neural networks are leveraged to model text normalization as a sequence to sequence task where the source and target sequences are non-standard words and spoken-form words respectively~\cite{sproat2016rnn,mansfield2019neural,zhang2019neural}. Recently, some works~\cite{zhang2020hybrid} propose to combine the advantages of both rule-based and neural-based models to further improve the performance of text normalization. 
    \item Word segmentation. For character-based languages such as Chinese, word segmentation~\cite{xue2003chinese,zheng2013deep,pei2014max} is necessary to detect the word boundary from raw text, which is important to ensure the accuracy for later POS tagging, prosody prediction, and grapheme-to-phoneme conversion process. 
    \item Part-of-speech tagging. The part-of-speech (POS) of each word, such as noun, verb, preposition, is also important for grapheme-to-phoneme conversion and prosody prediction in TTS. Several works have investigated POS tagging in speech synthesis~\cite{schlunz2010effects,sun2011improved,mamateli2011morphological,zheng2013deep,janicki2004application}. 
    \item Prosody prediction. The prosody information, such as rhythm, stress, and intonation of speech, corresponds to the variations in syllable duration, loudness and pitch, which plays an important perceptual role in human speech communication. Prosody prediction relies on tagging systems to label each kind of prosody. Different languages have different prosody tagging systems and tools~\cite{silverman1992tobi,rosenberg2010autobi,taylor1998tilt,hirst2001automatic,obin2014slam}. For English, ToBI (tones and break indices)~\cite{silverman1992tobi,rosenberg2010autobi} is a popular tagging system, which describes the tags for tones (e.g., pitch accents, phrase accents, and boundary tones) and break (how strong the break is between words). For example, in this sentence ``Mary went to the store ?'', ``Mary'' and ``store'' can be emphasized, and this sentence is raising tone. A lot of works~\cite{sridhar2007exploiting,levow2008automatic,jeon2009automatic,qian2010automatic} investigate different models and features to predict the prosody tags based on ToBI. For Chinese speech synthesis, the typical prosody boundary labels consist of prosodic word (PW), prosodic phrase (PPH) and intonational phrase (IPH), which can construct a three-layer hierarchical prosody tree~\cite{chu2001locating,sun2009chinese,ding2015automatic}. Some works~\cite{chu2001locating,ai2020neural,ding2015automatic,sun2009chinese,lu2019self,lu2019implementing} investigate different model structures such as CRF~\cite{lafferty2001conditional}, RNN~\cite{hochreiter1997long}, and self-attention~\cite{vaswani2017attention} for prosody prediction in Chinese. 
    \item Grapheme-to-phoneme (G2P) conversion. Converting character (grapheme) into pronunciation (phoneme) can greatly ease speech synthesis. For example, the word ``speech'' is converted into ``s p iy ch''. A manually collected grapheme-to-phoneme lexicon is usually leveraged for conversion. However, for alphabetic languages like English, lexicon cannot cover the pronunciations of all the words. Thus, the G2P conversion for English is mainly responsible to generate the pronunciations of out-of-vocabulary words~\cite{chen2003conditional,bisani2008joint,yao2015sequence,rao2015grapheme,chae2018convolutional,sun2019token}. For languages like Chinese, although the lexicon can cover nearly all the characters, there are a lot of polyphones that can be only decided according to the context of a character\footnote{A lot of languages including English have polyphones. For example, ``resume'' in English can be pronounced as ``ri$'$zju:m$'$'' (means to go on or continue after interruption) or ``$'$rezjumei'' (means curriculum vitae).}. Thus, G2P conversion in this kind of languages is mainly responsible for polyphone disambiguation, which decides the appropriate pronunciation based on the current word context~\cite{zhang2002efficient,xu2004grapheme,mao2007inequality,shan2016bi,sun2019knowledge,cai2019polyphone,park2020g2pm}.
\end{itemize}

After the above text analyses, we can further construct linguistic features and take them as input to the later part of TTS pipeline, e.g., acoustic models in SPSS or vocoders~\cite{oord2016wavenet}. Usually, we can construct linguistic features by aggregating the results of text analysis from different levels including phoneme, syllable, word, phrase and sentence levels~\cite{tokuda2013speech}. 

\paragraph{Discussions} Although text analysis seems to receive less attention in neural TTS compared to SPSS, it has been incorporated into neural TTS in various ways: 1) Multi-task and unified frontend model. Recently, \citet{pan2020unified,zhang2020unified} design unified models to cover all the tasks in text analysis in a multi-task paradigm and achieve good results. 2) Prosody prediction. Prosody is critical for the naturalness of speech synthesis. Although neural TTS models simplify the text analysis module, some features for prosody prediction are incorporated into text encoder, such as the prediction of pitch~\cite{ren2021fastspeech}, duration~\cite{ren2019fastspeech}, phrase break~\cite{liu2020modeling}, breath or filled pauses~\cite{yan2021adaspeech3} are built on top of the text (character or phoneme) encoder in TTS models. Some other ways to incorporate prosody features include 1) reference encoders that learn the prosody representations from reference speech; 2) text pre-training that learns good text representations with implicit prosody information through self-supervised pre-training~\cite{hayashi2019pre,guo2019exploiting}; and 3) incorporating syntax information through dedicated modeling methods such as graph networks~\cite{liu2020graphspeech}.

\subsection{Acoustic Models} 
\label{sec_funda_am}
In this section, we review the works on acoustic models, which generate acoustic features from linguistic features or directly from phonemes or characters. As the development of TTS, different kinds of acoustic models have been adopted, including the early HMM and DNN based models in statistical parametric speech synthesis (SPSS)~\cite{yoshimura1999simultaneous,tokuda2000speech,zen2013statistical,qian2014training,fan2014tts,zen2015unidirectional}, and then the sequence to sequence models based on encoder-attention-decoder framework (including LSTM, CNN and self-attention)~\cite{wang2017tacotron,shen2018natural,ping2018deep,li2019neural}, and the latest feed-forward networks (CNN or self-attention)~\cite{ren2019fastspeech,peng2020non} for parallel generation.

Acoustic models aim to generate acoustic features that are further converted into waveform using vocoders. The choice of acoustic features largely determines the types of TTS pipeline. Different kinds of acoustic features have been tried, such as mel-cepstral coefficients (MCC)~\cite{fukada1992adaptive}, mel-generalized coefficients (MGC)~\cite{tokuda1994mel}, band aperiodicity (BAP)~\cite{kawahara1999restructuring,kawahara2001aperiodicity}, fundamental frequency (F0), voiced/unvoiced (V/UV), bark-frequency cepstral coefficients (BFCC), and the most widely used mel-spectrograms. Accordingly, we can divide the acoustic models into two periods: 1) acoustic models in SPSS, which typically predict acoustic features such as MGC, BAP and F0 from linguistic features, and 2) acoustic models in neural based end-to-end TTS, which predict acoustic features such as mel-spectrograms from phonemes or characters.


\subsubsection{Acoustic Models in SPSS} 
In SPSS~\cite{zen2009statistical,tokuda2013speech}, statistical models such as HMM~\cite{yoshimura1999simultaneous,tokuda2000speech}, DNN~\cite{zen2013statistical,qian2014training} or RNN~\cite{fan2014tts,zen2015unidirectional} are leveraged to generate acoustic features (speech parameters) from linguistic features, where the generated speech parameters are converted into speech waveform using a vocoder such as STRAIGHT~\cite{kawahara2006straight} and WORLD~\cite{morise2016world}. The developments of these acoustic models are driven by several considerations: 1) taking more context information as input; 2) modeling the correlation between output frames; 3) better combating the over-smoothing prediction problem~\cite{zen2009statistical}, since the mapping from linguistic features to acoustic features is one-to-many. We briefly review some works as follows.

HMM~\cite{rabiner1986introduction} is leveraged to generate speech parameters in ~\citet{yoshimura1999simultaneous,tokuda2000speech}, where the observation vectors of HMM consist of spectral parameter vectors such as mel-cepstral coefficients (MCC) and F0. Compared to previous concatenative speech synthesis, HMM-based parametric synthesis is more flexible in changing speaker identities, emotions, and speaking styles~\cite{tokuda2000speech}. Readers can refer to ~\citet{zen2015acoustic,zen2009statistical,tokuda2013speech} for some analyses on the advantages and drawbacks of HMM-based SPSS. One major drawback of HMM-based SPSS is that the quality of the synthesized speech is not good enough~\cite{zen2009statistical,tokuda2013speech}, mainly due to two reasons: 1) the accuracy of acoustic models is not good, and the predicted acoustic features are over-smoothing and lack of details, and 2) the vocoding techniques are not good enough. The first reason is mainly due to the lack of modeling capacity in HMM. Thus, DNN-based acoustic models~\cite{zen2013statistical} are proposed in SPSS, which improve the synthesized quality of HMM-based models. Later, in order to better model the long time span contextual effect in a speech utterance, LSTM-based recurrent neural networks~\cite{fan2014tts} are leveraged to better model the context dependency. As the development of deep learning, some advanced network structure such as CBHG~\cite{wang2017tacotron} are leveraged to better predict acoustic features~\cite{li2018emphasis}. VoiceLoop~\cite{taigman2018voiceloop} adopts a working memory called phonological loop to generate acoustic features (e.g., F0, MGC, BAP) from phoneme sequence, and then uses a WORLD~\cite{morise2016world} vocoder to synthesize waveform from this acoustic features. \citet{yang2017statistical} leverage GAN~\cite{goodfellow2014generative} to improve the generation quality of acoustic features. \citet{wang2016first} explore a more end-to-end way that leverages an attention-based recurrent sequence transducer model to directly generate acoustic features from phoneme sequence, which can avoid the frame-by-frame alignments required in previous neural network-based acoustic models. ~\citet{wang2018comparison} conduct thorough experimental studies on different acoustic models. Some acoustic models in SPSS are summarized in Table~\ref{tab_am_summary}.


\begin{table}[h!]
\small
	\caption{A list of acoustic models and their corresponding characteristics. ``Ling'' stands for linguistic features, ``Ch'' stands for character, ``Ph'' stands for phoneme, ``MCC'' stands for mel-cepstral coefficients~\cite{fukada1992adaptive}, ``MGC'' stands for mel-generalized coefficients~\cite{tokuda1994mel}, ``BAP'' stands for band aperiodicities~\cite{kawahara1999restructuring,kawahara2001aperiodicity}, ``LSP'' stands for line spectral pairs~\cite{itakura1975line}, ``LinS'' stands for linear-spectrograms, and ``MelS'' stands for mel-spectrograms. ``NAR*'' means the model uses autoregressive structures upon non-autoregressive structures and is not fully parallel. }
	\centering
	\begin{tabular}{l | l l l l }
	\toprule
		Acoustic Model & Input$\rightarrow$Output & AR/NAR & Modeling & Structure \\
		\midrule
		HMM-based \cite{yoshimura1999simultaneous,tokuda2000speech}  &Ling$\rightarrow$MCC+F0   & / & / & HMM   \\
		DNN-based~\cite{zen2013statistical} & Ling$\rightarrow$MCC+BAP+F0 & NAR & / & DNN \\
		LSTM-based~\cite{fan2014tts} & Ling$\rightarrow$LSP+F0 & AR & / & RNN\\
		EMPHASIS~\cite{li2018emphasis} & Ling$\rightarrow$LinS+CAP+F0 & AR & / & Hybrid \\
	    ARST~\cite{wang2016first} & Ph$\rightarrow$LSP+BAP+F0 & AR & Seq2Seq & RNN \\
	    VoiceLoop~\cite{taigman2018voiceloop} & Ph$\rightarrow$MGC+BAP+F0 & AR & / & hybrid \\
	   	\midrule
	   	\midrule
        Tacotron~\cite{wang2017tacotron} & Ch$\rightarrow$LinS & AR & Seq2Seq & Hybrid/RNN \\ 
        Tacotron 2~\cite{shen2018natural} & Ch$\rightarrow$MelS & AR & Seq2Seq & RNN \\
        DurIAN~\cite{yu2020durian}& Ph$\rightarrow$MelS & AR & Seq2Seq & RNN \\
        Non-Att Tacotron~\cite{shen2020non} & Ph$\rightarrow$MelS & AR & / & Hybrid/CNN/RNN   \\
        Para. Tacotron 1/2~\cite{elias2020parallel,elias2021parallel} & Ph$\rightarrow$MelS & NAR & / & Hybrid/Self-Att/CNN  \\
        MelNet~\cite{vasquez2019melnet} & Ch$\rightarrow$MelS & AR & / & RNN \\
        \midrule
        DeepVoice~\cite{arik2017deep} & Ch/Ph$\rightarrow$MelS & AR & / & CNN \\
        DeepVoice 2~\cite{gibiansky2017deep} & Ch/Ph$\rightarrow$MelS & AR & / & CNN \\
        DeepVoice 3~\cite{ping2018deep} & Ch/Ph$\rightarrow$MelS & AR & Seq2Seq & CNN \\
        ParaNet~\cite{peng2020non}  & Ph$\rightarrow$MelS & NAR & Seq2Seq & CNN \\
        DCTTS~\cite{tachibana2018efficiently} & Ch$\rightarrow$MelS & AR & Seq2Seq & CNN \\
        SpeedySpeech~\cite{vainer2020speedyspeech} & Ph$\rightarrow$MelS & NAR & / & CNN  \\
        TalkNet 1/2~\cite{beliaev2020talknet,beliaev2021talknet} & Ch$\rightarrow$MelS & NAR & / & CNN  \\
        \midrule
        TransformerTTS~\cite{li2019neural} & Ph$\rightarrow$MelS & AR & Seq2Seq & Self-Att \\
        MultiSpeech~\cite{chen2020multispeech} & Ph$\rightarrow$MelS & AR & Seq2Seq & Self-Att  \\
        FastSpeech 1/2~\cite{ren2019fastspeech,ren2021fastspeech} & Ph$\rightarrow$MelS & NAR & Seq2Seq & Self-Att \\
        AlignTTS~\cite{zeng2020aligntts}  & Ch/Ph$\rightarrow$MelS & NAR & Seq2Seq & Self-Att \\
        JDI-T~\cite{lim2020jdi}  & Ph$\rightarrow$MelS & NAR & Seq2Seq & Self-Att  \\
        FastPitch~\cite{lancucki2020fastpitch} & Ph$\rightarrow$MelS & NAR & Seq2Seq & Self-Att  \\
        AdaSpeech 1/2/3~\cite{chen2021adaspeech,yan2021adaspeech,yan2021adaspeech3} & Ph$\rightarrow$MelS & NAR & Seq2Seq & Self-Att  \\
        DenoiSpeech~\cite{zhang2020denoising} & Ph$\rightarrow$MelS & NAR & Seq2Seq & Self-Att  \\
        DeviceTTS~\cite{huang2020devicetts} & Ph$\rightarrow$MelS & NAR & / &  Hybrid/DNN/RNN   \\
        LightSpeech~\cite{luo2021lightspeech}  & Ph$\rightarrow$MelS & NAR & / & Hybrid/Self-Att/CNN  \\
        \midrule
        Flow-TTS~\cite{miao2020flow}  & Ch/Ph$\rightarrow$MelS & NAR* & Flow & Hybrid/CNN/RNN \\
        Glow-TTS~\cite{kim2020glow} & Ph$\rightarrow$MelS & NAR & Flow & Hybrid/Self-Att/CNN \\
        Flowtron~\cite{valle2020flowtron} & Ph$\rightarrow$MelS & AR & Flow & Hybrid/RNN \\
        EfficientTTS~\cite{miao2020efficienttts} & Ch$\rightarrow$MelS & NAR & Flow & Hybrid/CNN  \\
        \midrule
        GMVAE-Tacotron~\cite{hsu2018hierarchical} & Ph$\rightarrow$MelS & AR & VAE &  Hybrid/RNN \\
        VAE-TTS~\cite{zhang2019learningb} & Ph$\rightarrow$MelS & AR & VAE &  Hybrid/RNN \\
        BVAE-TTS~\cite{lee2020bidirectional} & Ph$\rightarrow$MelS & NAR & VAE &  CNN \\
        \midrule
        GAN exposure~\cite{guo2019new} & Ph$\rightarrow$MelS & AR & GAN &  Hybrid/RNN \\
        TTS-Stylization~\cite{ma2018neural} & Ch$\rightarrow$MelS & AR & GAN &  Hybrid/RNN \\
        Multi-SpectroGAN~\cite{lee2020multi} & Ph$\rightarrow$MelS & NAR & GAN &  Hybrid/Self-Att/CNN \\
        \midrule
        Diff-TTS~\cite{jeong2021diff} & Ph$\rightarrow$MelS & NAR* & Diffusion &  Hybrid/CNN \\
        Grad-TTS~\cite{popov2021grad} & Ph$\rightarrow$MelS & NAR & Diffusion &  Hybrid/Self-Att/CNN \\
        PriorGrad~\cite{lee2021priorgrad} & Ph$\rightarrow$MelS & NAR & Diffusion &  Hybrid/Self-Att/CNN \\
        \bottomrule
	\end{tabular}
	\vspace{0.3cm}
	\label{tab_am_summary}
\end{table}

\subsubsection{Acoustic Models in End-to-End TTS} 
Acoustic models in neural-based end-to-end TTS have several advantages compared to those in SPSS: 1) Conventional acoustic models require alignments between linguistic and acoustic features, while sequence to sequence based neural models implicitly learn the alignments through attention or predict the duration jointly, which are more end-to-end and require less preprocessing. 2) As the increasing modeling power of neural networks, the linguistic features are simplified into only character or phoneme sequence, and the acoustic features have changed from low-dimensional and condensed cepstrums (e.g., MGC) to high-dimensional mel-spectrograms or even more high-dimensional linear-spectrograms. In the following paragraphs, we introduce some representative acoustic models in neural TTS\footnote{We mainly review the acoustic models according to different network structures such as RNN, CNN and Transformer (self-attention), while review the vocoders according to different generative models such as autoregressive-based, flow-based, GAN-based, Diffusion-based, as shown in Section~\ref{sec_funda_voc}. However, it is not the only perspective, since acoustic models also cover different generative models while vocoders also cover different network structures.}, and provide a comprehensive list of acoustic models in Table~\ref{tab_am_summary}.  

\paragraph{RNN-based Models (e.g., Tacotron Series)}
Tacotron~\cite{wang2017tacotron} leverages an encoder-attention-decoder framework and takes characters as input\footnote{Although either characters or phonemes are taken as input in neural TTS, we do not explicitly differentiate them mainly for two considerations: 1) To ensure high pronunciation accuracy for product usage, phonemes are necessary especially for those languages (e.g., Chinese) where graphemes and phonemes have large difference. 2) For the models directly taking characters as input, there is no specific design for characters input, and thus one can easily change characters into phonemes. It is worth to mention that there are some works~\cite{ping2018deep,peng2020non,kastner2019representation} using mixed representations of characters and phonemes as input to address the data sparsity problem.} and outputs linear-spectrograms, and uses Griffin-Lim algorithm~\cite{griffin1984signal} to generate waveform. Tacotron 2~\cite{shen2018natural} is proposed to generate mel-spectrograms and convert mel-spectrograms into waveform using an additional WaveNet~\cite{oord2016wavenet} model. Tacotron 2 greatly improves the voice quality over previous methods including concatenative TTS, parametric TTS, neural TTS such as Tacotron.

Later, a lot of works improve Tacotron from different aspects: 1) Using a reference encoder and style tokens to enhance the expressiveness of speech synthesis, such as GST-Tacotron~\cite{wang2018style} and Ref-Tacotron~\cite{skerry2018towards}. 2) Removing the attention mechanism in Tacotron, and instead using a duration predictor for autoregressive prediction, such as DurIAN~\cite{yu2020durian} and Non-attentative Tacotron~\cite{shen2020non}. 3) Changing the autoregressive generation in Tacotron to non-autoregressive generation, such as Parallel Tacotron 1/2~\cite{elias2020parallel,elias2021parallel}. 4) Building end-to-end text-to-waveform models based on Tacotron, such as Wave-Tacotron~\cite{weiss2020wave}.

\paragraph{CNN-based Models (e.g., DeepVoice Series)}
DeepVoice~\cite{arik2017deep} is actually an SPSS system enhanced with convolutional neural networks. After obtaining linguistic features through neural networks, DeepVoice leverages a WaveNet~\cite{oord2016wavenet} based vocoder to generate waveform. DeepVoice 2~\cite{gibiansky2017deep} follows the basic data conversion flow of DeepVoice and enhances DeepVoice with improved network structures and multi-speaker modeling. Furthermore, DeepVoice 2 also adopts a Tacotron + WaveNet model pipeline, which first generates linear-spectrograms using Tacotron and then generates waveform using WaveNet. DeepVoice 3~\cite{ping2018deep} leverage a fully-convolutional network structure for speech synthesis, which generates mel-spectrograms from characters and can scale up to real-word multi-speaker datasets. DeepVoice 3 improves over previous DeepVoice 1/2 systems by using a more compact sequence-to-sequence model and directly predicting mel-spectrograms instead of complex linguistic features. 

Later, ClariNet~\cite{ping2018clarinet} is proposed to generate waveform from text in a fully end-to-end way. ParaNet~\cite{peng2020non} is a fully convolutional based non-autoregressive model that can speed up the mel-spectrogram generation and obtain reasonably good speech quality. DCTTS~\cite{tachibana2018efficiently} shares similar data conversion pipeline with Tacotron, and leverages a fully convolutional based encoder-attention-decoder network to generate mel-spectrograms from character sequence. It then uses a spectrogram super-resolution network to obtain linear-spectrograms, and synthesizes waveform using Griffin-Lim~\cite{griffin1984signal}.

\paragraph{Transformer-based Models (e.g., FastSpeech Series)}
TransformerTTS~\cite{li2019neural} leverages Transformer~\cite{vaswani2017attention} based encoder-attention-decoder architecture to generate mel-spectrograms from phonemes. They argue that RNN-based encoder-attention-decoder models like Tacotron 2 suffer from the following two issues: 1) Due to the recurrent nature, both the RNN-based encoder and decoder cannot be trained in parallel, and the RNN-based encoder cannot be parallel in inference, which affects the efficiency both in training and inference. 2) Since the text and speech sequences are usually very long, RNN is not good at modeling the long dependency in these sequences. TransformerTTS adopts the basic model structure of Transformer and absorbs some designs from Tacotron 2 such as decoder pre-net/post-net and stop token prediction. It achieves similar voice quality with Tacotron 2 but enjoys faster training time. However, compared with RNN-based models such as Tacotron that leverage stable attention mechanisms such as location-sensitive attention, the encoder-decoder attentions in Transformer are not robust due to parallel computation. Thus, some works propose to enhance the robustness of Transformer-based acoustic models. For example, MultiSpeech~\cite{chen2020multispeech} improves the robustness of the attention mechanism through encoder normalization, decoder bottleneck, and diagonal attention constraint, and RobuTrans~\cite{li2020robutrans} leverages duration prediction to enhance the robustness in autoregressive generation. 

Previous neural-based acoustic models such as Tacotron 1/2~\cite{wang2017tacotron,shen2018natural}, DeepVoice 3~\cite{ping2018deep} and TransformerTTS~\cite{li2019neural} all adopt autoregressive generation, which suffer from several issues: 1) Slow inference speed. The autoregressive mel-spectrogram generation is slow especially for long speech sequence (e.g., for 1 second speech, there are nearly 500 frames of mel-spectrogram if hop size is 10ms, which is a long sequence). 2) Robust issues. The generated speech usually has a lot of word skipping and repeating and issues, which is mainly caused by the inaccurate attention alignments between text and mel-spectrograms in encoder-attention-decoder based autoregressive generation. Thus, FastSpeech~\cite{ren2019fastspeech} is proposed to solve these issues: 1) It adopts a feed-forward Transformer network to generate mel-spectrograms in parallel, which can greatly speed up inference. 2) It removes the attention mechanism between text and speech to avoid word skipping and repeating issues and improve robustness. Instead, it uses a length regulator to bridge the length mismatch between the phoneme and mel-spectrogram sequences. The length regulator leverages a duration predictor to predict the duration of each phoneme and expands the phoneme hidden sequence according to the phoneme duration, where the expanded phoneme hidden sequence can match the length of mel-spectrogram sequence and facilitate the parallel generation. FastSpeech enjoys several advantages~\cite{ren2019fastspeech}: 1) extremely fast inference speed (e.g., 270x inference speedup on mel-spectrogram generation, 38x speedup on waveform generation); 2) robust speech synthesis without word skipping and repeating issues; and 3) on par voice quality with previous autoregressive models. FastSpeech has been deployed in Microsoft Azure Text to Speech Service\footnote{https://azure.microsoft.com/en-us/services/cognitive-services/text-to-speech/} to support all the languages and locales in Azure TTS\footnote{https://techcommunity.microsoft.com/t5/azure-ai/neural-text-to-speech-extends-support-to-15-more-languages-with/ba-p/1505911}. 

FastSpeech leverages an explicit duration predictor to expand the phoneme hidden sequence to match to the length of mel-spectrograms. How to get the duration label to train the duration predictor is critical for the prosody and quality of generated voice. We briefly review the TTS models with duration prediction in Section~\ref{sec_advanced_robust_duration}. In the next, we introduce some other improvements based on FastSpeech. FastSpeech 2~\cite{ren2021fastspeech} is proposed to further enhance FastSpeech, mainly from two aspects: 1) Using ground-truth mel-spectrograms as training targets, instead of distilled mel-spectrograms from an autoregressive teacher model. This simplifies the two-stage teacher-student distillation pipeline in FastSpeech and also avoids the information loss in target mel-spectrograms after distillation. 2) Providing more variance information such as pitch, duration, and energy as decoder input, which eases the one-to-many mapping problem~\cite{jayne2012one,gadermayr2021asymmetric,wang2017tacotron,zhu2017toward} in text to speech\footnote{One-to-many mapping in TTS refers to that there are multiple possible speech sequences corresponding to a text sequence due to variations in speech, such as pitch, duration, sound volume, and prosody, etc.}. FastSpeech 2 achieves better voice quality than FastSpeech and maintains the advantages of fast, robust, and controllable speech synthesis in FastSpeech\footnote{FastSpeech 2s~\cite{ren2021fastspeech} is proposed together with FastSpeech 2. Since it is a fully end-to-end text-to-waveform model, we introduce it in Section~\ref{sec_funda_e2e}.}. FastPitch~\cite{lancucki2020fastpitch} improves FastSpeech by using pitch information as decoder input, which shares similar idea of variance predictor in FastSpeech 2.

\paragraph{Other Acoustic Models (e.g., Flow, GAN, VAE, Diffusion)} 
Besides the above acoustic models, there are a lot of other acoustic models~\cite{vasquez2019melnet,bi2018deep,huang2020devicetts,lee2020bidirectional,cong2021glow}, as shown in Table~\ref{tab_am_summary}. Flow-based models have long been used in neural TTS. After the early successful applications on vocoders (e.g., Parallel WaveNet~\cite{oord2018parallel}, WaveGlow~\cite{prenger2019waveglow}, FloWaveNet~\cite{kim2019flowavenet}), flow-based models are also applied in acoustic models, such as Flowtron~\cite{valle2020flowtron} that is an autoregressive flow-based mel-spectrogram generation model, Flow-TTS~\cite{miao2020flow} and Glow-TTS~\cite{kim2020glow} that leverage generative flow for non-autoregressive mel-spectrogram generation. Besides flow-based models, other generative models have also been leveraged in acoustic models. For example, 1) GMVAE-Tacotron~\cite{hsu2018hierarchical}, VAE-TTS~\cite{zhang2019learningb}, and BVAE-TTS~\cite{lee2020bidirectional} are based on   VAE~\cite{kingma2013auto}; 2) GAN exposure~\cite{guo2019new}, TTS-Stylization~\cite{ma2018neural}, and Multi-SpectroGAN~\cite{lee2020multi} are based on GAN~\cite{goodfellow2014generative}; 3) Diff-TTS~\cite{jeong2021diff}, Grad-TTS~\cite{popov2021grad}, and PriorGrad~\cite{lee2021priorgrad} are based on diffusion model~\cite{sohl2015deep,ho2020denoising}. 



\begin{table}[h!]
\small
	\caption{A list of vocoders and their corresponding characteristics. }
	\centering
	\begin{tabular}{l | l l l l }
	\toprule
		Vocoder & Input & AR/NAR & Modeling & Architecture \\
		\midrule
        WaveNet~\cite{oord2016wavenet} & Linguistic Feature  & AR & / & CNN \\
        SampleRNN~\cite{mehri2016samplernn} & /  & AR & / & RNN \\
        WaveRNN~\cite{kalchbrenner2018efficient} & Linguistic Feature & AR & / & RNN \\
        LPCNet~\cite{valin2019lpcnet}& BFCC  & AR & / & RNN \\
        Univ. WaveRNN~\cite{lorenzo2019towards} &  Mel-Spectrogram   & AR & / & RNN \\
        SC-WaveRNN~\cite{paul2020speaker} &  Mel-Spectrogram   & AR & / & RNN \\
        MB WaveRNN~\cite{yu2020durian} &  Mel-Spectrogram   & AR & / & RNN \\
        FFTNet~\cite{jin2018fftnet}& Cepstrum  & AR & / & CNN \\
        \midrule
        Par. WaveNet~\cite{oord2018parallel} & Linguistic Feature & NAR & Flow & CNN \\
        WaveGlow~\cite{prenger2019waveglow}& Mel-Spectrogram  & NAR & Flow & Hybrid/CNN \\
        FloWaveNet~\cite{kim2019flowavenet}& Mel-Spectrogram  & NAR & Flow & Hybrid/CNN \\
        WaveFlow~\cite{ping2020waveflow} & Mel-Spectrogram  & AR & Flow & Hybrid/CNN \\
        SqueezeWave~\cite{zhai2020squeezewave} & Mel-Spectrogram  & NAR & Flow & CNN \\
        \midrule
        WaveGAN~\cite{donahue2018adversarial}& /  & NAR & GAN & CNN \\
        GELP~\cite{juvela2019gelp} & Mel-Spectrogram  & NAR & GAN & CNN \\
        GAN-TTS~\cite{binkowski2019high} & Linguistic Feature & NAR & GAN & CNN \\
        MelGAN~\cite{kumar2019melgan} & Mel-Spectrogram  & NAR & GAN & CNN \\
        Par. WaveGAN~\cite{yamamoto2020parallel} & Mel-Spectrogram  & NAR & GAN & CNN \\
        HiFi-GAN~\cite{kong2020hifi} & Mel-Spectrogram  & NAR & GAN & Hybrid/CNN \\
        VocGAN~\cite{yang2020vocgan} & Mel-Spectrogram  & NAR & GAN & CNN \\
        GED~\cite{gritsenko2020spectral} & Linguistic Feature & NAR & GAN & CNN \\
        Fre-GAN~\cite{kim2021fre} & Mel-Spectrogram  & NAR & GAN & CNN \\
        \midrule
        Wave-VAE~\cite{peng2020non} & Mel-Spectrogram  & NAR & VAE & CNN \\
        \midrule
        WaveGrad~\cite{chen2020wavegrad} & Mel-Spectrogram  & NAR & Diffusion & Hybrid/CNN \\
        DiffWave~\cite{kong2020diffwave} & Mel-Spectrogram  & NAR & Diffusion & Hybrid/CNN \\
        PriorGrad~\cite{lee2021priorgrad} & Mel-Spectrogram  & NAR & Diffusion & Hybrid/CNN \\
        \bottomrule
	\end{tabular}
	\vspace{0.3cm}
	\label{tab_vocoder_summary}
\end{table}

\subsection{Vocoders}
\label{sec_funda_voc}
Roughly speaking, the development of vocoders can be categorized into two stages: the vocoders used in statistical parametric speech synthesis (SPSS)~\cite{kawahara2006straight,morise2016world,ai2020neural}, and the neural network-based vocoders~\cite{oord2016wavenet,sotelo2017char2wav,kalchbrenner2018efficient,prenger2019waveglow,kim2019flowavenet}. Some popular vocoders in SPSS include STRAIGHT~\cite{kawahara2006straight} and WORLD~\cite{morise2016world}. We take the WORLD vocoder as an example, which consists of vocoder analysis and vocoder synthesis steps. In vocoder analysis, it analyzes the speech and gets acoustic features such as mel-cepstral coefficients~\cite{fukada1992adaptive}, band aperiodicity~\cite{kawahara1999restructuring,kawahara2001aperiodicity} and F0. In vocoder synthesis, it generates speech waveform from these acoustic features. In this section, we mainly review the works on neural-based vocoders due to their high voice quality. 

Early neural vocoders such as WaveNet~\cite{oord2016wavenet,oord2018parallel}, Char2Wav~\cite{sotelo2017char2wav}, WaveRNN~\cite{kalchbrenner2018efficient} directly take linguistic features as input and generate waveform. Later, ~\citet{prenger2019waveglow,kim2019flowavenet,kumar2019melgan,yamamoto2020parallel} take mel-spectrograms as input and generate waveform. Since speech waveform is very long, autoregressive waveform generation takes much inference time. Thus, generative models such as Flow~\cite{dinh2014nice,kingma2016improved,kingma2018glow}, GAN~\cite{goodfellow2014generative}, VAE~\cite{kingma2013auto}, and DDPM (Denoising Diffusion Probabilistic Model, Diffusion for short)~\cite{sohl2015deep,ho2020denoising} are used in waveform generation. Accordingly, we divide the neural vocoders into different categories: 1) Autoregressive vocoders, 2) Flow-based vocoders, 3) GAN-based vocoders, 4) VAE-based vocoders, and 5) Diffusion-based vocoders. We list some representative vocoders in Table~\ref{tab_vocoder_summary} and describe them as follows.

\paragraph{Autoregressive Vocoders}
WaveNet~\cite{oord2016wavenet} is the first neural-based vocoder, which leverages dilated convolution to generate waveform points autoregressively. Unlike the vocoder analysis and synthesis in SPSS~\cite{fukada1992adaptive,tokuda1994mel,kawahara1999restructuring,itakura1975line,kawahara2006straight,morise2016world}, WaveNet incorporates almost no prior knowledge about audio signals, and purely relies on end-to-end learning. The original WaveNet, as well as some following works that leverage WaveNet as vocoder~\cite{arik2017deep,gibiansky2017deep}, generate speech waveform conditioned on linguistic features, while WaveNet can be easily adapted to condition on linear-spectrograms~\cite{gibiansky2017deep} and mel-spectrograms~\cite{tamamori2017speaker,ping2018deep,shen2018natural}. Although WaveNet achieves good voice quality, it suffers from slow inference speed. Therefore, a lot of works~\cite{paine2016fast,hsu2020wg,mehri2016samplernn} investigate lightweight and fast vocoders. SampleRNN~\cite{mehri2016samplernn} leverages a hierarchical recurrent neural network for unconditional waveform generation, and it is further integrated into Char2Wav~\cite{sotelo2017char2wav} to generate waveform conditioned on acoustic features. Further, WaveRNN~\cite{zhang2002efficient} is developed for efficient audio synthesis, using a recurrent neural network and leveraging several designs including dual softmax layer, weight pruning, and subscaling techniques to reduce the computation. \citet{lorenzo2019towards,paul2020speaker,jiao2021universal} further improve the robustness and universality of the vocoders. LPCNet~\cite{valin2019lpcnet,valin2019real} introduces conventional digital signal processing into neural networks, and uses linear prediction coefficients to calculate the next waveform point while leveraging a lightweight RNN to compute the residual. LPCNet generates speech waveform conditioned on BFCC (bark-frequency cepstral coefficients) features, and can be easily adapted to condition on mel-spectrograms. Some following works further improve LPCNet from different perspectives, such as reducing complexity for speedup~\cite{vipperla2020bunched,popov2020gaussian,kanagawa2020lightweight}, and improving stability for better quality~\cite{hwang2020improving}.

\paragraph{Flow-based Vocoders}
Normalizing flow~\cite{dinh2014nice,dinh2016density,rezende2015variational,kingma2016improved,kingma2018glow} is a kind of generative model. It transforms a probability density with a sequence of invertible mappings~\cite{rezende2015variational}. Since we can get a standard/normalized probability distribution (e.g., Gaussion) through the sequence of invertible mappings based on the change-of-variables rules, this kind of flow-based generative model is called as a normalizing flow. During sampling, it generates data from a standard probability distribution through the inverse of these transforms. The flow-based models used in neural TTS can be divided into two categories according to the two different techniques~\cite{papamakarios2019normalizing}: 1) autoregressive transforms~\cite{kingma2016improved} (e.g., inverse autoregressive flow used in Parallel WaveNet~\cite{oord2018parallel}), and 2) bipartite transforms (e.g., Glow~\cite{kingma2018glow} used in WaveGlow~\cite{prenger2019waveglow}, and RealNVP~\cite{dinh2016density} used in FloWaveNet~\cite{kim2019flowavenet}), as shown in Table~\ref{tab_flow_summary}.

\begin{table}[h!]
\small
	\caption{Several representative flow-based models and their formulations~\cite{ping2020waveflow}.}
	\centering
	\begin{tabular}{l |l | l |l}
	\toprule
		\multicolumn{2}{l|}{Flow} & Evaluation $z=f^{-1}(x)$ & Synthesis $x=f(z)$\\
		\midrule
		\multirow{3}{*}{AR} &  AF~\cite{papamakarios2017masked} & $z_t=x_t\cdot \sigma_t(x_{<t}; \theta) + \mu_t (x_{<t};\theta)$ &  $x_t=\frac{z_t-u_t(x_{<t};\theta)}{\sigma_t(x_{<t};\theta)}$   \\
		\cmidrule{2-4}
		          &  IAF~\cite{kingma2016improved} & $z_t=\frac{x_t-\mu_t(z_{<t};\theta)}{\sigma_t(z_{<t};\theta)}$ &$x_t=z_t\cdot \sigma_t(z_{<t}; \theta) + \mu_t (z_{<t};\theta)$   \\
		\midrule
		\multirow{3}{*}{Bipartite} & RealNVP~\cite{dinh2016density} & $z_a = x_a$,  & $x_a = z_a$,   \\
		\cmidrule{2-2}
		          &  Glow~\cite{kingma2018glow} & $z_b=x_b\cdot \sigma_b(x_a;\theta) + \mu_b(x_a;\theta)$ &		          $x_b=\frac{z_b-\mu_b(x_a;\theta)}{\sigma_b(x_a;\theta)}$ \\
        \bottomrule
	\end{tabular}
	\vspace{0.3cm}
	\label{tab_flow_summary}
\end{table}

\begin{itemize}[leftmargin=*]
\item Autoregressive transforms, e.g., inverse autoregressive flow (IAF)~\cite{kingma2016improved}. IAF can be regarded as a dual formulation of autoregressive flow (AF)~\cite{papamakarios2017masked,huang2018neural}. The training of AF is parallel while the sampling is sequential. In contrast, the sampling in IAF is parallel while the inference for likelihood estimation is sequential. Parallel WaveNet~\cite{oord2018parallel} leverages probability density distillation to marry the efficient sampling of IAF with the efficient training of AR modeling. It uses an autoregressive WaveNet as the teacher network to guide the training of the student network (Parallel WaveNet) to approximate the data likelihood. Similarly, ClariNet~\cite{ping2018clarinet} uses IAF and teacher distillation, and leverages a closed-form KL divergence to simplify and stabilize the distillation process. Although Parallel Wavenet and ClariNet can generate speech in parallel, it relies on sophisticated teacher-student training and still requires large computation.

\item Bipartite transforms, e.g., Glow~\cite{kingma2018glow} or RealNVP~\cite{dinh2016density}. To ensure the transforms to be invertible, bipartite transforms leverage the affine coupling layers that ensure the output can be computed from the input and vice versa. Some vocoders based on bipartite transforms include WaveGlow~\cite{prenger2019waveglow} and FloWaveNet~\cite{kim2019flowavenet}, which achieve high voice quality and fast inference speed. 
\end{itemize}

Both autoregressive and bipartite transforms have their advantages and disadvantages~\cite{ping2020waveflow}: 1) Autoregressive transforms are more expressive than bipartite transforms by modeling dependency between data distribution $x$ and standard probability distribution $z$, but require teacher distillation that is complicated in training. 2) Bipartite transforms enjoy much simpler training pipeline, but usually require larger number of parameters (e.g., deeper layers, larger hidden size) to reach comparable capacities with autoregressive models. To combine the advantages of both autoregressive and bipartite transforms, WaveFlow~\cite{ping2020waveflow} provides a unified view of likelihood-based models for audio data to explicitly trade inference parallelism for model capacity. In this way, WaveNet, WaveGlow, and FloWaveNet can be regarded as special cases of WaveFlow.


\paragraph{GAN-based Vocoders}
Generative adversarial networks (GANs)~\cite{goodfellow2014generative} have been widely used in data generation tasks, such as image generation~\cite{goodfellow2014generative,zhu2017unpaired}, text generation~\cite{yu2017seqgan}, and audio generation~\cite{donahue2018adversarial}. GAN consists a generator for data generation, and a discriminator to judge the authenticity of the generated data. A lot of vocoders leverage GAN to ensure the audio generation quality, including WaveGAN~\cite{donahue2018adversarial}, GAN-TTS~\cite{binkowski2019high}, MelGAN~\cite{kumar2019melgan}, Parallel WaveGAN~\cite{yamamoto2020parallel}, HiFi-GAN~\cite{kong2020hifi}, and other GAN-based vocoders~\cite{yamamoto2019probability,wu2020quasi,song2021improved,you2021gan,wang2021improve,jang2020universal}. 


\begin{table}[h!]
\small
	\caption{Several representative GAN based vocoders and their characteristics.}
	\centering
	\begin{tabular}{l |c | c |c}
	\toprule
	GAN & Generator & Discriminator & Loss \\
	\midrule
	WaveGAN~\cite{donahue2018adversarial} & DCGAN~\cite{radford2015unsupervised} & / & WGAN-GP~\cite{gulrajani2017improved}\\
	\midrule
    GAN-TTS~\cite{binkowski2019high} & /  & Random Window D & Hinge-Loss GAN~\cite{lim2017geometric} \\
    \midrule
    \multirow{2}{*}{MelGAN~\cite{kumar2019melgan}}  & \multirow{2}{*}{/}   &  \multirow{2}{*}{Multi-Scale D} & LS-GAN~\cite{mao2017least}\\
            &&& Feature Matching Loss~\cite{larsen2016autoencoding}  \\ 
    \midrule
    \multirow{2}{*}{Par.WaveGAN~\cite{yamamoto2020parallel}} & \multirow{2}{*}{WaveNet~\cite{oord2016wavenet}} & \multirow{2}{*}{/} &  LS-GAN, \\
        &&& Multi-STFT Loss\\
    \midrule
    \multirow{2}{*}{HiFi-GAN~\cite{kong2020hifi}} & \multirow{2}{*}{\shortstack{Multi-Receptive \\Field Fusion}} & Multi-Period D,  &  LS-GAN, STFT Loss,   \\
    && Multi-Scale D &  Feature Matching Loss\\
    \midrule
    \multirow{2}{*}{VocGAN~\cite{yang2020vocgan}} & \multirow{2}{*}{Multi-Scale G} & \multirow{2}{*}{Hierarchical D} & LS-GAN, Multi-STFT Loss,\\
    &&&Feature Matching Loss \\
    \midrule
    \multirow{2}{*}{GED~\cite{gritsenko2020spectral}} & \multirow{2}{*}{/}  & \multirow{2}{*}{Random Window D} & Hinge-Loss GAN, \\
    &&& Repulsive loss \\
    \bottomrule
	\end{tabular}
	\vspace{0.3cm}
	\label{tab_gan_summary}
\end{table}

We summarize the characteristics according to the generators, discriminators, and losses used in each vocoder in Table~\ref{tab_gan_summary}.
\begin{itemize}[leftmargin=*]
\item Generator. Most GAN-based vocoders use dilated convolution to increase the receptive field to model the long-dependency in waveform sequence, and transposed convolution to upsample the condition information (e.g., linguistic features or mel-spectrograms) to match the length of waveform sequence. \citet{yamamoto2020parallel} choose to upsample the conditional information one time, and then perform dilated convolution to ensure model capacity. However, this kind of upsampling increases the sequence length too early, resulting larger computation cost. Therefore, some vocoders~\cite{kumar2019melgan,kong2020hifi} choose to iteratively upsample the condition information and perform dilated convolution, which can avoid too long sequence in the lower layers. Specifically, VocGAN~\cite{yang2020vocgan} proposes a multi-scale generator that can gradually output waveform sequence at different scales, from coarse-grained to fine-grained. HiFi-GAN~\cite{kong2020hifi} processes different patterns of various lengths in parallel through a multi-receptive field fusion module, and also has the flexibility to trade off between synthesis efficiency and sample quality.

\item Discriminator. The research efforts~\cite{binkowski2019high,kumar2019melgan,kong2020hifi,yang2020vocgan} on discriminators focus on how to design models to capture the characteristics of waveform, in order to provide better guiding signal for the generators. We review these efforts as follows: 1) Random window discriminators, proposed in GAN-TTS~\cite{binkowski2019high}, which use multiple discriminators, where each is feeding with different random windows of waveform with and without conditional information. Random window discriminators have several benefits, such as evaluating audios in different complementary way, simplifying the true/false judgements compared with full audio, and acting as a data augmentation effect, etc. 2) Multi-scale discriminators, proposed in MelGAN~\cite{kumar2019melgan}, which use multiple discriminators to judge audios in different scales (different downsampling ratios compared with original audio). The advantage of multi-scale discriminators is that the discriminator in each scale can focus on the characteristics in different frequency ranges. 3) Multi-period discriminators, proposed in HiFi-GAN~\cite{kong2020hifi}, which leverage multiple discriminators, where each accepts equally spaced samples of an input audio with a period. Specifically, the 1D waveform sequence with a length of $T$ is reshaped into a 2D data $[p, T/p]$ where $p$ is the period, and processed by a 2D convolution. Multi-period discriminators can capture different implicit structures by looking at different parts of an input audio in different periods. 4) Hierarchical discriminators, leveraged in VocGAN~\cite{yang2020vocgan} to judge the generated waveform in different resolutions from coarse-grained to fine-grained, which can guide the generator to learn the mapping between the acoustic features and waveform in both low and high frequencies.

\item Loss. Except for the regular GAN losses such as WGAN-GP~\cite{gulrajani2017improved}, hinge-loss GAN~\cite{lim2017geometric}, and LS-GAN~\cite{mao2017least}, other specific losses such as STFT loss~\cite{arik2018fast,yamamoto2019probability} and feature matching loss~\cite{larsen2016autoencoding} are also leveraged. These additional losses can improve the stability and efficiency of adversarial training~\cite{yamamoto2020parallel}, and improve the perceptual audio quality. \citet{gritsenko2020spectral} propose a generalized energy distance with a repulsive term to better capture the multi-modal waveform distribution.
\end{itemize}

\paragraph{Diffusion-based Vocoders} Recently, there are some works leveraging denoising diffusion probabilistic models (DDPM or Diffusion)~\cite{ho2020denoising} for vocoders, such as DiffWave~\cite{kong2020diffwave}, WaveGrad~\cite{chen2020wavegrad}, and PriorGrad~\cite{lee2021priorgrad}. The basic idea is to formulate the mapping between data and latent distributions with diffusion process and reverse process: in the diffusion process, the waveform data sample is gradually added with some random noises and finally becomes Gaussion noise; in the reverse process, the random Gaussion noise is gradually denoised into waveform data sample step by step. Diffusion-based vocoders can generate speech with very high voice quality, but suffer from slow inference speed due to the long iterative process. Thus, a lot of works on diffusion models~\cite{song2020denoising,lee2021priorgrad,watson2021learning,kong2021fast} are investigating how to reduce inference time while maintaining generation quality.

\paragraph{Other Vocoders}
Some works leverage neural-based source-filter model for waveform generation~\cite{wang2019neural_b,wang2019neural,wangneural,liu2020neural,juvela2019gelp,juvela2019glotnet,engel2019ddsp,song2020neural,yoneyama2021unified}, aiming to achieve high voice quality while maintaining controllable speech generation. \citet{govalkar2019comparison} conduct a comprehensive study on different kinds of vocoders. \citet{hsu2019towards} study the robustness of vocoders by evaluating several common vocoders with comprehensive experiments.

\paragraph{Discussions} We summarize the characteristics of different kinds of generative models used in vocoders, as shown in Table~\ref{tab_generative_summary}: 1) In terms of mathematical simplicity, autoregressive (AR) based models are easier than other generative models such as VAE, Flow, Diffusion, and GAN. 2) All the generative models except AR can support parallel speech generation. 3) Except for AR models, all generative models can support latent manipulations to some extent (some GAN-based vocoders do not take random Gaussian noise as model input, and thus cannot support latent manipulation). 4) GAN-based models cannot estimate the likelihood of data samples, while other models enjoy this benefit. 



\begin{table}[h!]
\small
	\caption{Some characteristics of several representative generative models used in vocoders.}
	\centering
	\begin{tabular}{l | c| c |c c | c| c }
	\toprule
		Generative Model & AR & VAE & Flow/AR & Flow/Bipartite & Diffusion & GAN \\
		\midrule
		Vocoder (e.g.)  & WaveNet & WaveVAE & Par.WaveNet & WaveGlow & DiffWave & MelGAN \\
		\midrule
		 Simple  & Y       & N         & N    &N    &  N        & N \\
		 Parallel& N       & Y         & Y    &Y    &  Y        & Y \\
		 Latent Manipulate & N & Y   & Y   & Y    &  Y        & Y* \\
		 Likelihood Estimate & Y&Y   & Y   & Y    &  Y        & N \\
        \bottomrule
	\end{tabular}
	\vspace{0.3cm}
	\label{tab_generative_summary}
\end{table}

\subsection{Towards Fully End-to-End TTS}
\label{sec_funda_e2e}

\begin{figure} [t!]
\begin{minipage}[b]{1.0\linewidth} 
\centering 
\includegraphics[scale=0.47]{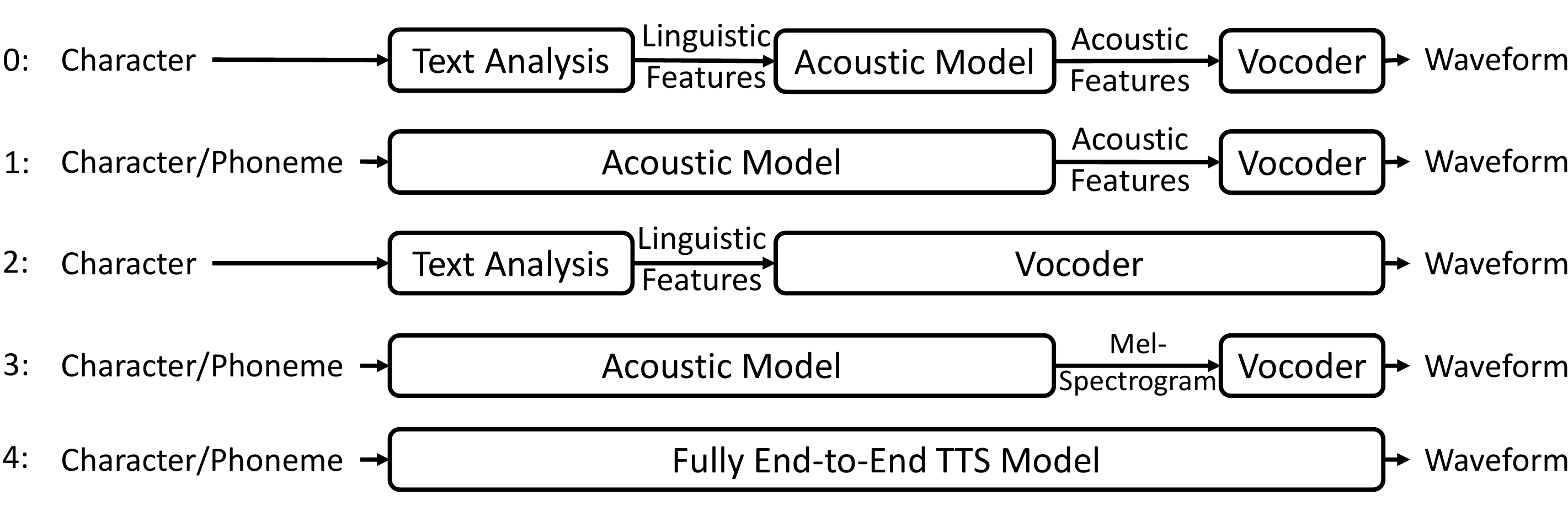}
\end{minipage} 
\begin{minipage}[b]{1.0\linewidth} 
\centering 
\small
\begin{tabular}{c|l} 
\toprule
Stage &  Models \\
\midrule
0 & SPSS~\cite{yoshimura1999simultaneous,tokuda2000speech,yoshimura2002simultaneous,zen2009statistical,tokuda2013speech} \\
1 & ARST~\cite{wang2016first} \\
2 & WaveNet~\cite{oord2016wavenet}, DeepVoice 1/2~\cite{arik2017deep,gibiansky2017deep}, Par. WaveNet~\cite{oord2018parallel}, WaveRNN~\cite{kalchbrenner2018efficient}, HiFi-GAN~\cite{binkowski2019high}  \\
3 & DeepVoice 3~\cite{ping2018deep}, Tacotron 2~\cite{shen2018natural}, FastSpeech 1/2~\cite{ren2019fastspeech,ren2021fastspeech}, WaveGlow~\cite{prenger2019waveglow}, FloWaveNet~\cite{kim2019flowavenet} \\
4 & Char2Wav~\cite{sotelo2017char2wav}, ClariNet~\cite{ping2018clarinet}, FastSpeech 2s~\cite{ren2021fastspeech}, EATS~\cite{donahue2020end}, Wave-Tacotron~\cite{weiss2020wave}, VITS~\cite{kim2021conditional}  \\
\bottomrule
\end{tabular} 
\captionof{figure}{The progressively end-to-end process for TTS models.} 
\label{fig_progress_e2e}
\end{minipage} 
\end{figure}

Fully end-to-end TTS models can generate speech waveform from character or phoneme sequence directly, which have the following advantages: 1) It requires less human annotation and feature development (e.g., alignment information between text and speech); 2) The joint and end-to-end optimization can avoid error propagation in cascaded models (e.g., Text Analysis + Acoustic Model + Vocoder); 3) It can also reduce the training, development and deployment cost. 

However, there are big challenges to train TTS models in an end-to-end way, mainly due to the different modalities between text and speech waveform, as well as the huge length mismatch between character/phoneme sequence and waveform sequence. For example, for a speech with a length of 5 seconds and about 20 words, the length of the phoneme sequence is just about 100, while the length of the waveform sequence is 80k (if the sample rate is 16kHz). It is hard to put the waveform points of the whole utterance into model training, due to the limit of memory. It is hard to capture the context representations if only using a short audio clip for the end-to-end training.

Due to the difficulty of fully end-to-end training, the development of neural TTS follows a progressive process towards fully end-to-end models. Figure~\ref{fig_progress_e2e} illustrates this progressive process starting from early statistic parametric synthesis~\cite{yoshimura1999simultaneous,tokuda2000speech,yoshimura2002simultaneous,zen2009statistical,tokuda2013speech}. The process towards fully end-to-end models typically contains these upgrades: 1) Simplifying text analysis module and linguistic features. In SPSS, text analysis module contains different functionalities such as text normalization, phrase/word/syllable segmentation, POS tagging, prosody prediction, grapheme-to-phoneme conversion (including polyphone disambiguation). In end-to-end models, only the text normalization and grapheme-to-phoneme conversion are retained to convert characters into phonemes, or the whole text analysis module is removed by directly taking characters as input. 2) Simplifying acoustic features, where the complicated acoustic features such as MGC, BAP and F0 used in SPSS are simplified into mel-spectrograms. 3) Replacing two or three modules with a single end-to-end model. For example, the acoustic models and vocoders can be replaced with a single vocoder model such as WaveNet. Accordingly, we illustrate the progressive process in Figure~\ref{fig_progress_e2e} and describe it as follows.

\begin{table}[h!]
\small
	\caption{A list of fully end-to-end TTS models. }
	\centering
	\begin{tabular}{l | l l l l }
	\toprule
		Model & One-Stage Training & AR/NAR & Modeling & Architecture \\
		\midrule
        Char2Wav~\cite{sotelo2017char2wav} & N  & AR & Seq2Seq & RNN \\
        ClariNet~\cite{ping2018clarinet} & N  & AR & Flow & CNN \\
        FastSpeech 2s~\cite{ren2021fastspeech} & Y  & NAR & GAN & Self-Att/CNN \\
        EATS~\cite{donahue2020end} & Y  & NAR & GAN & CNN \\
        Wave-Tacotron~\cite{weiss2020wave} & Y  & AR & Flow & CNN/RNN/Hybrid \\
        EfficientTTS-Wav~\cite{miao2020efficienttts} & Y  & NAR & GAN & CNN \\
        VITS~\cite{kim2021conditional} & Y  & NAR & VAE+Flow & CNN/Self-Att/Hybrid \\
        \bottomrule
	\end{tabular}
	\vspace{0.3cm}
	\label{tab_e2e_summary}
\end{table}

\begin{itemize}[leftmargin=*]
    \item Stage 0. Statistic parametric synthesis~\cite{yoshimura1999simultaneous,tokuda2000speech,yoshimura2002simultaneous,zen2009statistical,tokuda2013speech} uses three basic modules, where text analysis module converts characters into linguistic features, and acoustic models generate acoustic features from linguistic features (where the target acoustic features are obtained through vocoder analysis), and then vocoders synthesize speech waveform from acoustic features through parametric calculation. 
    \item Stage 1. \citet{wang2016first} in statistic parametric synthesis explore to combine the text analysis and acoustic model into an end-to-end acoustic model that directly generates acoustic features from phoneme sequence, and then uses a vocoder in SPSS to generate waveform. 
    \item Stage 2. WaveNet~\cite{oord2016wavenet} is first proposed to directly generate speech waveform from linguistic features, which can be regarded as a combination of an acoustic model and a vocoder. This kind of models~\cite{oord2016wavenet,oord2018parallel,kalchbrenner2018efficient,binkowski2019high} still require a text analysis module to generate linguistic features. 
    \item Stage 3. Tacotron~\cite{wang2017tacotron} is further proposed to simplify linguistic and acoustic features, which directly predicts linear-spectrograms from characters/phonemes with an encoder-attention-decoder model, and converts linear-spectrograms into waveform with Griffin-Lim~\cite{griffin1984signal}. The following works such as DeepVoice 3~\cite{ping2018deep}, Tacotron 2~\cite{shen2018natural}, TransformerTTS~\cite{li2019neural}, and FastSpeech 1/2~\cite{ren2019fastspeech,ren2021fastspeech} predict mel-spectrograms from characters/phonemes and further use a neural vocoder such as WaveNet~\cite{oord2016wavenet}, WaveRNN~\cite{kalchbrenner2018efficient}, WaveGlow~\cite{prenger2019waveglow}, FloWaveNet~\cite{kim2019flowavenet}, and Parallel WaveGAN~\cite{yamamoto2020parallel} to generate waveform.
    \item Stage 4. Some fully end-to-end TTS models are developed for direct text to waveform synthesis, as listed in Table~\ref{tab_e2e_summary}. Char2Wav~\cite{sotelo2017char2wav} leverages an RNN-based encoder-attention-decoder model to generate acoustic features from characters, and then uses SampleRNN~\cite{mehri2016samplernn} to generate waveform. The two models are jointly tuned for direct speech synthesis. Similarly, ClariNet~\cite{ping2018clarinet} jointly tunes an autoregressive acoustic model and a non-autoregressive vocoder for direct waveform generation. FastSpeech 2s~\cite{ren2021fastspeech} directly generate speech from text with a fully parallel structure, which can greatly speed up inference. To alleviate the difficulty of joint text-to-waveform training, it leverages an auxiliary mel-spectrogram decoder to help learn the contextual representations of phoneme sequence. A concurrent work called EATS~\cite{donahue2020end} also directly generates waveform from characters/phonemes, which leverages duration interpolation and soft dynamic time wrapping loss for end-to-end alignment learning. Wave-Tacotron~\cite{weiss2020wave} builds a flow-based decoder on Tacotron to directly generate waveform, which uses parallel waveform generation in the flow part but still autoregressive generation in the Tacotron part. 
\end{itemize}

\tikzstyle{leaf}=[mybox,minimum height=1.2em,
fill=hidden-orange!50, text width=5em,  text=black,align=left,font=\footnotesize,
inner xsep=4pt,
inner ysep=1pt,
]

\begin{figure*}[h]
  \centering
\begin{forest}
  forked edges,
  for tree={
  grow=east,
  reversed=true,  
  anchor=base west,
  parent anchor=east,
  child anchor=west,
  base=left,
  font=\normalsize,
  rectangle,
  draw=hiddendraw,
  rounded corners,
  align=left,
  text width=8em,
  minimum width=2.5em,
  inner xsep=4pt,
  inner ysep=0pt,
  },
  where level=1{text width=3.5em,font=\footnotesize}{},
  where level=2{font=\footnotesize,yshift=0.0pt}{},
  [AR or NAR
            [AR
                [\textcolor{black}{WaveNet}~\cite{oord2016wavenet}{,}
                 \textcolor{black}{SampleRNN}~\cite{mehri2016samplernn}{,}
                 \textcolor{black}{WaveRNN}~\cite{kalchbrenner2018efficient}{,}\\
                 \textcolor{black}{DeepVoice 1/2/3}~\cite{arik2017deep,gibiansky2017deep,ping2018deep}{,}
                 \textcolor{black}{Tacotron 1/2}~\cite{wang2017tacotron,shen2018natural}{,} \\
                 \textcolor{black}{TransformerTTS}~\cite{li2019neural},leaf,text width=22em
                ]
            ]
            [NAR
                [\textcolor{black}{Par.WaveNet}~\cite{oord2018parallel}{,}
                 \textcolor{black}{WaveGlow}~\cite{prenger2019waveglow}{,}
                 FloWaveNet~\cite{kim2019flowavenet}{,}\\
                 \textcolor{black}{MelGAN}~\cite{kumar2019melgan}{,} 
                 \textcolor{black}{FastSpeech 1/2/2s}~\cite{ren2019fastspeech,ren2021fastspeech}{,}
                 \textcolor{black}{EATS}~\cite{donahue2020end}{,}\\
                 \textcolor{black}{ParaNet}~\cite{peng2020non}{,} 
                 VITS~\cite{kim2021conditional}{,}
                 \textcolor{black}{HiFi-GAN}~\cite{kong2020hifi},leaf,text width=22em
                ]
            ]
    ]
\end{forest}
\vspace{0.3cm}

\begin{forest}
  forked edges,
  for tree={
  grow=east,
  reversed=true,  
  anchor=base west,
  parent anchor=east,
  child anchor=west,
  base=left,
  font=\normalsize,
  rectangle,
  draw=hiddendraw,
  rounded corners,
  align=left,
  minimum width=2.5em,
  text width=8em,
  inner xsep=4pt,
  inner ysep=0pt,
  },
  where level=1{text width=3.5em,font=\footnotesize}{},
  where level=2{font=\footnotesize,yshift=0.0pt}{},
    [Generative Model
            [Seq2Seq
                [\textcolor{black}{WaveNet}~\cite{oord2016wavenet}{,}
                \textcolor{black}{DeepVoice 1/2/3}~\cite{arik2017deep,gibiansky2017deep,ping2018deep}{,} \\
                 \textcolor{black}{Tacotron 1/2}~\cite{wang2017tacotron,shen2018natural}{,}
                 \textcolor{black}{WaveRNN}~\cite{kalchbrenner2018efficient}{,} \\
                 \textcolor{black}{FastSpeech 1/2}~\cite{ren2019fastspeech,ren2021fastspeech},leaf,text width=22em
                ]
            ]
            [Flow
                [\textcolor{black}{Par.WaveNet}~\cite{oord2018parallel}{,}
                 \textcolor{black}{ClariNet}~\cite{ping2018clarinet}{,}
                  FloWaveNet~\cite{kim2019flowavenet}{,}\\
                 \textcolor{black}{WaveGlow}~\cite{prenger2019waveglow}{,} 
                 \textcolor{black}{WaveFlow}~\cite{ping2020waveflow}{,}
                 \textcolor{black}{Glow-TTS}~\cite{kim2020glow}{,}\\ Flow-TTS~\cite{miao2020flow}{,} 
                 Flowtron~\cite{valle2020flowtron}{,} 
                 \textcolor{black}{Wave-Tacotron}~\cite{weiss2020wave},leaf,text width=22em
                ]
            ]
            [GAN
                [\textcolor{black}{WaveGAN}~\cite{donahue2018adversarial}{,} \textcolor{black}{GAN-TTS}~\cite{binkowski2019high}{,} \textcolor{black}{MelGAN}~\cite{kumar2019melgan}{,}\\
                 \textcolor{black}{Par.WaveGAN}~\cite{yamamoto2020parallel}{,}
                 \textcolor{black}{FastSpeech 2s}~\cite{ren2021fastspeech}{,} \textcolor{black}{EATS}~\cite{donahue2020end}{,}\\
                 \textcolor{black}{HiFi-GAN}~\cite{kong2020hifi}{,} VocGAN~\cite{yang2020vocgan}{,} Multi-SpectroGAN~\cite{lee2020multi},leaf,text width=22em
                ]
            ]
            [VAE
                [\textcolor{black}{WaveVAE}~\cite{peng2020non}{,} VAE-TTS~\cite{zhang2019learningb}{,} GMVAE-Tacotron~\cite{hsu2018hierarchical},leaf,text width=22em
                ]
            ]
            [Diffusion
                [\textcolor{black}{DiffWave}~\cite{kong2020diffwave}{,}
                 \textcolor{black}{WaveGrad}~\cite{chen2020wavegrad}{,} Diff-TTS~\cite{jeong2021diff}{,}\\
                 Grad-TTS~\cite{popov2021grad}{,} PriorGrad~\cite{lee2021priorgrad},leaf,text width=22em
                ]
            ]
    ]
\end{forest}

\vspace{0.3cm}
\begin{forest}
  forked edges,
  for tree={
  grow=east,
  reversed=true,  
  anchor=base west,
  parent anchor=east,
  child anchor=west,
  base=left,
  font=\normalsize,
  rectangle,
  draw=hiddendraw,
  rounded corners,
  align=left,
  text width=8em,
  minimum width=2.5em,
  inner xsep=4pt,
  inner ysep=0pt,
  },
  where level=1{text width=3.5em,font=\footnotesize}{},
  where level=2{font=\footnotesize,yshift=0.0pt}{},
   [Network Structure
            [CNN
                [\textcolor{black}{WaveNet}~\cite{oord2016wavenet}{,}
                \textcolor{black}{Par.WaveNet}~\cite{oord2018parallel}{,}
                \textcolor{black}{DeepVoice 3}~\cite{ping2018deep}{,} \\
                \textcolor{black}{ClariNet}~\cite{ping2018clarinet}{,}
                \textcolor{black}{ParaNet}~\cite{peng2020non}{,}
                 \textcolor{black}{MelGAN}~\cite{kumar2019melgan}{,} 
                 \textcolor{black}{EATS}~\cite{donahue2020end}{,} \\
                 DCTTS~\cite{tachibana2018efficiently}{,}
                \textcolor{black}{WaveGlow}~\cite{prenger2019waveglow}{,}
                 FloWaveNet~\cite{kim2019flowavenet},leaf,text width=22em
                ]
            ]
            [RNN
                [\textcolor{black}{SampleRNN}~\cite{mehri2016samplernn}{,}
                \textcolor{black}{Tacotron 2}~\cite{shen2018natural}{,}
                \textcolor{black}{WaveRNN}~\cite{kalchbrenner2018efficient}{,}\\
                \textcolor{black}{LPCNet}~\cite{valin2019lpcnet}{,} DurIAN~\cite{yu2020durian},leaf,text width=22em
                ]
            ]
            [Self-Att
                [\textcolor{black}{TransformerTTS}~\cite{li2019neural}{,}
                \textcolor{black}{FastSpeech 1/2/2s}~\cite{ren2019fastspeech,ren2021fastspeech}{,}\\
                \textcolor{black}{AlignTTS}~\cite{zeng2020aligntts}{,}
                FastPitch~\cite{lancucki2020fastpitch}{,}
                \textcolor{black}{JDI-T}~\cite{lim2020jdi},leaf,text width=22em
                ]
            ]
            [Hybrid
                [DeepVoice 1/2~\cite{arik2017deep,gibiansky2017deep}{,}
                Tacotron~\cite{wang2017tacotron}{,}
                DurIAN~\cite{yu2020durian}{,}\\
                LightSpeech~\cite{luo2021lightspeech}{,}
                Wave-Tacotron~\cite{weiss2020wave},leaf,text width=22em
                ]
            ]
    ]
\end{forest}
\caption{Some other taxonomies of neural TTS from the perspectives of AR/NAR, generative model, and network structure.}
\label{other_taxonomy_of_tts}
\end{figure*}
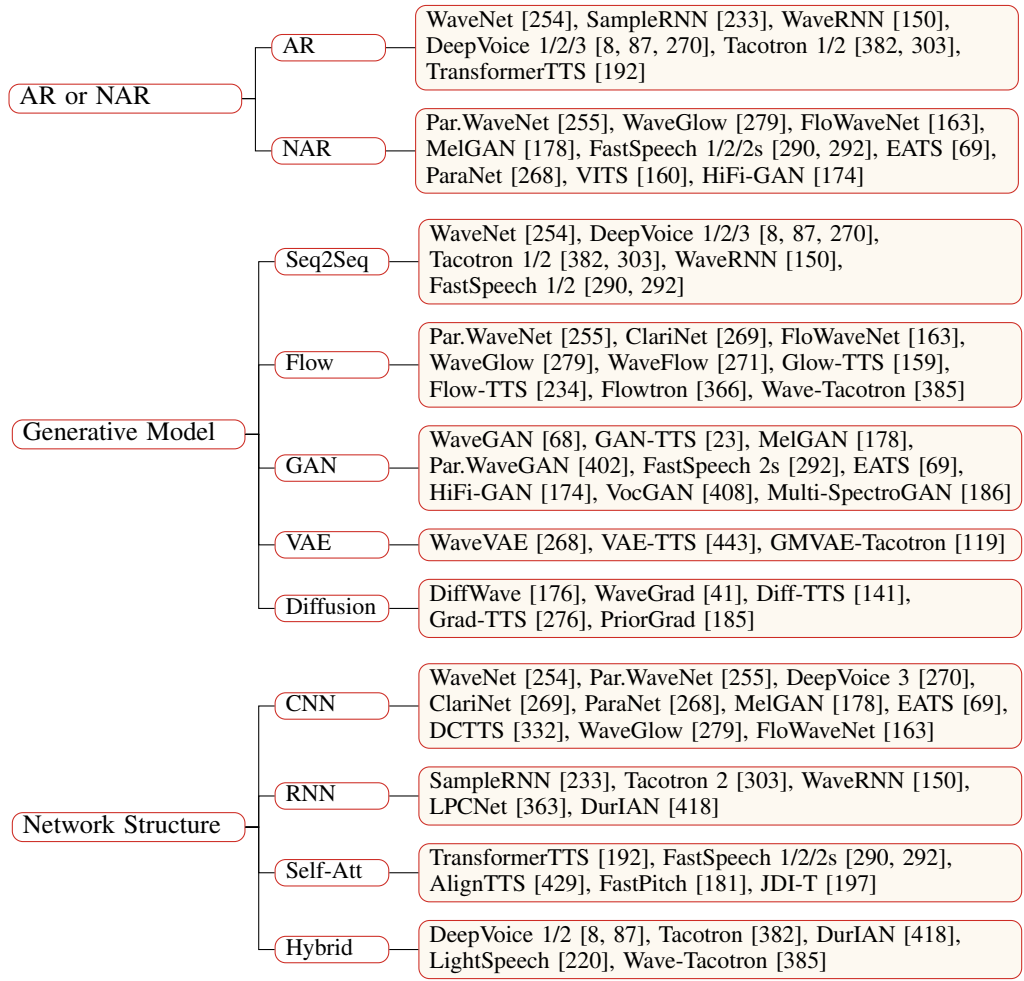

\begin{figure}[h]
  \centering
  \includegraphics[scale=0.6]{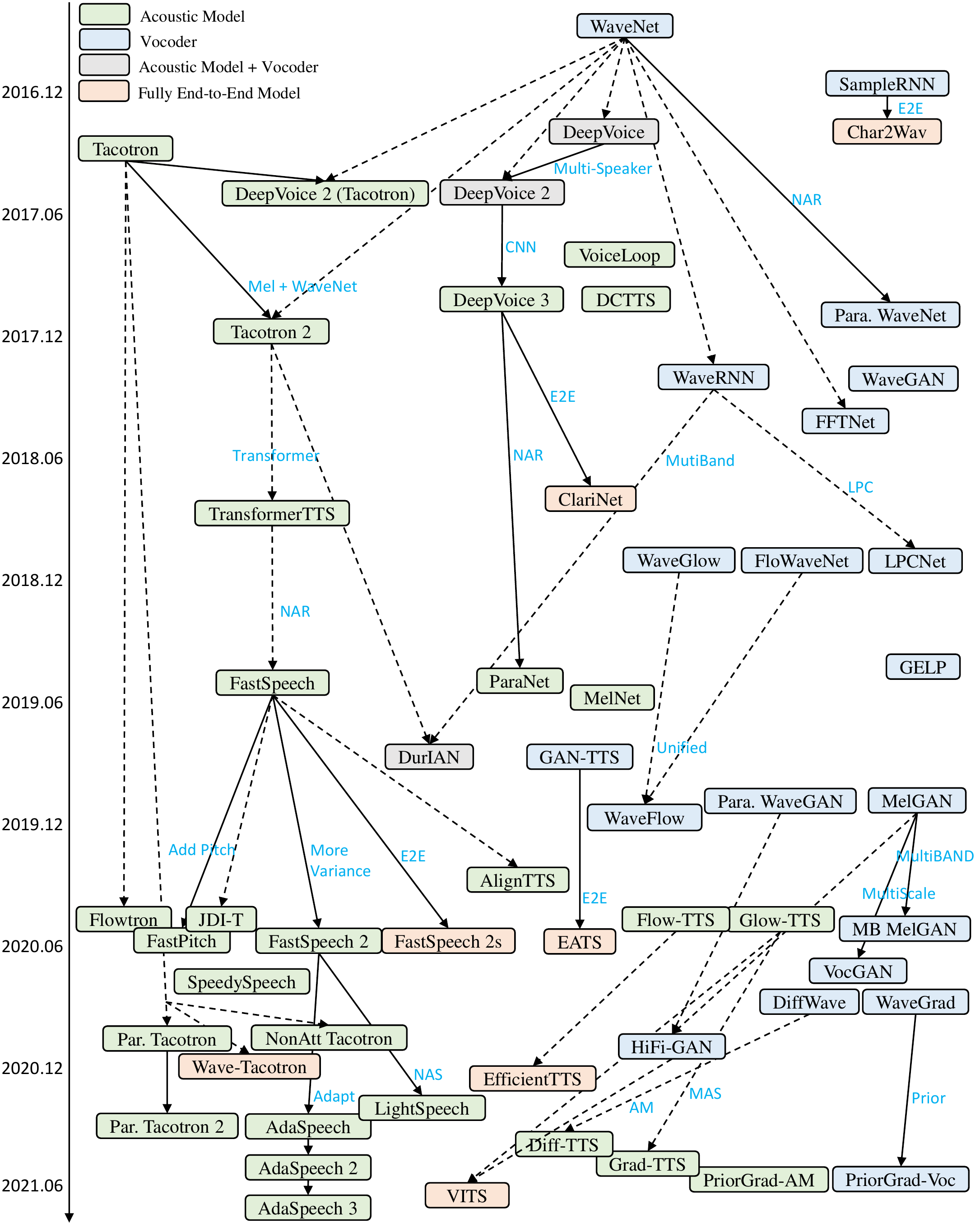}
  \caption{The evolution of neural TTS models.}
  \label{fig_tts_timeline}
\end{figure}

\subsection{Other Taxonomies}
\label{sec_funda_other}
Besides the main taxonomy from the perspective of key components and data flow as shown in Figure~\ref{main_taxonomy}, we can also categorize TTS works from several different taxonomies, as shown in Figure~\ref{other_taxonomy_of_tts}: 1) \textbf{Autoregressive or non-autoregressive}. We can divide these works into autoregressive and non-autoregressive generation models. 2) \textbf{Generative model}. Since TTS is a typical sequence generation task and can be modeled through typical generative models, we can categorize in terms of different generative models: normal sequence generation model, flow, GAN, VAE, and diffusion model. 3) \textbf{Network structure}. We can divide the works according to their network structures, such as CNN, RNN, self-attention, and hybrid structures (which contains more than one type of structures, such as CNN+RNN, CNN+self-attention). 

\paragraph{Evolution of Neural TTS Models} In order to better understand the development of various research works on neural TTS and their relationships, we illustrate the evolution of neural TTS models, as shown in Figure~\ref{fig_tts_timeline}. Note that we organize the research works according to the time that the paper is open to the public (e.g., put on arXiv), but not formally published later. We choose the early time since we appreciate researchers making their paper public early to encourage knowledge sharing. Since the research works on neural TTS are so abundant, we only choose some representative works in Figure~\ref{fig_tts_timeline}, and list more works in Table~\ref{tab_all_paper_list}.

\section{Advanced Topics in TTS}
\label{sec_advanced}

\tikzstyle{leaf}=[mybox,minimum height=1.2em,
fill=hidden-orange!50, text width=5em,  text=black,align=left,font=\footnotesize,
inner xsep=4pt,
inner ysep=1pt,
]

\begin{figure*}[thp]
 \centering
\begin{forest}
  forked edges,
  for tree={
  grow=east,
  reversed=true,  
  anchor=base west,
  parent anchor=east,
  child anchor=west,
  base=left,
  font=\normalsize,
  rectangle,
  draw=hiddendraw,
  rounded corners,
  align=left,
  minimum width=2.5em,
  inner xsep=4pt,
  inner ysep=0pt,
  },
  where level=1{text width=4.2em,font=\normalsize}{},
  where level=2{text width=9em,font=\footnotesize}{},
  where level=3{font=\footnotesize,yshift=0.0pt}{},
    [TTS
        [Fast
            [Parallel Generation
                [FastSpeech 1/2~\cite{ren2019fastspeech,ren2021fastspeech}{,}  Par.WaveNet~\cite{oord2018parallel}\\
                WaveGlow~\cite{prenger2019waveglow}{,} FloWaveNet~\cite{kim2019flowavenet}\\ 
                ParaNet~\cite{peng2020non}{,} MelGAN~\cite{kumar2019melgan}{,} HiFi-GAN~\cite{kong2020hifi}\\
                DiffWave~\cite{kong2020diffwave}{,} WaveGrad~\cite{chen2020wavegrad}{,}  \cite{lee2021priorgrad,yamamoto2020parallel, kim2021conditional},leaf,text width=18.3em
                ]
            ]
            [Lightweight Model
                [WaveRNN~\cite{kalchbrenner2018efficient}{,} LightSpeech~\cite{luo2021lightspeech}\\
                SqueezeWave~\cite{zhai2020squeezewave}{,} \cite{kanagawa2020lightweight,hsu2020wg,huang2020devicetts,zeng2021lvcnet},leaf,text width=18.3em
                ]
            ]
            [Speedup with Domain\\Knowledge
                [LPCNet~\cite{valin2019lpcnet}{,} Multi-Band~\cite{yu2020durian,yang2020multi,okamoto2018investigation,cui2020efficient}\\
                FFTNet~\cite{jin2018fftnet}{,} Streaming~\cite{ellinas2020high,ma2020incremental,stephenson2020future,yanagita2019neural},leaf,text width=18.3em
                ]
            ]
        ]
        [Low-\\Resource
            [Self-Supervised Training
                [\cite{chung2019semi,wang2015word,zhang2019joint,fang2019towards,jia2021png,tjandra2019vqvae,liu2020towards,tu2020semi,dunbar2019zero,chen2021speech},leaf,text width=18.3em
                ]
            ]
            [Cross-Lingual Transfer
                [LRSpeech~\cite{xu2020lrspeech}{,} \cite{chen2019end,azizah2020hierarchical,de2020efficient,yang2020towards,prajwal2021data,he2021multilingual},leaf,text width=18.3em
                ]
            ]
            [Cross-Speaker Transfer
                [\cite{luong2019training,huybrechts2020low,yang2020towards,dai2020noise,chen2021adaspeech},leaf,text width=18.3em
                ]
            ]
            [Speech Chain/\\Back Transformation
                [SpeechChain~\cite{tjandra2017listening,tjandra2018machine}{,} LRSpeech~\cite{xu2020lrspeech,ren2019almost},leaf,text width=18.3em
                ]
            ]
            [Dataset Mining in the \\Wild
                [\cite{cooper2019text,hu2019neural,cooper2020pretraining},leaf,text width=18.3em]
            ]
        ]
        [Robust
            [Enhancing Attention          
                [Tacotron 2~\cite{wang2017tacotron}{,} DCTTS~\cite{tachibana2018efficiently}{,} SMA~\cite{he2019robust}\\ MultiSpeech~\cite{chen2020multispeech}{,} \cite{sotelo2017char2wav,shen2018natural,zhang2018forward,tachibana2018efficiently,ping2018deep,peng2020non},leaf,text width=18.3em
                ]
            ]
            [Replacing Attention\\with Duration
                [FastSpeech~\cite{ren2019fastspeech}{,} DurIAN~\cite{yu2020durian}{,} EATS~\cite{donahue2020end}\\
                Glow-TTS~\cite{kim2020glow}{,} \cite{li2020robutrans,beliaev2020talknet,vainer2020speedyspeech,zeng2020aligntts,elias2021parallel,shen2020non},leaf,text width=18.3em
                ]
            ]
            [Enhancing AR
                [GAN Exposure~\cite{guo2019new}{,} \cite{liu2019new,liu2020teacher,ren2019almost},leaf,text width=18.3em
                ]
            ]
            [Replacing AR with NAR
                [FastSpeech 1/2~\cite{ren2019fastspeech,ren2021fastspeech}{,}  ParaNet~\cite{peng2020non} \\ 
                EATS~\cite{donahue2020end}{,} Flow-TTS~\cite{miao2020flow}{,} \cite{kim2020glow,liu2021vara,vainer2020speedyspeech,elias2021parallel},leaf,text width=18.3em
                ]
            ]
        ]
        [Expressive
            [Modeling Variation \\Information
                [GST-Tacotron~\cite{wang2018style}{,} Ref-Tacotron~\cite{skerry2018towards}{,} \cite{chen2021adaspeech}\\
                Multi-SpectroGAN~\cite{lee2020multi}{,} PriorGrad~\cite{lee2021priorgrad}{,} \cite{choi2020attentron}\\
                VAE-TTS~\cite{zhang2019learningb}{,} Glow-TTS~\cite{kim2020glow}{,} \cite{hsu2018hierarchical,hsu2019disentangling}\\
                \cite{gururani2019prosody,akuzawa2018expressive,valle2020flowtron,habib2019semi,sun2020generating,sun2020fully,valle2020flowtron,du2021mixture,jeong2021diff},leaf,text width=18.3em
                ]
            ]
            [Disentangling/ \\Controlling/Transferring
                [GMVAE-Tacotron~\cite{hsu2018hierarchical}{,} DenoiSpeech~\cite{zhang2020denoising}\\
                \cite{ma2018neural,hsu2019disentangling,qian2020unsupervised,um2020emotional,lee2021styler,neekhara2021expressive,bae2020speaking,polyak2021speech,tits2021analysis}\\
                \cite{li2021towards,karlapati2020copycat,inoue2021model,liu2018improving,whitehill2020multi,liu2020expressive,cai2020speaker,li2021controllable,habib2019semi},leaf,text width=18.3em
                ]
            ]
        ]
        [Adaptive
            [General Adaptation
                [AdaSpeech 1/3~\cite{chen2021adaspeech,yan2021adaspeech3}{,} \cite{cooper2020can,paul2020enhancing,hu2021whispered,chen2019cross,liu2019cross},leaf,text width=18.3em
                ]
            ]
            [Efficient Adaptation
                [SEA-TTS~\cite{chen2018sample}{,} DV3-Clone~\cite{arik2018neural}{,} \cite{kons2019high,zhang2020adadurian,luong2020nautilus}  \\
                SV-Tacotron~\cite{jia2018transfer}{,} AdaSpeech 1/2~\cite{chen2021adaspeech,yan2021adaspeech}{,} \cite{cooper2020zero},leaf,text width=18.3em
                ]
            ]
        ]
    ]
\end{forest}
\caption{Overview of the advanced topics in neural TTS as described in Section~\ref{sec_advanced}.}
\label{taxonomy_of_advanced_tts}
\end{figure*}
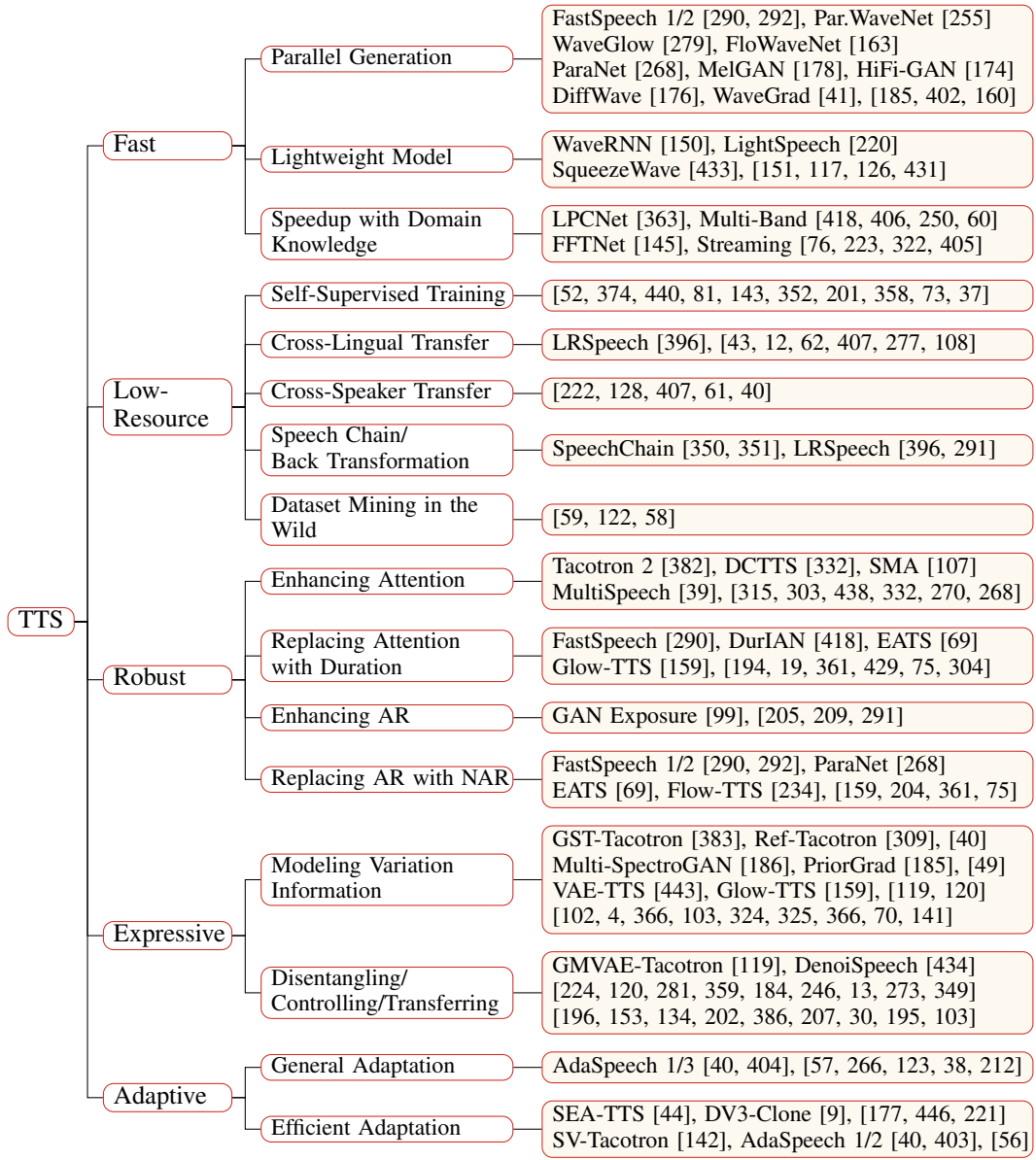

\subsection{Background and Taxonomy}
\label{sec_advance_taxonomy}

In previous section, we have introduced neural TTS in terms of basic model components. In this section, we review some advanced topics in neural TTS that aim to push the frontier and cover more practical product usage. Specifically, as TTS is a typical sequence to sequence generation task with slow autoregressive generation, how to speed up the autoregressive generation or reduce the model size for fast speech synthesis is a hot research topic (Section~\ref{sec_advance_fast}). A good TTS system should generate both natural and intelligible speech and a lot of TTS research works aim to improve the intelligibility and naturalness of speech synthesis. For example, in low-resource scenarios where the data to train a TTS model is not enough, the synthesized speech may have both low intelligibility and naturalness. Therefore, a lot of works aim to build data efficient TTS models under low-resource settings (Section~\ref{sec_advance_low}). Since TTS models are prone to suffer from robust issues where the generated speech usually has word skipping and repeating problems that affect the intelligibility, a lot of works aim to improve the robustness of speech synthesis (Section~\ref{sec_advance_robust}). To improve the naturalness, a lot of works aim to model, control, and transfer the style/prosody of speech in order to generate expressive speech (Section~\ref{sec_advance_express}). Adapting TTS models to support the voice of any target speakers is very helpful for broad usage of TTS. Therefore, efficient voice adaptation with limited adaptation data and parameters and with high-quality voice is critical for practical usage (Section~\ref{sec_advance_adapt}). A taxonomy of these advanced topics are shown in Figure~\ref{taxonomy_of_advanced_tts}.

\subsection{Fast TTS}
\label{sec_advance_fast}
Text to speech synthesis systems are usually deployed in cloud server or embedded devices, which require fast synthesis speed. However, early neural TTS models usually adopt autoregressive mel-spectrogram and waveform generation, which are very slow considering the long speech sequence (e.g., a 1 second speech usually has 500 mel-spetrograms if hop size is 10ms, and 24k waveform points if sampling rate is 24kHz). To solve this problem, different techniques have been leveraged to speed up the inference of TTS models, including 1) non-autoregressive generation that generates mel-spectrograms and waveform in parallel; 2) lightweight and efficient model structure; 3) techniques leveraging the domain knowledge of speech for fast speech synthesis. We introduce these techniques as follows.

\begin{table}[h]
\small
	\caption{The time complexity of different TTS models in training and inference with regard to sequence length $N$. $T$ is the number of steps/iterations in flow/diffusion based models. }
	\centering
	\begin{tabular}{l l  l  l }
		\toprule
		Modeling Paradigm & TTS Model & Training &  Inference  \\
		\midrule
		AR (RNN) & Tacotron 1/2, SampleRNN, LPCNet & $\mathcal{O}(N)$ & $\mathcal{O}(N)$ \\
		AR (CNN/Self-Att) & DeepVoice 3, TransformerTTS, WaveNet & $\mathcal{O}(1)$ & $\mathcal{O}(N)$ \\
		NAR (CNN/Self-Att) & FastSpeech 1/2, ParaNet & $\mathcal{O}(1)$ & $\mathcal{O}(1)$  \\
		NAR (GAN/VAE) & MelGAN, HiFi-GAN, FastSpeech 2s, EATS & $\mathcal{O}(1)$ & $\mathcal{O}(1)$ \\
		Flow (AR)  & Par. WaveNet, ClariNet, Flowtron & $\mathcal{O}(1)$ & $\mathcal{O}(1)$  \\
		Flow (Bipartite) & WaveGlow, FloWaveNet, Glow-TTS  & $\mathcal{O}(T)$ & $\mathcal{O}(T)$ \\
		Diffusion & DiffWave, WaveGrad, Grad-TTS, PriorGrad  & $\mathcal{O}(T)$ & $\mathcal{O}(T)$ \\
		\bottomrule
	\end{tabular}
	\label{tab_time_complexity}
\end{table}

\paragraph{Parallel Generation} Table~\ref{tab_time_complexity} summarizes typical modeling paradigms, the corresponding TTS models, and time complexity in training and inference. As can be seen, TTS models that use RNN-based autoregressive models~\cite{wang2017tacotron,shen2018natural,mehri2016samplernn,valin2019lpcnet} are slow in both training and inference, with $\mathcal{O}(N)$ computation, where $N$ is the sequence length. To avoid the slow training time caused by RNN structure, DeepVoice 3~\cite{ping2018deep} and TransformerTTS~\cite{li2019neural} leverage CNN or self-attention based structure that can support parallel training but still require autoregressive inference. To speed up inference, FastSpeech 1/2~\cite{ren2019fastspeech,ren2021fastspeech} design a feed-forward Transformer that leverages self-attention structure for both parallel training and inference, where the computation is reduced to $\mathcal{O}(1)$. Most GAN-based models for mel-spectrogram and waveform generation~\cite{kumar2019melgan,kong2020hifi,ren2021fastspeech,donahue2020end} are non-autoregressive, with $\mathcal{O}(1)$ computation in both training and inference.  Parallel WaveNet~\cite{oord2018parallel} and ClariNet~\cite{ping2018clarinet} leverage inverse autoregressive flow~\cite{kingma2016improved}, which enable parallel inference but require teacher distillation for parallel training. WaveGlow~\cite{prenger2019waveglow} and FloWaveNet~\cite{kim2019flowavenet} leverage generative flow for parallel training and inference. However, they usually need to stack multiple flow iterations $T$ to ensure the quality of the mapping between data and prior distributions. Similar to flow-based models, diffusion-based models~\cite{chen2020wavegrad,kong2020diffwave,lee2021priorgrad,jeong2021diff,popov2021grad} also require multiple diffusion steps $T$ in the forward and reverse process, which increase the computation.

\paragraph{Lightweight Model}
While non-autoregressive generation can fully leverage the parallel computation for inference speedup, the number of model parameters and total computation cost are not reduced, which make it slow when deploying on the mobile phones or embedded devices since the parallel computation capabilities in these devices are not powerful enough. Therefore, we need to design lightweight and efficient models with less computation cost for inference speedup, even using autoregressive generation. Some widely used techniques for designing lightweight models include pruning, quantization, knowledge distillation~\cite{hinton2015distilling}, and neural architecture search~\cite{luo2021lightspeech,xu2021bert}, etc. WaveRNN~\cite{kalchbrenner2018efficient} uses techniques like dual softmax, weight pruning, subscale prediction to speed up inference. LightSpeech~\cite{luo2021lightspeech} leverages neural architecture search~\cite{zoph2016neural,luo2020neural} to find lightweight architectures to further speed up the inference of FastSpeech 2~\cite{ren2021fastspeech} by 6.5x, while maintaining voice quality. SqueezeWave~\cite{zhai2020squeezewave} leverages waveform reshaping to reduce the temporal length and replaces the 1D convolution with depthwise separable convolution to reduce computation cost while achieving similar audio quality. \citet{kanagawa2020lightweight} compress the model parameters of LPCNet with tensor decomposition. \citet{hsu2020wg} propose a heavily compressed flow-based model to reduce computational resources, and a WaveNet-based post-filter to maintain audio quality. DeviceTTS~\cite{huang2020devicetts} leverages the model structure of DFSMN~\cite{zhang2018deep} and mix-resolution decoder to predict multiple frames in one decoding step to speed up inference. LVCNet~\cite{zeng2021lvcnet} adopts a location-variable convolution for different waveform intervals, where the convolution coefficients are predicted from mel-spectrograms. It speeds up the Parallel WaveGAN vocoder by 4x without any degradation in sound quality. \citet{wangfcl} propose a semi-autoregressive mode for mel-spectrogram generation, where the mel-spectrograms are generated in an autoregressive mode for individual phoneme and non-autoregressive mode for different phonemes.

\paragraph{Speedup with Domain Knowledge}
Domain knowledge from speech can be leveraged to speed up inference, such as linear prediction~\cite{valin2019lpcnet}, multiband modeling~\cite{yu2020durian,yang2020multi,cui2020efficient}, subscale prediction~\cite{kalchbrenner2018efficient}, multi-frame prediction~\cite{zen2016fast,wang2017tacotron,wangfcl,huang2020devicetts,liu2021fasttalker}, streaming synthesis~\cite{ellinas2020high}, etc. LPCNet~\cite{valin2019lpcnet} combines digital signal processing with neural networks, by using linear prediction coefficients to calculate the next waveform and a lightweight model to predict the residual value, which can speed the inference of autoregressive waveform generation. Another technique that is widely used to speed up the inference of vocoders is subband modeling, which divides the waveform into multiple subbands for fast inference. Typical models include DurIAN~\cite{yu2020durian}, multi-band MelGAN~\cite{yang2020multi}, subband WaveNet~\cite{okamoto2018investigation}, and multi-band LPCNet~\cite{tian2020featherwave,cui2020efficient}. Bunched LPCNet~\cite{vipperla2020bunched} reduces the computation complexity of LPCNet with sample bunching and bit bunching, achieving more than 2x speedup. Streaming TTS~\cite{ellinas2020high,ma2020incremental,stephenson2020future,yanagita2019neural,stephenson2021alternate,mohan2020incremental} synthesizes speech once some input tokens are comming, without waiting for the whole input sentence, which can also speed up inference. FFTNet~\cite{jin2018fftnet} uses a simple architecture to mimic the Fast Fourier Transform (FFT), which can generate audio samples in real-time. \citet{okamoto2018improving} further enhance FFTNet with noise shaping and subband techniques, improving the voice quality while keeping small model size. \citet{popov2020fast} propose frame splitting and cross-fading to synthesize some parts of the waveform in parallel and then concatenate the synthesized waveforms together to ensure fast synthesis on low-end devices. \citet{kang2021fast} accelerate DCTTS~\cite{tachibana2018efficiently} with network reduction and fidelity improvement techniques such as group highway activation, which can synthesize speech in real time with a single CPU thread.

\subsection{Low-Resource TTS}
\label{sec_advance_low}

Building high-quality TTS systems usually requires a large amount of high-quality paired text and speech data. However, there are more than 7,000 languages in the world\footnote{https://www.ethnologue.com/browse}, and most languages are lack of training data for developing TTS systems. As a result, popular commercialized speech services\footnote{For example, Microsoft Azure, Google Cloud, and Amazon AWS.} can only support dozens of languages for TTS. Supporting TTS for low-resource languages can not only have business value, but is also beneficial for social good. Thus, a lot of research works build TTS system under low data resource scenarios. We summarize some representative techniques for low-resource TTS in Table~\ref{tab_low_resouce_tts}, and introduce these techniques as follows.
\begin{table}[h]
\small
	\caption{Some representative techniques for low-resource TTS.}
	\centering
	\begin{tabular}{l l  l }
		\toprule
		Techniques & Data & Work  \\
		\midrule
		Self-supervised Training &  Unpaired text or speech & ~\cite{chung2019semi,wang2015word,zhang2019joint,fang2019towards,jia2021png,tjandra2019vqvae,liu2020towards,tu2020semi,dunbar2019zero} \\
		Cross-lingual Transfer & Paired text and speech & ~\cite{chen2019end,xu2020lrspeech,azizah2020hierarchical,yang2020towards,de2020efficient,prajwal2021data,he2021multilingual} \\
		Cross-speaker Transfer & Paired text and speech & ~\cite{luong2019training,huybrechts2020low,dai2020noise,yang2020towards,chen2021adaspeech} \\
		Speech chain/Back transformation & Unpaired text or speech & ~\cite{ren2019almost,xu2020lrspeech,tjandra2017listening,tjandra2018machine}  \\
		Dataset mining in the wild & Paired text and speech &~\cite{cooper2019text,hu2019neural,cooper2020pretraining} \\
		\bottomrule
	\end{tabular}
	\label{tab_low_resouce_tts}
\end{table}
\begin{itemize}[leftmargin=*]
    \item Self-supervised training. Although paired text and speech data are hard to collect, unpaired speech and text data (especially text data) are relatively easy to obtain. Self-supervised pre-training methods can be leveraged to enhance the language understanding or speech generation capabilities~\cite{chung2019semi,wang2015word,zhang2019joint,fang2019towards}. For example, the text encoder in TTS can be enhanced by the pre-trained BERT models~\cite{chung2019semi,fang2019towards,jia2021png}, and the speech decoder in TTS can be pre-trained through autoregressive mel-spectrogram prediction~\cite{chung2019semi} or jointed trained with voice conversion task~\cite{zhang2019joint}. Besides, speech can be quantized into discrete token sequence to resemble the phoneme or character sequence~\cite{tjandra2019vqvae}. In this way, the quantized discrete tokens and the speech can be regarded as pseudo paired data to pre-train a TTS model, which is then fine-tuned on few truly paired text and speech data~\cite{liu2020towards,tu2020semi,zhang2020unsupervised}.   
    \item Cross-lingual transfer. Although paired text and speech data is scarce in low-resource languages, it is abundant in rich-resource languages. Since human languages share similar vocal organs, pronunciations~\cite{wind1989evolutionary} and semantic structures~\cite{tan2019multilingual}, pre-training the TTS models on rich-resource languages can help the mapping between text and speech in low-resource languages~\cite{chen2019end,xu2020lrspeech,azizah2020hierarchical,de2020efficient,guo2018dnn,tan2019study,zhang2019deep,nekvinda2020one,yang2020towards,zhang2020uwspeech}. Usually, there are different phoneme sets between rich- and low-resource languages. Thus, \citet{chen2019end} propose to map the embeddings between the phoneme sets from different languages, and LRSpeech~\cite{xu2020lrspeech} discards the pre-trained phoneme embeddings and initializes the phoneme embeddings from scratch for low-resource languages. International phonetic alphabet (IPA)~\cite{hemati2020using} or byte representation~\cite{he2021multilingual} is adopted to support arbitrary texts in multiple languages. Besides, language similarity~\cite{tan2019multilingual} can also be considered when conducting the cross-lingual transfer.  
    \item Cross-speaker transfer. When a certain speaker has limited speech data, the data from other speakers can be leveraged to improve the synthesis quality of this speaker. This can be achieved by converting the voice of other speakers into this target voice through voice conversion to increase the training data~\cite{huybrechts2020low}, or by adapting the TTS models trained on other voices to this target voice through voice adaptation or voice cloning~\cite{chen2018sample,chen2021adaspeech} that are introduced in Section~\ref{sec_advance_adapt}. 
    \item Speech chain/Back transformation. Text to speech (TTS) and automatic speech recognition (ASR) are two dual tasks~\cite{qin2020dual} and can be leveraged together to improve each other. Techniques like speech chain~\cite{tjandra2017listening,tjandra2018machine} and back transformation~\cite{ren2019almost,xu2020lrspeech} leverage additional unpaired text and speech data to boost the performance of TTS and ASR.
    \item Dataset mining in the wild. In some scenarios, there may exist some low-quality paired text and speech data in the Web. \citet{cooper2019text,hu2019neural} propose to mine this kind of data and develop sophisticated techniques to train a TTS model. Some techniques such as speech enhancement~\cite{valentini2018speech}, denoising~\cite{zhang2020denoising}, and disentangling~\cite{wang2018style,hsu2019disentangling} can be leveraged to improve the quality of the speech data mined in the wild.
\end{itemize}

\subsection{Robust TTS}
\label{sec_advance_robust}
A good TTS system should be robust to always generate ``correct'' speech according to text even when encountering corner cases. In neural TTS, robust issues such as word skipping, repeating, and attention collapse\footnote{Attention collapse means the generated speech has unintelligible gibberish, which is usually caused by the not focused attention on a single input token~\cite{he2019robust}.} often happen in acoustic models\footnote{Robust issues can also happened in neural vocoders, where the generated waveform could have some glitches such as hoarseness, metallic noise, jitter, or pitch breaking. However, they are not so severe as in acoustic models, and the reasons causing these issues are not clear and more likely to be repaired by universal vocoder modeling~\cite{lorenzo2019towards,paul2020speaker,jang2020universal,jiao2021universal} or sophisticated designs~\cite{chen2020hifisinger}. Thus, we mainly introduce the works addressing the robust issues in acoustic models in this survey.} when generating mel-spectrogram sequence from character/phoneme sequence. Basically speaking, the causes of these robust issues are from two categories: 1) The difficulty in learning the alignments between characters/phonemes and mel-spectrograms; 2) The exposure bias and error propagation problems incurred in autoregressive generation. Vocoders do not face severely robust issues, since the acoustic features and waveform are already aligned frame-wisely (i.e., each frame of acoustic features correspond to a certain number (hop size) of waveform points). Therefore, existing works on robust TTS address the above two problems respectively\footnote{There are some other reasons that can cause robust issues, such as the test domain is not well covered by the training domain. Research works that scale to unseen domain can alleviate this issue, such as increasing the amount and diversity of the training data~\cite{hwang2020tts}, adopting relative position encoding to support long sequence unseen in training~\cite{battenberg2020location,zeng2020prosody}, etc.}. 
 
\begin{itemize}[leftmargin=*]
    \item For the alignment learning between characters/phonemes and mel-spectrograms, the works can be divided into two aspects: 1) enhancing the robustness of attention mechanism~\cite{wang2017tacotron,sotelo2017char2wav,shen2018natural,zhang2018forward,tachibana2018efficiently,he2019robust,chen2020multispeech}, and 2) removing attention and instead predicting duration explicitly to bridge the length mismatch between text and speech~\cite{ren2019fastspeech,yu2020durian,donahue2020end,elias2021parallel}.
    \item For the exposure bias and error propagation problems in autoregressive generation, the works can also be divided into two aspects: 1) improving autoregressive generation to alleviate the exposure bias and error propagation problems~\cite{guo2019new,liu2019new,liu2020teacher,ren2019almost}, and 2) removing autoregressive generation and instead using non-autoregressive generation~\cite{ren2019fastspeech,ren2021fastspeech,peng2020non,donahue2020end}.
\end{itemize}

\begin{table}[h]
\small
	\caption{Categorization of the methods for robust TTS.}
	\centering
	\begin{tabular}{l | l  l }
		\toprule
		Category & Technique & Work  \\
		\midrule
		\multirow{7}{*}{Enhancing Attention} & Content-based attention & \cite{wang2017tacotron,li2019neural}\\
		& Location-based attention & \cite{sotelo2017char2wav,taigman2018voiceloop,vasquez2019melnet,battenberg2020location} \\ 
		& Content/Location hybrid attention  & \cite{shen2018natural} \\
		& Monotonic attention & \cite{zhang2018forward,he2019robust,yasuda2019initial} \\
		& Windowing or off-diagonal penalty & \cite{tachibana2018efficiently,zhang2018forward,ping2018deep,chen2020multispeech}  \\
		& Enhancing enc-dec connection & \cite{wang2017tacotron,shen2018natural,ping2018deep,liu2019maximizing,chen2020multispeech} \\
		& Positional attention & \cite{peng2020non,miao2020flow,liu2021vara} \\
		\midrule
		\multirow{6}{*}{\shortstack{Replacing Attention with \\ Duration Prediction}} & Label from encoder-decoder attention & \cite{ren2019fastspeech,vainer2020speedyspeech,lim2020jdi,lancucki2020fastpitch}  \\
		& Label from CTC alignment & \cite{beliaev2020talknet} \\
		& Label from HMM alignment & \cite{ren2021fastspeech,yu2020durian,li2020robutrans,okamoto2019tacotron,elias2020parallel,shen2020non} \\
		& Dynamic programming & \cite{zeng2020aligntts,li2020moboaligner,miao2020efficienttts} \\
		& Monotonic alignment search & \cite{kim2020glow}\\
		& Monotonic interpolation with soft DTW & \cite{donahue2020end,elias2021parallel} \\
		\midrule
		\midrule
		\multirow{4}{*}{Enhancing AR} & Professor forcing &  \cite{guo2019new,liu2019new}  \\
		& Reducing training/inference gap & \cite{vainer2020speedyspeech}\\
		& Knowledge distillation & \cite{liu2020teacher}   \\
		& Bidirectional regularization &  \cite{ren2019almost,zheng2019forward} \\
		\midrule
		\multirow{1}{*}{Replacing AR with NAR} & Parallel generation  &  \cite{ren2019fastspeech,ren2021fastspeech,peng2020non,donahue2020end}  \\
		\bottomrule
	\end{tabular}
	\label{tab_robustness}
\end{table}

We summarize some popular techniques in these categories to improve robustness, as shown in Table~\ref{tab_robustness}. The works addressing the two problems may have overlapping, e.g., some works may enhance the attention mechanism in AR or NAR generation, and similarly, the duration prediction can be applied in both AR and NAR generation. We review these categories as follows.

\subsubsection{Enhancing Attention} 
In autoregressive acoustic models, a lot of word skipping/repeating and attention collapse issues are caused by the incorrect attention alignments learned in encoder-decoder attention. To alleviate this problem, some properties of the alignments between text (characters/phonemes) sequence and mel-spectrogram sequence are considered~\cite{he2019robust}: 1) Local: one character/phoneme token can be aligned to one or multiple consecutive mel-spectrogram frames, while one mel-spectrogram frame can only be aligned to a single character/phoneme token, which can avoid the blurry attention and attention collapse; 2) Monotonic: if character A is behind character B, the mel-spectrogram corresponding to A is also behind that corresponding to B, which can avoid word repeating; 3) Complete: each character/phoneme token must be covered by at least one mel-spectrogram frame, which can avoid word skipping. We analyze the techniques to enhance attention (from Table~\ref{tab_robustness}) according to whether they satisfy the above three properties and list them in Table~\ref{tab_enhance_attention}. We describe these techniques as follows.

\begin{table}[h]
\small
	\caption{The techniques on enhancing attention and whether they  satisfy the three properties (local/monotonic/complete).}
	\centering
	\begin{tabular}{l | c | c | c }
		\toprule
	    Techniques & Local & Monotonic & Complete \\
		\midrule
	    Content-based attention & \texttimes & \texttimes & \texttimes \\
	    Location-based attention & \texttimes & \checkmark & \texttimes \\ 
		Content/Location hybrid attention & \texttimes & \checkmark & \texttimes \\ 
		Monotonic attention & \checkmark & \checkmark & \texttimes \\
		Stepwise monotonic attention & \checkmark & \checkmark & \checkmark \\
		Windowing or off-diagonal penalty & \texttimes & \texttimes & \texttimes \\ 
		Enhancing enc-dec connection & \texttimes & \texttimes & \texttimes \\ 
		Positional attention & \texttimes & \texttimes & \texttimes \\ 
		\midrule
		Predicting duration & \checkmark & \checkmark & \checkmark \\
		\bottomrule
	\end{tabular}
	\label{tab_enhance_attention}
\end{table}

\begin{itemize}[leftmargin=*]
    \item Content-based attention. The early attention mechanisms adopted in TTS (e.g. Tacotron~\cite{wang2017tacotron}) are content-based~\cite{bahdanau2014neural}, where the attention distributions are determined by the degree of match between the hidden representations from the encoder and decoder. Content-based attention is suitable for the tasks such as neural machine translation~\cite{bahdanau2014neural,vaswani2017attention} where the alignments between the source and target tokens are purely based semantic meaning (content). However, for the tasks like automatic speech recognition~\cite{chorowski2015attention,chan2016listen,chiu2018state} and text to speech synthesis~\cite{wang2017tacotron}, the alignments between text and speech have some specific properties. For example, in TTS~\cite{he2019robust}, the attention alignments should be local, monotonic, and complete. Therefore, advanced attention mechanisms should be designed to better leverage these properties.  
    \item Location-based attention. Considering the alignments between text and speech are depending on their positions, location-based attention~\cite{graves2013generating,battenberg2020location} is proposed to leverage the positional information for alignment. Several TTS models such as Char2Wav~\cite{sotelo2017char2wav}, VoiceLoop~\cite{taigman2018voiceloop}, and MelNet~\cite{vasquez2019melnet} adopt the location-based attention. As we summarize in Table~\ref{tab_enhance_attention}, location-based attention can ensure the monotonicity property if properly handled.  
    \item Content/Location-based hybrid attention. To combine the advantages of content and location based attentions, \citet{chorowski2015attention,shen2018natural} introduce location sensitive attention: when calculating the current attention alignment, the previous attention alignment is used. In this way, the attention would be more stable due to monotonic alignment. 
    \item Monotonic attention. For monotonic attention~\cite{raffel2017online,chiu2018monotonic,he2019robust,yasuda2019initial,tian2020feathertts}, the attention position is monotonically increasing, which also leverages the prior that the alignments between text and speech are monotonic. In this way, it can avoid the skipping and repeating issues. However, the completeness property cannot be guaranteed in the above monotonic attention. Therefore, \citet{he2019robust} propose stepwise monotonic attention, where in each decoding step, the attention alignment position moves forward at most one step, and is not allowed to skip any input unit. 
    \item Windowing or off-diagonal penalty. Since attention alignments are monotonic and diagonal, \citet{chorowski2015attention,tachibana2018efficiently,zhang2018forward,ping2018deep,chen2020multispeech} propose to restrict the attention on the source sequence into a window subset. In this way, the learning flexibility and difficulty are reduced. \citet{chen2020multispeech} use penalty loss for off-diagonal attention weights, by constructing a band mask and encouraging the attention weights to be distributed in the diagonal band. 
    \item Enhancing encoder-decoder connection. Since speech has more correlation among adjacent frames, the decoder itself contains enough information to predict next frame, and thus tends to ignore the text information from encoder. Therefore, some works propose to enhance the connection between encoder and decoder, and thus can improve attention alignment. \citet{wang2017tacotron,shen2018natural} use multi-frame prediction that generates multiple non-overlapping output frames at each decoder step. In this way, in order to predict consecutive frames, the decoder is forced to leverage information from the encoder side, which can improve the alignment learning. Other works also use a large dropout in the prenet before the decoder~\cite{wang2017tacotron,shen2018natural,chen2020multispeech}, or a small hidden size in the prenet as a bottleneck~\cite{chen2020multispeech}, which can prevent simply copying the previous speech frame when predicting the current speech frame. The decoder will get more information from the encoder side, which benefits the alignment learning. 
    \citet{ping2018deep,chen2020multispeech} propose to enhance the connection of the positional information between source and target sequences, which benefits the attention alignment learning. \citet{liu2019maximizing} leverage CTC~\cite{graves2006connectionist} based ASR as a cycle loss to encourage the generated mel-spectrograms to contain text information, which can also enhance the encoder-decoder connection for better attention alignment. 
    \item Positional attention. Some non-autoregressive generation models~\cite{peng2020non,miao2020flow} leverage position information as the query to attend the key and value from the encoder, which is another way to build the connection between encoder and decoder for parallel generation.
\end{itemize}

\subsubsection{Replacing Attention with Duration Prediction} 
\label{sec_advanced_robust_duration}
While improving the attention alignments between text and speech can alleviate the robust issues to some extent, it cannot totally avoid them. Thus, some works~\cite{ren2019fastspeech,yu2020durian,kim2020glow,donahue2020end} propose to totally remove the encoder-decoder attention, explicitly predict the duration of each character/phoneme, and expand the text hidden sequence according to the duration to match the length of mel-spectrogram sequence. After that, the model can generate mel-spectrogram sequence in an autoregressive or non-autoregressive manner. It is very interesting that the early SPSS uses duration for alignments, and then the sequence-to-sequence models remove duration but use attention instead, and the later TTS models discard attention and use duration again, which is a kind of technique renaissance. 

Existing works to investigate the duration prediction in neural TTS can be categorized from two perspectives: 1) Using external alignment tools or jointly training to get the duration label. 2) Optimizing the duration prediction in an end-to-end way or using ground-truth duration in training and predicted duration in inference. We summarize the works according to the two perspectives in Table~\ref{tab_robustness_duration}, and describe them as follows.

\begin{table}[h]
\small
	\caption{A category of neural TTS on duration prediction.}
	\centering
	\begin{tabular}{l | c | c }
		\toprule
	    Perspective & Category & Work  \\
		\midrule
		\multirow{2}{*}{External/Internal}  & External & FastSpeech 1/2~\cite{ren2019fastspeech,ren2021fastspeech}, DurIAN~\cite{yu2020durian}, TalkNet~\cite{beliaev2020talknet},
		~\cite{vainer2020speedyspeech,elias2020parallel,shen2020non}  \\
		& Internal & AlignTTS~\cite{zeng2020aligntts}, Glow-TTS~\cite{kim2020glow}, EATS~\cite{donahue2020end}, \cite{miao2020efficienttts,elias2021parallel} \\
		\midrule
	    \multirow{2}{*}{E2E Optimization} & Not E2E & ~\cite{ren2019fastspeech,vainer2020speedyspeech,beliaev2020talknet,ren2021fastspeech,yu2020durian,li2020robutrans,elias2020parallel,shen2020non,zeng2020aligntts,lim2020jdi,kim2020glow}  \\
	    & E2E & EATS~\cite{donahue2020end}, Parallel Tacotron 2~\cite{elias2021parallel} \\
		\bottomrule
	\end{tabular}
	\label{tab_robustness_duration}
\end{table}


\begin{itemize}[leftmargin=*]
    \item External alignment. The works leveraging external alignment tools~\cite{wightman1997aligner,graves2006connectionist,mcauliffe2017montreal,li2020moboaligner} can be divided into several categories according to the used alignment tools: 1) Encoder-decoder attention: FastSpeech~\cite{ren2019fastspeech} obtains the duration label from the attention alignments of an autoregressive acoustic model. SpeedySpeech~\cite{vainer2020speedyspeech} follows similar pipeline of FastSpeech to extract the duration from an autoregressive teacher model, but replaces the whole network structure with purely CNN. 2) CTC alignment. \citet{beliaev2020talknet} leverages a CTC~\cite{graves2006connectionist} based ASR model to provide the alignments between phoneme and mel-spectrogram sequence. 3) HMM alignment: FastSpeech 2~\cite{ren2021fastspeech} leverages the HMM based Montreal forced alignment (MFA)~\cite{mcauliffe2017montreal} to get the duration. Other works such as DurIAN~\cite{yu2020durian},  RobuTrans~\cite{li2020robutrans}, Parallel Tacotron~\cite{elias2020parallel}, and Non-Attentive Tacotron~\cite{shen2020non} use forced alignment or speech recognition tools to get the alignments. 
    \item Internal alignment. AlignTTS~\cite{zeng2020aligntts} follows the basic model structure of FastSpeech, but leverages a dynamic programming based method to learn the alignments between text and mel-spectrogram sequences with multi-stage training. JDI-T~\cite{lim2020jdi} follows FastSpeech to extract duration from an autoregressive teacher model, but jointly trains the autoregressive and non-autoregressive models, which does not need two-stage training. Glow-TTS~\cite{kim2020glow} leverages a novel monotonic alignment search to extract duration. EATS~\cite{donahue2020end} leverages the interpolation and soft dynamic time warping (DTW) loss to optimize the duration prediction in a fully end-to-end way.
    \item Non end-to-end optimization. Typical duration prediction methods~\cite{ren2019fastspeech,vainer2020speedyspeech,beliaev2020talknet,ren2021fastspeech,yu2020durian,li2020robutrans,elias2020parallel,shen2020non,zeng2020aligntts,lim2020jdi,kim2020glow} usually use duration obtained from external/internal alignment tools for training, and use predicted duration for inference. The predicted duration is not end-to-end optimized by receiving guiding signal (gradients) from the mel-spectrogram loss. 
    \item End-to-end optimization. In order to jointly optimize the duration to achieve better prosody, EATS~\cite{donahue2020end} predicts the duration using an internal module and optimizes the duration end-to-end with the help of duration interpolation and soft DTW loss. Parallel Tacotron 2~\cite{elias2021parallel} follows the practice of EATS to ensure differentiable duration prediction. Non-Attentive Tacotron~\cite{shen2020non} proposes a semi-supervised learning for duration prediction, where the predicted duration can be used for upsampling if no duration label available.
\end{itemize}

\subsubsection{Enhancing AR Generation}
Autoregressive sequence generation usually suffers from exposure bias and error propagation~\cite{bengio2015scheduled,wu2018beyond}. Exposure bias refers to that the sequence generation model is usually trained by taking previous ground-truth value as input (i.e., teacher-forcing), but generates the sequence autoregressively by taking previous predicted value as input in inference. The mismatch between training and inference can cause error propagation in inference, where the prediction errors can accumulate quickly along the generated sequence.

Some works have investigated different methods to alleviate the exposure bias and error propagation issues. \citet{guo2019new} leverage professor forcing~\cite{goyal2016professor} to alleviate the mismatch between the different distributions of real and predicted data. \citet{liu2020teacher} conduct teacher-student distillation~\cite{hinton2015distilling,kim2016sequence,tan2018multilingual} to reduce the exposure bias problem, where the teacher is trained with teacher-forcing mode, and the student takes the previously predicted value as input and is optimized to reduce the distance of hidden states between the teacher and student models. Considering the right part of the generated mel-spectrogram sequence is usually worse than than that in the left part due to error propagation, some works leverage both left-to-right and right-to-left generations~\cite{tan2019efficient} for data augmentation~\cite{ren2019almost} and regularization~\cite{zheng2019forward}. \citet{vainer2020speedyspeech} leverage some data augmentations to alleviate the exposure bias and error propagation issues, by adding some random Gaussian noises to each input spectrogram pixel to simulate the prediction errors, and degrading the input spectrograms by randomly replacing several frames with random frames to encourage the model to use temporally more distant frames.

\subsubsection{Replacing AR Generation with NAR Generation}
Although the exposure bias and error propagation problems in AR generation can be alleviated through the above methods, the problems cannot be addressed thoroughly. Therefore, some works directly adopt non-autoregressive generation to avoid these issues. They can be divided into two categories according to the use of attention or duration prediction. Some works such as ParaNet~\cite{peng2020non} and Flow-TTS~\cite{miao2020flow} uses positional attention~\cite{ping2018deep} for the text and speech alignment in parallel generation. The remaining works such as FastSpeech~\cite{ren2019fastspeech,ren2021fastspeech} and EATS~\cite{donahue2020end} use duration prediction to bridge the length mismatch between text and speech sequences. 

Based on the introductions in the above subsections, we have a new category of TTS according to the alignment learning and AR/NAR generation, as shown in Table~\ref{tab_robustness_taxonomy}: 1) AR + Attention, such as Tacotron~\cite{wang2017tacotron,shen2018natural}, DeepVoice 3~\cite{ping2018deep}, and TransformerTTS~\cite{li2019neural}. 2) AR + Non-Attention (Duration), such as DurIAN~\cite{yu2020durian}, RobuTrans~\cite{li2020robutrans}, and Non-Attentive Tacotron~\cite{shen2020non}. 3) Non-AR + Attention, such as ParaNet~\cite{peng2020non}, Flow-TTS~\cite{miao2020flow}, and VARA-TTS~\cite{liu2021vara}. 4) Non-AR + Non-Attention, such as FastSpeech 1/2~\cite{ren2019fastspeech,ren2021fastspeech}, Glow-TTS~\cite{kim2020glow}, and EATS~\cite{donahue2020end}.

\begin{table}[h]
\small
	\caption{A new category of TTS according to the alignment learning and AR/NAR generation.}
	\centering
	\begin{tabular}{l | c | c }
		\toprule
		\diagbox{Attention?}{AR?} & AR & Non-AR  \\
		\midrule
		 Attention & Tacotron 2~\cite{shen2018natural}, DeepVoice 3~\cite{ping2018deep} & ParaNet~\cite{peng2020non}, Flow-TTS~\cite{miao2020flow} \\
		\midrule
		Non-Attention & DurIAN~\cite{yu2020durian}, Non-Att Tacotron~\cite{shen2020non} & FastSpeech~\cite{ren2019fastspeech,ren2021fastspeech}, EATS~\cite{donahue2020end} \\
		\bottomrule
	\end{tabular}
	\label{tab_robustness_taxonomy}
\end{table}

\subsection{Expressive TTS}
\label{sec_advance_express}
The goal of text to speech is to synthesize intelligible and natural speech. The naturalness largely depends on the expressiveness of synthesized voice, which is determined by multiple characteristics, such as content, timbre, prosody, emotion, and style, etc. The research on expressiveness TTS covers broad topics including modeling, disentangling, controlling, and transferring the content, timbre, prosody, style, and emotion, etc. We review those topics in this subsection. 

A key for expressive speech synthesis is to handle the problem of one-to-many mapping, which refers to that there are multiple speech variations corresponding to the same text, in terms of duration, pitch, sound volume, speaker style, emotion, etc. Modeling the one-to-many mapping under the regular L1 loss~\cite{gazor2003speech,usman2018probabilistic} without enough input information will cause over-smoothing mel-spectrogram prediction~\cite{toda2007speech,takamichi2016postfilters}, e,g., predicting the average mel-spectrograms in the dataset instead of capturing the expressiveness of every single speech utterance, which leads to low-quality and less expressive speech. Therefore, providing these variation information as input and better modeling these variation information are important to alleviate this problem and improve the expressiveness of synthesized speech. Furthermore, by providing variation information as input, we can disentangle, control, and transfer the variation information: 1) by adjusting these variation information (any specific speaker timbre, style, accent, speaking rate, etc) in inference, we can control the synthesized speech; 2) by providing the variation information corresponding to another style, we can transfer the voice to this style; 3) in order to achieve fine-grained voice control and transfer, we need to disentangle different variation information, such as content and prosody, timbre and noise, etc. 

In the remaining parts of this subsection, we first conduct a comprehensive analysis on these variation information, and then introduce some advanced techniques for modeling, disentangling, controlling, and transferring these variation information.

\subsubsection{Categorization of Variation Information} We first categorize the information needed to synthesize a voice into four aspects: 
\begin{itemize}[leftmargin=*]
\item Text information, which can be characters or phonemes, represents the content of the synthesized speech (i.e., what to say). Some works improve the representation learning of text through enhanced word embeddings or text pre-training~\cite{fang2019towards,hayashi2019pre,xiao2020improving,jia2021png}, aiming to improve the quality and expressiveness of synthesized speech.
\item Speaker or timbre information, which represents the characteristics of speakers (i.e., who to say). Some multi-speaker TTS systems explicitly model the speaker representations through a speaker lookup table or speaker encoder~\cite{gibiansky2017deep,ping2018deep,jia2018transfer,moss2020boffin,chen2020multispeech}. 
\item Prosody, style, and emotion information, which covers the intonation, stress, and rhythm of speech and represents how to say the text~\cite{wagner2010experimental,ladd2008intonational}. Prosody/style/emotion is the key information to improve the expressiveness of speech and the vast majority of works on expressive TTS focus on improving the prosody/style/emotion of speech~\cite{skerry2018towards,wang2018style,stanton2018predicting,gao2020interactive,um2020emotional,sun2020generating}. 
\item Recording devices or noise environments, which are the channels to convey speech, and are not related to the content/speaker/prosody of speech, but will affect speech quality. Research works in this area focus on disentangling, controlling, and denoising for clean speech synthesis~\cite{hsu2019disentangling,chen2021adaspeech,zhang2020denoising}. 
\end{itemize}

\subsubsection{Modeling Variation Information}
Many methods have been proposed to model different types of variation information in different granularities, as shown in Table~\ref{tab_expressiveness_taxomomy}.

\begin{table}[h!]
\small
	\caption{Some perspectives of modeling variation information for expressive speech synthesis.}
	\centering
	\begin{tabular}{l | l | l | l}
		\toprule
		Perspective & Category & Description &  Work  \\
		\midrule
		\multirow{9}{*}{\shortstack{Information\\Type}}   & \multirow{3}{*}{Explicit} & Language/Style/Speaker ID &  \cite{zhang2019learning,nekvinda2020one,li2021controllable,kim2021expressive,chen2020multispeech}\\
		\cmidrule{3-4}
		&&  Pitch/Duration/Energy & \cite{ren2019fastspeech,ren2021fastspeech,lancucki2020fastpitch,kenter2019chive,morrison2020controllable,valle2020mellotron} \\ 
		\cmidrule{2-4}
		                                        & \multirow{5}{*}{Implicit} & Reference encoder & \cite{skerry2018towards,wang2018style,ma2018neural,jia2018transfer,arik2018neural,choi2020attentron,chen2021speech,chen2021adaspeech} \\
		\cmidrule{3-4}
		                                        && VAE &  \cite{hsu2018hierarchical,akuzawa2018expressive,zhang2019learningb,hsu2019disentangling,sun2020generating,sun2020fully,elias2020parallel} \\
		\cmidrule{3-4}
		                                        && GAN/Flow/Diffusion & \cite{ma2018neural,lee2020multi,valle2020flowtron,miao2020flow,kim2020glow,jeong2021diff} \\
		\cmidrule{3-4}
		                                        && Text pre-training & \cite{fang2019towards,hayashi2019pre,xiao2020improving,jia2021png} \\
		\midrule
		\multirow{9}{*}{\shortstack{Information\\Granularity}}& Language/Speaker Level & Multi-lingual/speaker TTS & \cite{zhang2019learning,nekvinda2020one,chen2020multispeech} \\
		\cmidrule{2-4}
	                                            & Paragraph Level & Long-form reading & \cite{aubin2019improving,xu2020improving,wang2020s} \\ 
	    \cmidrule{2-4}
	                                            & Utterance Level & Timbre/Prosody/Noise & \cite{skerry2018towards,wang2018style,jia2018transfer,stanton2018predicting,liu2020expressive,chen2021adaspeech}\\
	    \cmidrule{2-4}
	                                            & Word/Syllable Level & \multirow{4}{*}{Fine-grained information} & \cite{sun2020fully,hono2020hierarchical,chien2021hierarchical,talman2019predicting}\\
	    \cmidrule{2-2} \cmidrule{4-4}
	                                            & Character/Phoneme Level &  & \cite{lee2019robust,sun2020generating,zeng2020prosody,sun2020fully,chien2021hierarchical,chen2021adaspeech,lei2021fine}\\
	    \cmidrule{2-2} \cmidrule{4-4}
	                                            & Frame Level  & & \cite{lee2019robust,kenter2019chive,choi2020attentron,zhang2020denoising} \\
	\bottomrule

	\end{tabular}
	\label{tab_expressiveness_taxomomy}
\end{table}

\paragraph{Information Type} We can categorize the works according to the types of information being modeled: 1) explicit information, where we can explicitly get the labels of these variation information, and 2) implicit information, where we can only implicitly obtain these variation information. 

For explicit information, we directly use them as input to enhance the models for expressive synthesis. We can obtain these information through different ways: 1) Get the language ID, speaker ID, style, and prosody from labeling data~\cite{zhang2019learning,nekvinda2020one,li2021controllable,chen2020multispeech}. For example, the prosody information can be labeled according to some annotation schemas, such as  ToBI~\cite{silverman1992tobi}, AuToBI~\cite{rosenberg2010autobi}, Tilt~\cite{taylor1998tilt}, INTSINT~\cite{hirst2001automatic}, and SLAM~\cite{obin2014slam}. 2) Extract the pitch and energy information from speech and extract duration from paired text and speech data~\cite{ren2019fastspeech,ren2021fastspeech,lancucki2020fastpitch,kenter2019chive,morrison2020controllable,valle2020mellotron}.

In some situations, there are no explicit labels available, or explicit labeling usually causes much human effort and cannot cover the specific or fine-grained variation information. Thus, we can model the variation information implicitly from data. Typical implicit modeling methods include: 
\begin{itemize}[leftmargin=*]
\item Reference encoder~\cite{skerry2018towards,wang2018style,ma2018neural,jia2018transfer,arik2018neural,choi2020attentron,chen2021adaspeech,gururani2019prosody}. \citet{skerry2018towards} define the prosody as the variation in speech signals that remains after removing variation due to text content, speaker timbre, and channel effects, and model prosody through a reference encoder, which does not require explicit annotations. Specifically, it extracts prosody embeddings from a reference audio, and uses it as the input of decoder. During training, a ground-truth reference audio is used, and during inference, another refer audio is used to synthesize speech with similar prosody. \citet{wang2018style} extract embeddings from a reference audio and use them as the query to attend (through Q/K/V based attention~\cite{vaswani2017attention}) a banks of style tokens, and the attention results are used as the prosody condition of TTS models for expressive speech synthesis. The style tokens can increase the capacity and variation of TTS models to learn different kinds of styles, and enable the knowledge sharing across data samples in the dataset. Each token in the style token bank can learn different prosody representations, such as different speaking rates and emotions. During inference, it can use a reference audio to attend and extract prosody representations, or simply pick one or some style tokens to synthesize speech. 

\item Variational autoencoder~\cite{hsu2018hierarchical,akuzawa2018expressive,zhang2019learningb,hsu2019disentangling,habib2019semi,sun2020generating,sun2020fully,elias2020parallel}. \citet{zhang2019learningb} leverage VAE to model the variance information in the latent space with Gaussian prior as a regularization, which can enable expressive modeling and control on synthesized styles. Some works~\cite{akuzawa2018expressive,hsu2019disentangling,aggarwal2020using,elias2020parallel} also leverage the VAE framework to better model the variance information for expressive synthesis. 

\item Advanced generative models~\cite{ma2018neural,lee2020multi,valle2020flowtron,miao2020flow,kim2020glow,du2021mixture,jeong2021diff,lee2021priorgrad}. One way to alleviate the one-to-many mapping problem and combat over-smoothing prediction is to use advanced generative models to implicitly learn the variation information, which can better model the multi-modal distribution. 

\item Text pre-training~\cite{fang2019towards,hayashi2019pre,xiao2020improving,jia2021png,guo2019exploiting,zhou2021dependency}, which can provide better text representations by using pre-trained word embeddings or model parameters.


\end{itemize}

\paragraph{Information Granularity} Variation information can be modeled in different granularities. We describe these information from coarse-grained to fine-grained levels: 1) Language level and speaker level~\cite{zhang2019learning,nekvinda2020one,chen2020multispeech}, where multilingual and multispeaker TTS systems use language ID or speaker ID to differentiate languages and speakers. 2) Paragraph level~\cite{aubin2019improving,xu2020improving,wang2020s}, where a TTS model needs to consider the connections between utterances/sentences for long-form reading. 3) Utterance level~\cite{skerry2018towards,wang2018style,jia2018transfer,stanton2018predicting,liu2020expressive,chen2021adaspeech}, where a single hidden vector is extracted from the reference speech to represent the timber/style/prosody of this utterance. 4) Word/syllable level~\cite{sun2020fully,hono2020hierarchical,chien2021hierarchical,talman2019predicting}, which can model the fine-grained style/prosody information that cannot be covered by utterance level information. 5) Character/phoneme level~\cite{lee2019robust,sun2020generating,zeng2020prosody,sun2020fully,chien2021hierarchical,chen2021adaspeech,lei2021fine}, such as duration, pitch or prosody information. 6) Frame level~\cite{lee2019robust,kenter2019chive,choi2020attentron,zhang2020denoising}, the most fine-grained information. Some corresponding works on different granularities can be found in Table~\ref{tab_expressiveness_taxomomy}. 

Furthermore, modeling the variance information with hierarchical structure that covers different granularities is helpful for expressive synthesis. \citet{suni2017hierarchical} demonstrate that hierarchical structures of prosody intrinsically exist in spoken languages. \citet{kenter2019chive} predict prosody features from frame and phoneme levels to syllable level, and concatenate with word- and sentence-level features. \citet{hono2020hierarchical} leverage a multi-grained VAE to obtain different time-resolution latent variables and sample finer-level latent variables from coarser-level ones (e.g., from utterance level to phrase level and then to word level). \citet{sun2020fully} use VAE to model variance information on both phoneme and word levels and combine them together to feed into the decoder. \citet{chien2021hierarchical} study on prosody prediction and propose a hierarchical structure from the word to phoneme level to improve the prosody prediction.

\subsubsection{Disentangling, Controlling and Transferring}

In this subsection, we review techniques on disentangling~\cite{ma2018neural,hsu2019disentangling,qian2020unsupervised}, controlling~\cite{um2020emotional,lee2021styler,neekhara2021expressive,bae2020speaking,polyak2021speech,tits2021analysis,li2021towards}, and transferring~\cite{karlapati2020copycat,inoue2021model,xue2021cycle,an2021improving} variation information, as shown in Table~\ref{tab_expressiveness_technique}.

\begin{table}[h!]
\small
	\caption{Some representative techniques for disentangling, controlling, and transferring in expressive speech synthesis.}
	\centering
	\begin{tabular}{ l | l | l}
	\toprule
	Technique & Description &  Work  \\
	\midrule
    Disentangling with Adversarial Training & Disentanglement for control   & \cite{ma2018neural,hsu2019disentangling,qian2020unsupervised,zhang2020denoising}\\
    Cycle Consistency/Feedback for Control & Enhance style/timbre generation & \cite{liu2018improving,whitehill2020multi,liu2020expressive,cai2020speaker,li2021controllable} \\
    Semi-Supervised Learning for Control & Use VAE and adversarial training & \cite{habib2019semi,hsu2018hierarchical,hsu2019disentangling,zhang2020denoising,shechtman2021supervised}  \\
    Changing Variance Information for Transfer & Different information in inference  & \cite{skerry2018towards,wang2018style,jia2018transfer,zhang2019learningb,chen2021adaspeech} \\
	\bottomrule
	\end{tabular}
	\label{tab_expressiveness_technique}
\end{table}

\paragraph{Disentangling with Adversarial Training} When multiple styles or prosody information are entangled together, it is necessary to disentangle them during training for better expressive speech synthesis and control. \citet{ma2018neural} enhance the content-style disentanglement ability and controllability with adversarial and collaborative games. \citet{hsu2019disentangling} leverage the VAE framework with adversarial training to disentangle noise from speaker information. \citet{qian2020unsupervised} propose speechflow to disentangle the rhythm, pitch, content, and timbre using three bottleneck reconstructions. \citet{zhang2020denoising} propose to disentangle noise from speaker with frame-level noise modeling and adversarial training.

\paragraph{Cycle Consistency/Feedback Loss for Control} When providing variance information such as style tag as input, the TTS models are supposed to synthesize speech with the corresponding style. However, if no constraint is added, the TTS models tend to ignore the variance information and the synthesized speech that does not follow the style. To enhance the contollability of the TTS models, some works propose to use cycle consistency or feedback loss to encourage the synthesized speech to contain the variance information in the input. \citet{li2021controllable} conduct controllable emotional transfer by adding an emotion style classifier with feedback cycle, where the classifier encourages the TTS model to synthesize speech with specific emotion. \citet{whitehill2020multi} use style classifier to provide the feedback loss to encourage the speech synthesis of a given style. Meanwhile, it incorporates adversarial learning between different style classifiers to ensure the preservation of different styles from multiple reference audios. \citet{liu2018improving} use ASR to provide the feedback loss to train the unmatched text and speech, which aims to reduce the mismatch between training and inference, since random chosen audio is used as the reference in inference. Other works~\cite{nachmani2018fitting,liu2020expressive,cai2020speaker,shi2020aishell,xue2021cycle,an2021improving} leverage the feedback loss to ensure the controllability on style and speaker embeddings, etc.

\paragraph{Semi-Supervised Learning for Control}
Some attributes used to control the speech include pitch, duration, energy, prosody, emotion, speaker, noise, etc. If we have the label for each attribute, we can easily control the synthesized speech, by using the tag as input for model training and using the corresponding tag to control the synthesized speech in inference. However, when there is no tag/label available, or only a part is available, how to disentangle and control these attributes are challenging. When partial label is available, \citet{habib2019semi} propose semi-supervised learning method to learn the latent of VAE model, in order to control attributes such as affect or speaking rate. When no label available, \citet{hsu2018hierarchical} propose Gaussian mixture VAE models to disentangle different attributes, and~\citet{hsu2019disentangling,zhang2020denoising} leverage gradient reversal or adversarial training to disentangle speaker timbre from noise in order to synthesize clean speech for noisy speakers.

\paragraph{Changing Variance Information for Transfer} We can transfer the style of synthesized speech by changing the variation information to different styles. If the variation information is provided in the labeled tag, we can use the speech and the corresponding tag in training, and transfer the style with corresponding tags in inference~\cite{zhang2019learning,nekvinda2020one,li2021controllable,chen2020multispeech}. Alternatively, if we do not have labeled tag for the variation information, we can get the variation information from speech during training, no matter through explicit or implicit modeling as introduced above: Pitch, duration and energy can be explicitly extracted from speech, and some latent representations can be implicitly extracted by reference encoder or VAE. In this way, in order to achieve style transfer in inference, we can obtain the variation information in three ways: 1) extracting from reference speech~\cite{skerry2018towards,wang2018style,jia2018transfer,zhang2019learningb,choi2020attentron,chen2021adaspeech,xue2021cycle,an2021improving}; 2) predicting from text~\cite{stanton2018predicting,ren2019fastspeech,sun2020generating,zeng2020prosody,ren2021fastspeech,chen2021adaspeech}; 3) obtaining by sampling from the latent space ~\cite{wang2018style,zhang2019learningb,hsu2018hierarchical}.

\subsection{Adaptive TTS}
\label{sec_advance_adapt}

Adaptive TTS\footnote{Here we mainly discuss adaptive TTS for different voices, instead of languages, styles, domains, etc.} is an important feature for TTS that can synthesize voice for any user. It is known as different terms in academia and industry, such as voice adaptation~\cite{chen2018sample}, voice cloning~\cite{arik2018neural}, custom voice~\cite{chen2021adaspeech}, etc. Adaptive TTS has been a hot research topic, e,g., a lot of works in statistic parametric speech synthesis have studied voice adaptation~\cite{fan2015multi,wu2015study,zhao2016speaker,fan2016speaker,doddipatla2017speaker,huang2018linear}, and the recent voice cloning challenge also attracts a lot of participants~\cite{xie2021multi,hu2021nu,tan2021cuhk,chien2021investigating}. In adaptive TTS scenario, a source TTS model (usually trained on a multi-speaker speech dataset) is usually adapted with few adaptation data for each target voice. 

We review the works on adaptive TTS from two perspectives: 1) General adaptation setting, which covers the improvements of generalization of source TTS model to support new speakers, and the adaptation to different domains. 2) Efficient adaptation setting, which covers the reduction of adaptation data and adaptation parameters for each target speaker. We summarize the works in the two perspectives in Table~\ref{tab_voice_adaptation} and introduce these works as follows.


\begin{table}[h]
\small
	\caption{The research works in adaptive TTS from two perspectives.}
	\centering
	\begin{tabular}{l | l  l }
		\toprule
		Category & Topic & Work   \\
		\midrule
		\multirow{6}{*}{General Adaptation} & Modeling Variation Information  & \cite{chen2021adaspeech} \\
		& Increasing Data Coverage & \cite{cooper2020can,yang2020towards} \\
        \cmidrule{2-3}
		& Cross-Acoustic Adaptation & \cite{chen2021adaspeech,cong2020data} \\
		& Cross-Style Adaptation & \cite{yan2021adaspeech3,paul2020enhancing,hu2021whispered} \\
		& Cross-Lingual Adaptation & \cite{zhang2019learning,chen2019cross,liu2019cross} \\
		\midrule
		\multirow{4}{*}{Efficient Adaptation} & Few-Data Adaptation & \cite{chen2018sample,arik2018neural,kons2019high,moss2020boffin,zhang2020adadurian,choi2020attentron,chen2021adaspeech,min2021meta} \\
		& Untranscribed Data Adaptation & \cite{yan2021adaspeech,inoue2020semi,luong2020nautilus} \\
	    & Few-Parameter Adaptation  & \cite{arik2018neural,chen2018sample,chen2021adaspeech} \\
		& Zero-Shot Adaptation & \cite{arik2018neural,chen2018sample,jia2018transfer,cooper2020zero} \\
	
		\bottomrule
	\end{tabular}
	\label{tab_voice_adaptation}
\end{table}


\subsubsection{General Adaptation}

\paragraph{Source Model Generalization} The works in this category aim to improve the generalization of source TTS model. In source model training, the source text does not contain enough acoustic information such as prosody, speaker timbre, and recording environments to generate target speech. As a result, the TTS model is prone to overfit on the training data and has poor generalization for new speakers in adaptation. \citet{chen2021adaspeech} propose acoustic condition modeling to provide necessary acoustic information as model input to learn the text-to-speech mapping with better generalization instead of memorizing. Another way to improve the generalization of source TTS model is to increase the amount and diversity of training data. \citet{cooper2020can} leverage speaker augmentation to increase the number of speakers when training source TTS model, which can generalize well to unseen speakers in adaptation. \citet{yang2020towards} train a universal TTS model with multiple speakers in 50 language locales, which increase the generalization when adapting to a new speaker.

\paragraph{Cross-Domain Adaptation} In adaptive TTS, an important factor is that the adaptation speech has different acoustic conditions or styles with the speech data used to train the source TTS model. In this way, special designs need to be considered to improve the generalization of source TTS model and support the styles in target speakers. AdaSpeech~\cite{chen2021adaspeech} designs acoustic condition modeling to better model the acoustic conditions such as recording devices, environment noise, accents, speaker rates, speaker timbre, etc. In this way, the model tends to generalize instead of memorizing the acoustic conditions, and can be well adapted to the speech data with different acoustic conditions. AdaSpeech 3~\cite{yan2021adaspeech3} adapts a reading-style TTS model to spontaneous style, by designing specific filled pauses adaptation, rhythm adaptation, and timbre adaptation. Some other works~\cite{paul2020enhancing,hu2021whispered} consider the adaptation across different speaking styles, such as Lombard~\cite{paul2020enhancing} or whisper~\cite{hu2021whispered}. Some works~\cite{zhang2019learning,chen2019cross,liu2019cross,zhao2020towards,himawan2020speaker,staib2020phonological,maiti2020generating,zhou2020end,hemati2020using} propose to transfer voices across languages, e.g., synthesize Mandarin speech using an English speaker, where the English speaker does not have any Mandarin speech data. 

\subsubsection{Efficient Adaptation}
Roughly speaking, more adaptation data will result in better voice quality, but incur high data collection cost. For adaptation parameters, the whole TTS model~\cite{chen2018sample,kons2019high}, or part of the model (e.g., decoder)~\cite{moss2020boffin,zhang2020adadurian}, or only speaker embedding~\cite{arik2018neural,chen2018sample,chen2021adaspeech} can be fine-tuned. Similarly, fine-tuning more parameters will result in good voice quality, but increase the memory and deployment cost. In practice, we aim to adapt as few data and parameters as possible while achieving high adaptation voice quality. We divide the works in this category into several aspects: 1) few data adaptation; 2) few parameter adaptation; 3) untranscribed data adaptation; 4) zero-shot adaptation. We introduce these works as follows.
\begin{itemize}[leftmargin=*]
    \item Few data adaptation. Some works~\cite{chen2018sample,arik2018neural,kons2019high,moss2020boffin,zhang2020adadurian,choi2020attentron,chien2021investigating,chen2021adaspeech,min2021meta} conduct few-shot adaptation that only uses few paired text and speech data, varying from several minutes to several seconds. \citet{chien2021investigating} explore different speaker embeddings for few-shot adaptation. \citet{yue2021exploring} leverage speech chain~\cite{tjandra2017listening} for few-shot adaptation. \citet{chen2021adaspeech,arik2018neural} compare the voice quality with different amounts of adaptation data and find that voice quality improves quickly with the increase of adaptation data when data size is small (less than 20 sentences) and improves slowly with dozens of adaptation sentences. 
    \item Few parameter adaptation. To support many users/customers, the adaptation parameters need to be small enough for each target speaker to reduce memory usage while maintaining high voice quality. For example, if each user/voice consumes 100MB parameters, the total memory storage equals to 100PB for 1M users, which is a huge memory cost. Some works propose to reduce the adaptation parameter as few as possible, while maintaining the adaptation quality. AdaSpeech~\cite{chen2021adaspeech} proposes conditional layer normalization to generate the scale and bias parameters in layer normalization from the speaker embeddings based on contextual parameter generation~\cite{platanios2018contextual} and only fine-tune the parameters related to the conditional layer normalization and speaker embeddings to achieve good adaptation quality. \citet{moss2020boffin} propose a fine-tuning method that selects different model hyperparameters for different speakers based on the Bayesian optimization, which achieves the goal of synthesizing the voice of a specific speaker with only a small number of speech samples.
    \item Untranscribed data adaptation. In many scenarios, only speech data can be collected such as in conversions or online meetings, without the corresponding transcripts. AdaSpeech 2~\cite{yan2021adaspeech} leverages untranscribed speech data for voice adaptation, with the help of speech reconstruction and latent alignments~\cite{luong2020nautilus}. \citet{inoue2020semi} use an ASR model to transcribe the speech data and use the transcribed paired data for voice adaptation.
    \item Zero-shot adaptation. Some works~\cite{arik2018neural,chen2018sample,jia2018transfer,cooper2020zero,casanova2021sc} conduct zero-shot adaptation, which leverage a speaker encoder to extract speaker embeddings given a reference audio. This scenario is quite appealing since no adaptation data and parameters are needed. However, the adaptation quality is not good enough especially when the target speaker is very different from the source speakers.
\end{itemize}

\section{Resources}
\label{sec_resource}
We collect some resources of TTS, including open-source implementations, TTS tutorials and keynotes, TTS challenges, and TTS corpora, as shown in Table~\ref{tab_resouce_tts}.

\begin{center}
\scriptsize
\begin{longtable}{l l l l l}
\caption{TTS resources.} 
\label{tab_resouce_tts}\\
\toprule
\multicolumn{5}{c}{\small Open-Source Implementations} \\
\midrule
ESPnet-TTS~\cite{hayashi2020espnet} & \multicolumn{4}{p{9.5cm}}{\url{https://github.com/espnet/espnet}} \\
Mozilla-TTS & \multicolumn{4}{p{9.5cm}}{\url{https://github.com/mozilla/TTS}} \\
TensorflowTTS & \multicolumn{4}{p{9.5cm}}{\url{https://github.com/TensorSpeech/TensorflowTTS}} \\
Coqui-TTS & \multicolumn{4}{p{9.5cm}}{\url{https://github.com/coqui-ai/TTS}} \\
Parakeet & \multicolumn{4}{p{9.5cm}}{\url{https://github.com/PaddlePaddle/Parakeet}} \\
NeMo & \multicolumn{4}{p{9.5cm}}{\url{https://github.com/NVIDIA/NeMo}} \\
WaveNet & \multicolumn{4}{p{9.5cm}}{\url{https://github.com/ibab/tensorflow-wavenet}} \\
WaveNet & \multicolumn{4}{p{9.5cm}}{\url{https://github.com/r9y9/wavenet_vocoder}} \\
WaveNet & \multicolumn{4}{p{9.5cm}}{\url{https://github.com/basveeling/wavenet}} \\
SampleRNN & \multicolumn{4}{p{9.5cm}}{\url{https://github.com/soroushmehr/sampleRNN_ICLR2017}} \\
Char2Wav & \multicolumn{4}{p{9.5cm}}{\url{https://github.com/sotelo/parrot}} \\
Tacotron & \multicolumn{4}{p{9.5cm}}{\url{https://github.com/keithito/tacotron}} \\
Tacotron & \multicolumn{4}{p{9.5cm}}{\url{https://github.com/Kyubyong/tacotron}} \\
Tacotron 2 & \multicolumn{4}{p{9.5cm}}{\url{https://github.com/Rayhane-mamah/Tacotron-2}} \\
Tacotron 2 & \multicolumn{4}{p{9.5cm}}{\url{https://github.com/NVIDIA/tacotron2}} \\
DeepVoice 3 & \multicolumn{4}{p{9.5cm}}{\url{https://github.com/r9y9/deepvoice3_pytorch}} \\
TransformerTTS & \multicolumn{4}{p{9.5cm}}{\url{https://github.com/as-ideas/TransformerTTS}} \\
FastSpeech & \multicolumn{4}{p{9.5cm}}{\url{https://github.com/xcmyz/FastSpeech}} \\
FastSpeech 2 & \multicolumn{4}{p{9.5cm}}{\url{https://github.com/ming024/FastSpeech2}} \\
MelGAN & \multicolumn{4}{p{9.5cm}}{\url{https://github.com/descriptinc/melgan-neurips}} \\
MelGAN & \multicolumn{4}{p{9.5cm}}{\url{https://github.com/seungwonpark/melgan}} \\
WaveRNN & \multicolumn{4}{p{9.5cm}}{\url{https://github.com/fatchord/WaveRNN}} \\
LPCNet & \multicolumn{4}{p{9.5cm}}{\url{https://github.com/mozilla/LPCNet}} \\
WaveGlow & \multicolumn{4}{p{9.5cm}}{\url{https://github.com/NVIDIA/WaveGlow}} \\
FloWaveNet & \multicolumn{4}{p{9.5cm}}{\url{https://github.com/ksw0306/FloWaveNet}} \\
WaveGAN & \multicolumn{4}{p{9.5cm}}{\url{https://github.com/chrisdonahue/wavegan}} \\
GAN-TTS & \multicolumn{4}{p{9.5cm}}{\url{https://github.com/r9y9/gantts}} \\
Parallel WaveGAN & \multicolumn{4}{p{9.5cm}}{\url{https://github.com/kan-bayashi/ParallelWaveGAN}} \\
HiFi-GAN & \multicolumn{4}{p{9.5cm}}{\url{https://github.com/jik876/hifi-gan}} \\
Glow-TTS & \multicolumn{4}{p{9.5cm}}{\url{https://github.com/jaywalnut310/glow-tts}} \\
Flowtron & \multicolumn{4}{p{9.5cm}}{\url{https://github.com/NVIDIA/flowtron}} \\
DiffWave & \multicolumn{4}{p{9.5cm}}{\url{https://github.com/lmnt-com/diffwave}} \\
WaveGrad & \multicolumn{4}{p{9.5cm}}{\url{https://github.com/ivanvovk/WaveGrad}} \\
VITS & \multicolumn{4}{p{9.5cm}}{\url{https://github.com/jaywalnut310/vits}} \\
TTS Samples & \multicolumn{4}{p{9.5cm}}{\url{https://github.com/seungwonpark/awesome-tts-samples}} \\
Software/Tool for Audio & \multicolumn{4}{p{8cm}}{\url{https://github.com/faroit/awesome-python-scientific-audio}} \\
\bottomrule
\toprule
\multicolumn{5}{c}{\small TTS Tutorials \& Keynotes} \\
\midrule
TTS Tutorial at ISCSLP 2014~\cite{qian2014tutorial}& \multicolumn{4}{p{9.5cm}}{\url{https://www.superlectures.com/iscslp2014/tutorial-4-deep-learning-for-speech-generation-and-synthesis}} \\
TTS Tutorial at ISCSLP 2016~\cite{ling2016deep} & \multicolumn{4}{p{9.5cm}}{\url{http://staff.ustc.edu.cn/~zhling/download/ISCSLP16_tutorial_DLSPSS.pdf}} \\
TTS Tutorial at IEICE~\cite{wang2019tutorial} & \multicolumn{4}{p{9.5cm}}{\url{https://www.slideshare.net/jyamagis/tutorial-on-endtoend-texttospeech-synthesis-part-1-neural-waveform-modeling}} \\
Generative Models for Speech~\cite{bengio2017tutorial} & \multicolumn{4}{p{9.5cm}}{\url{https://www.youtube.com/watch?v=vEAq_sBf1CA}} \\
Generative Model-Based TTS~\cite{zen2017tutorial} & \multicolumn{4}{p{9.5cm}}{\url{https://static.googleusercontent.com/media/research.google.com/en//pubs/archive/45882.pdf}} \\
Keynote at INTERSPEECH~\cite{tokuda2019statistical} & \multicolumn{4}{p{9.5cm}}{\url{http://www.sp.nitech.ac.jp/~tokuda/INTERSPEECH2019.pdf}} \\
TTS Tutorial at ISCSLP 2021~\cite{tan2021tutorial} &  \multicolumn{4}{p{9.5cm}}{\url{https://www.microsoft.com/en-us/research/uploads/prod/2021/02/ISCSLP2021-TTS-Tutorial.pdf}} \\
TTS Webinar~\cite{tan2021microsoft} & \multicolumn{4}{p{9.5cm}}{\url{https://www.youtube.com/watch?v=MA8PCvmr8B0}} \\
TTS Tutorial at IJCAI 2021~\cite{tan2021tutorial_ijcai2021} & \multicolumn{4}{p{9.5cm}}{\url{https://tts-tutorial.github.io/ijcai2021/}} \\
\bottomrule
\toprule
\multicolumn{5}{c}{\small  TTS Challenges} \\
\midrule 
Blizzard Challenge & \multicolumn{4}{p{9.5cm}}{\url{http://www.festvox.org/blizzard/}} \\
Zero Resource Speech Challenge & \multicolumn{4}{p{9.5cm}}{\url{https://www.zerospeech.com/}} \\
ICASSP2021 M2VoC & \multicolumn{4}{p{9.5cm}}{\url{http://challenge.ai.iqiyi.com/detail?raceId=5fb2688224954e0b48431fe0}} \\
Voice Conversion Challenge & \multicolumn{4}{p{9.5cm}}{\url{http://www.vc-challenge.org/}} \\
\bottomrule 
\toprule
\multicolumn{5}{c}{\small  TTS Corpora} \\
\midrule
Corpus & \#Hours & \#Speakers & Sampling Rate (kHz)  & Language\\
\midrule
ARCTIC~\cite{kominek2003cmu}   & 7  & 7   & 16 & English \\
VCTK~\cite{veaux2016superseded}   & 44 & 109 & 48 & English \\
Blizzard-2011~\cite{king2011blizzard} & 16.6 & 1 & 16 & English \\
Blizzard-2013~\cite{king2013blizzard}  & 319 & 1 & 44.1 &  English \\
LJSpeech~\cite{ljspeech17} & 25 & 1 & 22.05 & English\\
LibriSpeech~\cite{panayotov2015librispeech} & 982 & 2484 & 16  & English \\
LibriTTS~\cite{zen2019libritts}  & 586 & 2456 & 24 & English\\
VCC 2018~\cite{lorenzo2018voice}   & 1  & 12 & 22.05 & English \\
HiFi-TTS~\cite{bakhturina2021hi} & 300 & 11 & 44.1 & English \\
TED-LIUM~\cite{rousseau2012ted} & 118 & 666 & / & English \\
CALLHOME~\cite{callhome} & 60 & 120 & 8 & English \\
RyanSpeech~\cite{Zandie2021ryanspeech} &10 & 1& 44.1& English \\
CSMSC~\cite{databaker2017}   & 12 & 1  & 48 & Mandarin \\
HKUST~\cite{liu2006hkust}   & 200 & 2100 & 8 & Mandarin \\
AISHELL-1~\cite{bu2017aishell}  &170 &400 & 16 & Mandarin \\
AISHELL-2~\cite{du2018aishell} &1000 &1991 & 44.1 & Mandarin \\
AISHELL-3~\cite{shi2020aishell}  &85 &218 & 44.1 & Mandarin \\
DiDiSpeech-1~\cite{guo2020didispeech} &572 & 4500 & 48 & Mandarin \\
DiDiSpeech-2~\cite{guo2020didispeech} &227 & 1500 & 48 & Mandarin \\
JSUT~\cite{sonobe2017jsut} & 10& 1& 48& Japanese \\
KazakhTTS~\cite{mussakhojayeva2021kazakhtts} & 93 & 2 & 44.1/48 & Kazakh \\
Ruslan~\cite{gabdrakhmanov2019ruslan} &31 &1 &44.1& Russian \\
HUI-Audio-Corpus~\cite{puchtler2021hui}  & 326 & 122 & 44.1 & German \\
India Corpus~\cite{he2020open} & 39 & 253 & 48& Multilingual\\
M-AILABS \cite{mailabs19}  & 1000  & /  & 16  & Multilingual \\
MLS~\cite{pratap2020mls} & 51K & 6K & 16&Multilingual \\
CSS10~\cite{park2019css10} & 140 & 1 & 22.05& Multilingual \\
CommonVoice~\cite{ardila2020common} &2.5K &50K & 48 & Multilingual \\
\bottomrule
\end{longtable}
\end{center}

\section{Future Directions}
\label{sec_summary}
In this paper, we conducted a survey on neural text to speech and mainly focused on (1) the basic models of TTS including text analysis, acoustic models, vocoders, and fully end-to-end models, and (2) several advanced topics including fast TTS, low-resource TTS, robust TTS, expressive TTS, and adaptive TTS. As a quick summary, we list representative TTS algorithms in Table~\ref{tab_all_paper_list}. Due to page limitations, we only reviewed core algorithms of TTS; readers can refer to other papers for TTS related problems and applications, such as voice conversion~\cite{sisman2020overview}, singing voice synthesis~\cite{hono2019singing,lu2020xiaoicesing,chen2020hifisinger}, talking face synthesis~\cite{chen2020comprises}, etc. 

We point out some future research directions on neural TTS, mainly in two categories according to the end goals of TTS.

\paragraph{High-quality speech synthesis} The most important goal of TTS is to synthesize high-quality speech. The quality of speech is determined by many aspects that influence the perception of speech, including intelligibility, naturalness, expressiveness, prosody, emotion, style, robustness, controllability, etc. While neural approaches have significantly improved the quality of synthesized speech, there is still large room to make further improvements. 
\begin{itemize}[leftmargin=*]
    \item \textit{Powerful generative models}. TTS is a generation task, including the generation of waveform and/or acoustic features, which can be better handled by powerful generative models. Although advanced generative models based on VAE, GAN, flow, or diffusion have been adopted in acoustic models, vocoders and fully end-to-to models, research efforts on more powerful and efficient generative models are appealing to further improve the quality of synthesized speech. 
    
    \item \textit{Better representation learning}. Good representations of text and speech are beneficial for neural TTS models, which can improve the quality of synthesized speech. Some initial explorations on text pre-training indicate that better text representations can indeed improve the speech prosody. How to learn powerful representations for text/phoneme sequence and especially for speech sequence through unsupervised/self-supervised learning and pre-training is challenging and worth further explorations. 
    
    \item \textit{Robust speech synthesis}. While current TTS models eliminate word skipping and repeating issues caused by incorrect attention alignments, they still suffer from robustness issues when encountering corner cases that are not covered in the training set, such as longer text length, different text domains, etc. Improving the generalizability of the TTS model to different domains is critical for robust synthesis. 
    
    \item \textit{Expressive/controllable/transferrable speech synthesis}. The expressiveness, controllability and transferability of TTS models rely on better variation information modeling. Existing methods leverage reference encoder or explicit prosody features (e.g., pitch, duration, energy) for variation modeling, which enjoys good controllability and transferrability in inference but suffering from training/inference mismatch since ground-truth reference speech or prosody features used in training are usually unavailable in inference. Advanced TTS models capture the variation information implicitly, which enjoy good expressiveness in synthesized speech but perform not good in control and transfer, since sampling from latent space cannot explicitly and precisely control and transfer each prosody feature (e.g., pitch, style). How to design better methods for expressive/controllable/transferrable speech synthesis is also appealing.
    
    \item \textit{More human-like speech synthesis}. Current speech recordings used in TTS training are usually in formal reading styles, where no pauses, repeats, changing speeds, varying emotions, and errors are permitted. However, in casual or conversational talking, human seldomly speaks like standard reading. Therefore, better modelling the casual, emotional, and spontaneous styles is critical to improve the naturalness of synthesized speech. 
    
  \end{itemize}
    
\paragraph{Efficient speech synthesis} Once we can synthesize high-quality speech, the next most important task is efficient synthesis, i.e., how to reduce the cost of speech synthesis including the cost of collecting and labeling training data, training and serving TTS models, etc.
\begin{itemize}[leftmargin=*]
\item \textit{Data-efficient TTS}. Many low-resource languages are lack of training data. How to leverage unsupervised/semi-supervised learning and cross-lingual transfer learning to help the low-resource languages is an interesting direction. For example, the ZeroSpeech Challenge~\cite{zerospeechchallenge} is a good initiative to explore the techniques to learn only from speech, without any text or linguistic knowledge. Besides, in voice adaptation, a target speaker usually has little adaptation data, which is another application scenario for data-efficient TTS.

\item \textit{Parameter-efficient TTS}. Today’s neural TTS systems usually employ large neural networks with tens of millions of parameters to synthesize high-quality speech, which block the applications in mobile, IoT and other low-end devices due to their limited memory and power consumption. Designing compact and lightweight models with less memory footprints, power consumption and latency are critical for those application scenarios. 

\item \textit{Energy-efficient TTS}. Training and serving a high-quality TTS model consume a lot of energy and emit a lot of carbon. Improving energy efficiency, e.g., reducing the FLOPs in TTS training and inference, is important to let more populations to benefit from advanced TTS techniques while reducing carbon emissions to protect our environment.  
\end{itemize}

\begin{center}
\small
\begin{longtable}{l | l | l | l | l }
	\caption{Overview of TTS models. ``AM'' represents acoustic models, ``Voc'' represents vocoders, ``E2E'' represents fully end-to-end models, ``ling'' represents linguistic features, ``ch'' represents characters, ``ph'' represents phonemes, ``ceps'' represents cepstrums, ``linS'' represents linear-spectrograms, ``melS'' represents mel-spectrograms, ``wav'' represents waveform, ``FF'' represents feed-forward, ``AR'' represents autoregressive, ``$\emptyset$'' represents no conditional information, ``IS'' represents INTERSPEECH.} 
	\label{tab_all_paper_list} \\
		\toprule
		Model & AM/Voc & Data Flow & Publication  & Time  \\
		\midrule
		WaveNet~\cite{oord2016wavenet} & ~~~~~~~~Voc & {\color{green}ling}$\stackrel{\text{AR}}{\longrightarrow}$wav  & SSW16 &   2016.09 \\
		SampleRNN~\cite{mehri2016samplernn} & ~~~~~~~~Voc & $\emptyset\stackrel{\text{AR}}{\longrightarrow}$wav  & ICLR17 &  2016.12 \\
		Deep Voice~\cite{arik2017deep} & AM+Voc & {\color{blue}ch}$\rightarrow$ {\color{blue}ph}$\rightarrow${\color{green}ling}$\stackrel{\text{AR}}{\longrightarrow}$wav           & ICML17&  2017.02 \\
		Char2Wav~\cite{sotelo2017char2wav} & ~~~~E2E & {\color{blue}ch}$\stackrel{\text{AR}}{\longrightarrow}${\color{red}ceps}$\stackrel{\text{AR}}{\longrightarrow}$wav           & ICLR17 WS  & 2017.02        \\
		Tacotron~\cite{wang2017tacotron} & AM &  {\color{blue}ch/ph}$\stackrel{\text{AR}}{\longrightarrow}${\color{red}linS}$\stackrel{}{\longrightarrow}$wav             & IS17 & 2017.03        \\
		Deep Voice 2~\cite{gibiansky2017deep}           & AM+Voc       &  {\color{blue}ch}$\rightarrow${\color{blue}ph}$\stackrel{\text{FF}}{\longrightarrow}${\color{green}ling}$\stackrel{\text{AR}}{\longrightarrow}$wav   & NIPS17& 2017.05        \\
		DV2-Tacotron~\cite{gibiansky2017deep}           & AM+Voc       &  {\color{blue}ch}$\stackrel{\text{AR}}{\longrightarrow}${\color{red}linS}$\stackrel{\text{AR}}{\longrightarrow}$wav    & NIPS17 & 2017.05        \\
		VoiceLoop~\cite{taigman2018voiceloop}             & AM & {\color{blue}ph}$\rightarrow${\color{red}ceps}$\stackrel{}{\longrightarrow}$wav              & ICLR18 &  2017.07         \\
		Deep Voice 3~\cite{ping2018deep}           & AM       &  {\color{blue}ch/ph}$\stackrel{\text{AR}}{\longrightarrow}${\color{red}melS}$\stackrel{\text{AR}}{\longrightarrow}$wav     & ICLR18 & 2017.10        \\
		DCTTS~\cite{tachibana2018efficiently}           & AM       &  {\color{blue}ch}$\stackrel{\text{AR}}{\longrightarrow}${\color{red}melS}$\stackrel{}{\longrightarrow}$wav     & ICASSP18 & 2017.10        \\
		Par.WaveNet~\cite{oord2018parallel}       & ~~~~~~~~Voc &  {\color{blue}ling}$\stackrel{\text{FF}}{\longrightarrow}$wav     & ICML18 &  2017.11        \\
		Tacotron 2~\cite{shen2018natural}   & AM   &  {\color{blue}ch/ph}$\stackrel{\text{AR}}{\longrightarrow}${\color{red}melS}$\stackrel{\text{AR}}{\longrightarrow}$wav             & ICASSP18 & 2017.12        \\
        WaveGAN~\cite{donahue2018adversarial}                & ~~~~~~~~Voc      & $\emptyset\stackrel{\text{FF}}{\longrightarrow}$wav          & ICLR19 & 2018.02         \\
		WaveRNN~\cite{kalchbrenner2018efficient}                & ~~~~~~~~Voc      & {\color{green}ling}$\stackrel{\text{AR}}{\longrightarrow}$wav          & ICML18 & 2018.02          \\
		DV3-Clone~\cite{arik2018neural}   & AM   &  {\color{blue}ch/ph}$\stackrel{\text{AR}}{\longrightarrow}${\color{red}linS}$\rightarrow$wav             & NeurIPS18 &  2018.02        \\
		GST-Tacotron~\cite{wang2018style}   & AM   &  {\color{blue}ph}$\stackrel{\text{AR}}{\longrightarrow}${\color{red}melS}$\rightarrow$wav             & ICML18 & 2018.03        \\
		Ref-Tacotron~\cite{skerry2018towards}   & AM   &  {\color{blue}ph}$\stackrel{\text{AR}}{\longrightarrow}${\color{red}melS}$\rightarrow$wav             & ICML18 & 2018.03        \\
		FFTNet~\cite{jin2018fftnet}                 & ~~~~~~~~Voc      & {\color{red}ceps}$\stackrel{\text{AR}}{\longrightarrow}$wav          & ICASSP18&  2018.04          \\
		VAE-Loop~\cite{akuzawa2018expressive}               & AM &{\color{blue}ph}$\rightarrow${\color{red}ceps}$\stackrel{}{\longrightarrow}$wav       & IS18 &2018.04     \\
		SV-Tacotron~\cite{jia2018transfer}   & AM   &  {\color{blue}ch/ph}$\stackrel{\text{AR}}{\longrightarrow}${\color{red}melS}$\stackrel{\text{AR}}{\longrightarrow}$wav             & NeurIPS18 & 2018.06        \\
		ClariNet~\cite{ping2018clarinet}               & ~~~~E2E & {\color{blue}ch/ph}$\stackrel{\text{AR}}{\longrightarrow}$wav          & ICLR19 & 2018.07     \\
		ForwardAtt~\cite{zhang2018forward}               & AM & {\color{blue}ph}$\stackrel{\text{AR}}{\longrightarrow}${\color{red}linS}$\rightarrow$wav          & ICASSP18 & 2018.07     \\
		MCNN~\cite{arik2018fast}       & ~~~~~~~~Voc & {\color{red}linS}$\stackrel{\text{FF}}{\longrightarrow}$wav         & SPL18 & 2018.08     \\
		TransformerTTS~\cite{li2019neural} & AM  & {\color{blue}ph}$\stackrel{\text{AR}}{\longrightarrow}${\color{red}melS}$\stackrel{\text{AR}}{\longrightarrow}$wav  & AAAI19 &  2018.09        \\
		SEA-TTS~\cite{chen2018sample} & ~~~~~~~~Voc  & {\color{blue}ling}$\stackrel{\text{AR}}{\longrightarrow}$wav  & ICLR19 &  2018.09        \\
		GMVAE-Tacotron~\cite{hsu2018hierarchical} & AM  & {\color{blue}ph}$\stackrel{\text{AR}}{\longrightarrow}${\color{red}melS}$\stackrel{\text{AR}}{\longrightarrow}$wav  & ICLR19 &  2018.10        \\
		LPCNet~\cite{valin2019lpcnet}                 & ~~~~~~~~Voc      & {\color{red}ceps}$\stackrel{\text{AR}}{\longrightarrow}$wav           & ICASSP19 & 2018.10        \\
		WaveGlow~\cite{prenger2019waveglow}               & ~~~~~~~~Voc     & {\color{red}melS}$\stackrel{\text{FF}}{\longrightarrow}$wav   & ICASSP19 & 2018.10        \\
		FloWaveNet~\cite{kim2019flowavenet}             & ~~~~~~~~Voc      & {\color{red}melS}$\stackrel{\text{FF}}{\longrightarrow}$wav          &ICML19 & 2018.11        \\
		Univ. WaveRNN~\cite{lorenzo2019towards}             & ~~~~~~~~Voc      & {\color{red}melS}$\stackrel{\text{AR}}{\longrightarrow}$wav          &IS19 & 2018.11        \\
		VAE-TTS~\cite{zhang2019learningb}            & AM      & {\color{blue}ph}$\stackrel{\text{AR}}{\longrightarrow}${\color{red}melS}$\stackrel{\text{AR}}{\longrightarrow}$wav          &ICASSP19 & 2018.12        \\
		TTS-Stylization~\cite{ma2018neural}  & AM      & {\color{blue}ch}$\stackrel{\text{AR}}{\longrightarrow}${\color{red}melS}$\rightarrow$wav          &ICLR19 & 2018.12        \\
		AdVoc~\cite{neekhara2019expediting}                  & ~~~~~~~~Voc      & {\color{red}melS}$\stackrel{\text{FF}}{\longrightarrow}${\color{red}linS}$\rightarrow$wav          &IS19& 2019.04          \\
		GAN Exposure~\cite{guo2019new}           & AM      & {\color{blue}ph}$\stackrel{\text{AR}}{\longrightarrow}${\color{red}melS}$\stackrel{\text{AR}}{\longrightarrow}$wav           &IS19& 2019.04      \\
		GELP~\cite{juvela2019gelp}                   & ~~~~~~~~Voc      & {\color{red}melS}$\stackrel{\text{FF}}{\longrightarrow}$wav          &IS19& 2019.04         \\
		Almost Unsup~\cite{ren2019almost}             & AM       & {\color{blue}ph}$\stackrel{\text{AR}}{\longrightarrow}${\color{red}melS}$\rightarrow$wav               & ICML19 & 2019.05         \\
		FastSpeech~\cite{ren2019fastspeech}             & AM       & {\color{blue}ph}$\stackrel{\text{FF}}{\longrightarrow}${\color{red}melS}$\stackrel{\text{FF}}{\longrightarrow}$wav               & NeurIPS19 & 2019.05         \\
		ParaNet~\cite{peng2020non}                & AM     & {\color{blue}ph}$\stackrel{\text{FF}}{\longrightarrow}${\color{red}melS}$\stackrel{\text{FF}}{\longrightarrow}$wav               & ICML20 & 2019.05        \\
		WaveVAE~\cite{peng2020non}                & ~~~~~~~~Voc     & {\color{red}melS}$\stackrel{\text{FF}}{\longrightarrow}$wav               & ICML20 & 2019.05        \\
		MelNet~\cite{vasquez2019melnet}                 & AM       &{\color{blue}ch}$\stackrel{\text{AR}}{\longrightarrow}${\color{red}melS}$\stackrel{}{\longrightarrow}$wav               & arXiv19 & 2019.06        \\
		StepwiseMA~\cite{he2019robust}            & AM       &{\color{blue}ph}$\stackrel{\text{AR}}{\longrightarrow}${\color{red}melS}$\stackrel{AR}{\longrightarrow}$wav               & IS19 & 2019.06        \\
		GAN-TTS~\cite{binkowski2019high}                & ~~~~~~~~Voc      & {\color{green}ling}$\stackrel{\text{FF}}{\longrightarrow}$wav   & ICLR20 & 2019.09        \\
		DurIAN~\cite{yu2020durian}                 & AM & {\color{blue}ph}$\stackrel{\text{AR}}{\longrightarrow}${\color{red}melS}$\stackrel{\text{AR}}{\longrightarrow}$wav          &IS20& 2019.09         \\
		MB WaveRNN~\cite{yu2020durian}            & ~~~~~~~~Voc & {\color{red}melS}$\stackrel{\text{AR}}{\longrightarrow}$wav          &IS20& 2019.09         \\
		MelGAN~\cite{kumar2019melgan}                 & ~~~~~~~~Voc      & {\color{red}melS}$\stackrel{\text{FF}}{\longrightarrow}$wav   & NeurIPS19 & 2019.10         \\
		Para. WaveGAN~\cite{yamamoto2020parallel}       & ~~~~~~~~Voc      & {\color{red}melS}$\stackrel{\text{FF}}{\longrightarrow}$wav   & ICASSP20 & 2019.10        \\
		DCA-Tacotron~\cite{battenberg2020location}               & AM     &  {\color{blue}ph}$\stackrel{\text{AR}}{\longrightarrow}${\color{red}melS}$\stackrel{\text{AR}}{\longrightarrow}$wav               & ICASSP20 & 2019.10  \\
		WaveFlow~\cite{ping2020waveflow}               & ~~~~~~~~Voc      & {\color{red}melS}$\stackrel{\text{AR}}{\longrightarrow}$wav   & ICML20 &2019.12        \\
		SqueezeWave~\cite{zhai2020squeezewave}            & ~~~~~~~~Voc      & {\color{red}melS}$\stackrel{\text{FF}}{\longrightarrow}$wav          &arXiv20& 2020.01          \\
		AlignTTS~\cite{zeng2020aligntts}               & AM     &  {\color{blue}ch/ph}$\stackrel{\text{FF}}{\longrightarrow}${\color{red}melS}$\stackrel{\text{FF}}{\longrightarrow}$wav               & ICASSP20 & 2020.03         \\
		RobuTrans~\cite{li2020robutrans}              & AM     &  {\color{blue}ph}$\stackrel{\text{AR}}{\longrightarrow}${\color{red}melS}$\stackrel{\text{AR}}{\longrightarrow}$wav               & AAAI20 & 2020.04         \\
		Flow-TTS~\cite{miao2020flow}               & AM   &    {\color{blue}ch/ph}$\stackrel{\text{FF}}{\longrightarrow}${\color{red}melS}$\stackrel{\text{FF}}{\longrightarrow}$wav               & ICASSP20 & 2020.05        \\
		Flowtron~\cite{valle2020flowtron}               & AM       &   {\color{blue}ph}$\stackrel{\text{AR}}{\longrightarrow}${\color{red}melS}$\stackrel{\text{FF}}{\longrightarrow}$wav     &ICLR21 & 2020.05         \\
		Glow-TTS~\cite{kim2020glow}               & AM       & {\color{blue}ph}$\stackrel{\text{FF}}{\longrightarrow}${\color{red}melS}$\stackrel{\text{FF}}{\longrightarrow}$wav               & NeurIPS20& 2020.05        \\
		JDI-T~\cite{lim2020jdi}                  & AM       &  {\color{blue}ph}$\stackrel{\text{FF}}{\longrightarrow}${\color{red}melS}$\stackrel{\text{FF}}{\longrightarrow}$wav             &IS20& 2020.05     \\
		TalkNet~\cite{beliaev2020talknet}                  & AM       &  {\color{blue}ch}$\stackrel{\text{FF}}{\longrightarrow}${\color{red}melS}$\stackrel{\text{FF}}{\longrightarrow}$wav             &arXiv20& 2020.05     \\
		MB MelGAN~\cite{yang2020multi}      & Voc      & {\color{red}melS}$\stackrel{\text{FF}}{\longrightarrow}$wav   & SLT21 & 2020.05        \\
		MultiSpeech~\cite{chen2020multispeech}         & AM & {\color{blue}ph}$\stackrel{\text{AR}}{\longrightarrow}${\color{red}melS}$\stackrel{\text{FF}}{\longrightarrow}$wav              & IS20 & 2020.06        \\
		FastSpeech 2~\cite{ren2021fastspeech}         & AM & {\color{blue}ph}$\stackrel{\text{FF}}{\longrightarrow}${\color{red}melS}$\stackrel{\text{FF}}{\longrightarrow}$wav              & ICLR21& 2020.06        \\
		FastSpeech 2s~\cite{ren2021fastspeech}  & ~~~~E2E &{\color{blue}ph}$\stackrel{\text{FF}}{\longrightarrow}$wav   & ICLR21 & 2020.06         \\
		EATS~\cite{donahue2020end}  & ~~~~E2E & {\color{blue}ch/ph}$\stackrel{\text{FF}}{\longrightarrow}$wav   & ICLR21 & 2020.06  \\
		FastPitch~\cite{lancucki2020fastpitch}   & AM & {\color{blue}ph}$\stackrel{\text{FF}}{\longrightarrow}${\color{red}melS}$\stackrel{\text{FF}}{\longrightarrow}$wav               &  ICASSP21 & 2020.06        \\
		VocGAN~\cite{yang2020vocgan}                 & ~~~~~~~~Voc      & {\color{red}melS}$\stackrel{\text{FF}}{\longrightarrow}$wav          &IS20& 2020.07         \\
		LRSpeech~\cite{xu2020lrspeech}           & AM       &  {\color{blue}ch}$\stackrel{\text{AR}}{\longrightarrow}${\color{red}melS}$\stackrel{\text{FF}}{\longrightarrow}$wav             &KDD20& 2020.08         \\
		SpeedySpeech~\cite{vainer2020speedyspeech}           & AM       &  {\color{blue}ph}$\stackrel{\text{FF}}{\longrightarrow}${\color{red}melS}$\stackrel{\text{FF}}{\longrightarrow}$wav             &IS20& 2020.08         \\
		GED~\cite{gritsenko2020spectral}               & ~~~~~~~~Voc      & {\color{green}ling}$\stackrel{\text{FF}}{\longrightarrow}$wav          & NeurIPS20& 2020.08         \\
		SC-WaveRNN~\cite{paul2020speaker} &  ~~~~~~~~Voc  & {\color{red}melS}$\stackrel{\text{AR}}{\longrightarrow}$wav  & IS20 & 2020.08 \\
		WaveGrad~\cite{chen2020wavegrad}               & ~~~~~~~~Voc      & {\color{red}melS}$\stackrel{\text{FF}}{\longrightarrow}$wav          & ICLR21& 2020.09         \\
		DiffWave~\cite{kong2020diffwave}               & ~~~~~~~~Voc      & {\color{red}melS}$\stackrel{\text{FF}}{\longrightarrow}$wav          & ICLR21& 2020.09         \\
		HiFi-GAN~\cite{kong2020hifi}               & ~~~~~~~~Voc      & {\color{red}melS}$\stackrel{\text{FF}}{\longrightarrow}$wav   & NeurIPS20 & 2020.10         \\
		NonAtt Tacotron~\cite{shen2020non} & AM       & {\color{blue}ph}$\stackrel{\text{AR}}{\longrightarrow}${\color{red}melS}$\stackrel{\text{AR}}{\longrightarrow}$wav              &arXiv20& 2020.10       \\
		Para. Tacotron~\cite{elias2020parallel}      & AM       & {\color{blue}ph}$\stackrel{\text{FF}}{\longrightarrow}${\color{red}melS}$\stackrel{\text{AR}}{\longrightarrow}$wav              &arXiv20& 2020.10     \\
		DeviceTTS~\cite{huang2020devicetts}              & AM       &  {\color{blue}ph}$\stackrel{\text{AR}}{\longrightarrow}${\color{red}Ceps}$\rightarrow$wav              &arXiv20& 2020.10  \\
		Wave-Tacotron~\cite{weiss2020wave}          & ~~~~E2E & {\color{blue}ch/ph}$\stackrel{\text{AR}}{\longrightarrow}$wav  &ICASSP21& 2020.11    \\
		DenoiSpeech~\cite{zhang2020denoising}            & AM       &{\color{blue}ph}$\stackrel{\text{FF}}{\longrightarrow}${\color{red}melS}$\stackrel{\text{FF}}{\longrightarrow}$wav     & ICASSP21 & 2020.12        \\
		EfficientTTS~\cite{miao2020efficienttts}            & AM       &{\color{blue}ch}$\stackrel{\text{FF}}{\longrightarrow}${\color{red}melS}$\stackrel{\text{FF}}{\longrightarrow}$wav     & ICML21 & 2020.12        \\
		EfficientTTS-Wav~\cite{miao2020efficienttts}             & ~~~~E2E       &{\color{blue}ch}$\stackrel{\text{FF}}{\longrightarrow}$wav     & ICML21 & 2020.12        \\
		Multi-SpectroGAN~\cite{lee2020multi} & AM       &{\color{blue}ph}$\stackrel{\text{AR}}{\longrightarrow}${\color{red}melS} $\stackrel{\text{FF}}{\longrightarrow}$wav     & AAAI21 & 2020.12        \\
		LightSpeech~\cite{luo2021lightspeech}            & AM       &{\color{blue}ph}$\stackrel{\text{FF}}{\longrightarrow}${\color{red}melS}$\stackrel{\text{FF}}{\longrightarrow}$wav     & ICASSP21 & 2021.02        \\
		Para. Tacotron 2~\cite{elias2021parallel}    & AM       & {\color{blue}ph}$\stackrel{\text{FF}}{\longrightarrow}${\color{red}melS}$\stackrel{\text{AR}}{\longrightarrow}$wav                 &arXiv21 & 2021.03         \\
		AdaSpeech~\cite{chen2021adaspeech}            & AM       &{\color{blue}ph}$\stackrel{\text{FF}}{\longrightarrow}${\color{red}melS}$\stackrel{\text{FF}}{\longrightarrow}$wav     & ICLR21& 2021.03        \\
		BVAE-TTS~\cite{lee2020bidirectional}            & AM       &{\color{blue}ph}$\stackrel{\text{FF}}{\longrightarrow}${\color{red}melS}$\stackrel{\text{FF}}{\longrightarrow}$wav     & ICLR21& 2021.03        \\
		PnG BERT~\cite{jia2021png}            & AM       &{\color{blue}ph}$\stackrel{\text{AR}}{\longrightarrow}${\color{red}melS}$\stackrel{\text{AR}}{\longrightarrow}$wav     & IS21 & 2021.03        \\
        Fast DCTTS~\cite{kang2021fast} & AM       &{\color{blue}ch}$\stackrel{\text{AR}}{\longrightarrow}${\color{red}melS}$\stackrel{\text{FF}}{\longrightarrow}$wav     & ICASSP21 & 2021.04        \\
        AdaSpeech 2~\cite{yan2021adaspeech}           & AM       &{\color{blue}ph}$\stackrel{\text{FF}}{\longrightarrow}${\color{red}melS}$\stackrel{\text{FF}}{\longrightarrow}$wav     & ICASSP21 & 2021.04        \\
        TalkNet 2~\cite{beliaev2021talknet}              & AM       &  {\color{blue}ch}$\stackrel{\text{FF}}{\longrightarrow}${\color{red}melS}$\stackrel{\text{FF}}{\longrightarrow}$wav             &arXiv21& 2021.04     \\
        Triple M~\cite{lin2021triple}      & AM+Voc       &  {\color{blue}ch}$\stackrel{\text{AR}}{\longrightarrow}${\color{red}melS}$\stackrel{\text{AR}}{\longrightarrow}$wav             &arXiv21& 2021.04     \\
		Diff-TTS~\cite{jeong2021diff}            & AM       &{\color{blue}ph}$\stackrel{\text{FF}}{\longrightarrow}${\color{red}melS}$\stackrel{\text{FF}}{\longrightarrow}$wav     & arXiv21& 2021.04        \\
		Grad-TTS~\cite{popov2021grad}            & AM       &{\color{blue}ph}$\stackrel{\text{FF}}{\longrightarrow}${\color{red}melS}$\stackrel{\text{FF}}{\longrightarrow}$wav     & ICML21 & 2021.05        \\
		Fre-GAN~\cite{kim2021fre}               & ~~~~~~~~Voc      & {\color{red}melS}$\stackrel{\text{FF}}{\longrightarrow}$wav   &IS21  & 2021.06    \\
		VITS~\cite{kim2021conditional} & ~~~~E2E      &{\color{blue}ph}$\stackrel{\text{FF}}{\longrightarrow}$wav  & ICML21 & 2021.06 \\
		AdaSpeech 3~\cite{yan2021adaspeech3}  & AM       &{\color{blue}ph}$\stackrel{\text{FF}}{\longrightarrow}${\color{red}melS}$\stackrel{\text{FF}}{\longrightarrow}$wav     & IS21 & 2021.06        \\
		PriorGrad-AM~\cite{lee2021priorgrad}            & AM       &{\color{blue}ph}$\stackrel{\text{FF}}{\longrightarrow}${\color{red}melS}$\stackrel{\text{FF}}{\longrightarrow}$wav     &arXiv21  & 2021.06        \\
		PriorGrad-Voc~\cite{lee2021priorgrad}               & ~~~~~~~~Voc      & {\color{red}melS}$\stackrel{\text{FF}}{\longrightarrow}$wav   &arXiv21  & 2021.06         \\
		Meta-StyleSpeech~\cite{min2021meta} & AM       &{\color{blue}ph}$\stackrel{\text{FF}}{\longrightarrow}${\color{red}melS}$\stackrel{\text{FF}}{\longrightarrow}$wav     &ICML21  & 2021.06        \\
		\bottomrule
\end{longtable}
\end{center}

\bibliography{main}
\bibliographystyle{plainnat}

\end{document}